\documentstyle[12pt,epsf]{article}
\def \thesection {\arabic{section}.}
\def \thesubsection {\thesection\arabic{subsection}.}

\def \sect #1 {\setcounter{equation} 0\section{#1}}
\def \appendix #1#2 {\par\bigskip\bigskip\noindent
                    {\Large {\bf Appendix {#1}. {#2} }}
                    \def\thesubsection{}
                    \def\theequation{{#1}.\arabic{equation}}
                    \setcounter{equation} 0 \par\bigskip\noindent}
\def \theequation {\thesection\arabic{equation}}   
\def \be  {\begin{equation}}
\def \ee  {\end{equation}}
\def \ba  {\begin{eqnarray}}
\def \ea  {\end{eqnarray}}
\def \baa {\begin{eqnarray*}}
\def \eaa {\end{eqnarray*}}
\def \bb  {\begin{thebibliography}}
\def \eb  {\end{thebibliography}}
\newcommand \ci [1] {\cite{#1}}
\newcommand \bi [1] {\bibitem{#1}}
\def \lab #1 {\label{#1}}
\newcommand\re[1]{(\ref{#1})}
\def \qqquad {\qquad\quad}
\def \qqqquad {\qquad\qquad}
\newcommand\lr[1]{{\left({#1}\right)}}
\newcommand\lrs[1]{{\left[{#1}\right]}}

\newcommand \vev [1] {\langle{#1}\rangle}
\newcommand \VEV [1] {\left\langle{#1}\right\rangle}
\newcommand \ket [1] {|{#1}\rangle}

\def \e {\mbox{e}}
\def \CO {{\cal O}}

\def \CV {{\cal V}}

\renewcommand{\Re}{\mathop{\rm Re}\nolimits}
\newcommand\fra[2]{\mbox{\small $\frac{#1}{#2}$}}

\font\cmss=cmss10 \font\cmsss=cmss10 at 7pt
\def\inbar{\,\vrule height1.5ex width.4pt depth0pt}
\def\IC{\relax\hbox{$\inbar\kern-.3em{\rm C}$}}
\def\IZ{\relax\ifmmode\mathchoice
{\hbox{\cmss Z\kern-.4em Z}}{\hbox{\cmss Z\kern-.4em Z}}
{\lower.9pt\hbox{\cmsss Z\kern-.4em Z}}
{\lower1.2pt\hbox{\cmsss Z\kern-.4em Z}}\else{\cmss Z\kern-.4em Z}\fi}
\def\IR{{\hbox{{\rm I}\kern-.2em\hbox{\rm R}}}}
\def\IP{{\hbox{{\rm I}\kern-.2em\hbox{\rm P}}}}

\def\Im{\hbox{\rm Im}\,}
\newcommand{\as}{\ifmmode\alpha_{\rm s}\else{$\alpha_{\rm s}$}\fi}

\begin{document}

\def\thefootnote{\fnsymbol{footnote}}
\thispagestyle{empty}
\hfill\parbox{50mm}{{\sc ITP--SB--95--25}\par
                         LPTHE--Orsay--95--80 \par     
                         hep-th/9508025  \par
                         July, 1995}
\vspace*{38mm}
\begin{center}
{\LARGE Quasiclassical QCD Pomeron }
\par\vspace*{15mm}\par
{\large G.~P.~Korchemsky}%
\footnote{On leave from the Laboratory of Theoretical Physics,
          JINR, Dubna, Russia}
\footnote{Address after November 1, 1995: LPTHE, Universit\'e de Paris XI,
          b\^at 211, 91405 Orsay, France} 
\par\bigskip\par\medskip

{\em Institute for Theoretical Physics, \par
State University of New York at Stony Brook, \par
Stony Brook, New York 11794 -- 3840, U.S.A.}
\end{center}
\vspace*{15mm}

\begin{abstract}
The Regge behaviour of the scattering amplitudes in perturbative
QCD is governed in the generalized leading logarithmic approximation by the
contribution of the color--singlet compound states of Reggeized gluons.
The interaction between Reggeons is described
by the effective hamiltonian, which in the multi--color limit turns
out to be identical to the hamiltonian of the completely integrable
one--dimensional XXX Heisenberg magnet of noncompact spin $s=0$.
The spectrum of the color singlet Reggeon compound states, --
perturbative Pomerons and Odderons, is expressed by means of the
Bethe Ansatz in terms of the fundamental $Q-$function, which satisfies
the Baxter equation for the XXX Heisenberg magnet. The exact solution
of the Baxter equation is known only in the simplest case of the
compound state of two Reggeons, the BFKL Pomeron. For higher Reggeon
states the method is developed which allows to find its general solution
as an asymptotic series in powers of the inverse conformal weight of
the Reggeon states. The quantization conditions for the conserved
charges for interacting Reggeons are established and an agreement
with the results of numerical solutions is observed. The asymptotic
approximation of the energy of the Reggeon states is defined based on the
properties of the asymptotic series, and the intercept of the three--Reggeon
states, perturbative Odderon, is estimated.
\end{abstract}
\newpage
\def\thefootnote{\arabic{footnote}}
\setcounter{footnote} 0

\newpage

\sect{Introduction}

Recently, a lot of attention has been renewed to the old problem of
understanding QCD Pomeron. The interest was partially inspired by new
exciting experimental results, which confirmed a growth of the
structure function of deep inelastic scattering, $F_2(x,Q^2)$, at small
values of the Bjorken variable $x$ and increasing with energy of the total
hadronic cross-sections, $\sigma_{\rm tot}(s)$. Both phenomena allow us
to test QCD in the extreme limit of high energies and fixed transferred
momenta, the limit in which the famous Regge model emerges \ci{Col}.
Being one of the most popular subjects in 70's, the Regge theory interprets
the increasing of physical cross--sections with the energy by introducing the
concept of the Regge poles. The Regge poles can be classified into different
families according to their quantum numbers and their contribution to the
amplitude of elastic scattering of hadrons $h_1$ and $h_2$ in the Regge
limit, $s \gg -t$, has the form
$
A_{h_1h_2}(s,t)\sim s^{\alpha(t)}\,,
$
with  $\alpha=\alpha(t)$ being universal Regge pole trajectory independent
on particular choice of the scattered hadrons \ci{Col}.

Among all possible families of the Regge poles there is a special one,
with the quantum numbers of vacuum, the so--called Pomeron. The Pomeron
provides the leading contribution to the higher energy behaviour of the
total hadronic cross--section \ci{Col}
$$
\sigma^{\rm tot}_{h_1h_2}(s) = s^{-1} \Im A_{h_1h_2}(s,t=0)
\sim s^{\alpha_{_{\mbox{\tiny \IP}}}(0)-1}
$$
with $\alpha_{\mbox{\tiny \IP}}(0)$ being the Pomeron intercept.
Replacing one of incoming hadrons by virtual photon with invariant mass
$-Q^2$ we may find a similar expression for the asymptotics of the
structure function of deep inelastic scattering $\gamma^*(Q)+h\to X$ in
the Regge limit, $x=Q^2/s \ll 1$ and $Q^2={\rm fixed}$.
There is however an
important difference between these two processes from point of view of
underlying QCD dynamics. The Regge behavior of the total hadronic
cross--sections is essentially nonperturbative phenomenon and it is
governed by the ``soft'' QCD Pomeron \ci{softP} with the intercept
$\alpha_{\mbox{\tiny \IP}}(0)\approx 1.08$.
At the same time, the deep inelastic scattering can be analyzed for
large enough $Q^2$ within the framework of perturbative QCD by using
the operator product expansion, or factorization, and the small$-x$
asymptotics of the structure function is controlled by the ``hard''
Pomeron \ci{softP}, $F_2(x,Q^2) \sim x^{1-\alpha_{_{\mbox{\tiny \IP}}}(0)}$,
with larger intercept $\alpha_{\mbox{\tiny \IP}}(0)\approx 1.38$.

Although the Regge model is extremely successful in describing
a wealth of different data \ci{Col,softP} it is still unclear whether it
can be justified from the first principles of
QCD. Since we do not have a complete understanding of nonperturbative
QCD we have to restrict ourselves to processes involving ``hard''
Pomeron. As an example of such processes, one may consider the
scattering of two onium states each consisting of a heavy quark-antiquark
pair \ci{bfkl,onium}.
The first attempts to describe the ``hard'' Pomeron in QCD
revealed the remarkable property of gluon reggeization \ci{bfkl,Lip1}.
It was found
from the analysis of the Feynman diagram in the leading logarithmic
approximation (LLA), $\as \ln s \sim 1$ and $\as \ll 1$, that interacting
with each other gluons form a new
collective excitation, Reggeon, which according to its contribution
to the scattering amplitude can be interpreted as a Regge pole with
quantum numbers of a gluon. As a result, QCD can be replaced in the
Regge limit by an effective field theory \ci{Gr} in which Reggeons play a
role of new elementary fields while Pomerons appear as compound states
of interacting Reggeons.
The simplest example of such state built from
only two Reggeons, Balitsky--Fadin--Kuraev--Lipatov Pomeron, was
found many years ago \ci{bfkl} and its intercept was calculated in
the LLA as
\be
\alpha_{_{\rm BFKL}}=1+\frac{\as N}{\pi} 4\ln 2
\lab{bfkl}
\ee
where $\as$ is the QCD coupling constant and
$N$ is the number of quark colors. Apart from the BFKL Pomeron there is an
infinite number of another states with vacuum quantum numbers, which
are built from an arbitrary number of Reggeons $n>2$.

To identify higher $n$ Reggeon compound states we have to go beyond
the LLA in calculating the scattering
amplitudes. One possible way of doing this has been proposed in
\ci{CW,Bar}
and is called the generalized leading logarithmic approximation.
In this approximation, the amplitude of the onium-onium scattering, $A(s,t)$,
is given in the Regge limit by the sum of diagrams describing the
propagation of an arbitrary number $n$ of interacting Reggeons in the
$t-$channel
\be
A(s,t)=\sum_{n=2}^\infty \as^{n-2} A_n(s,t)\,.
\lab{A}
\ee
The Reggeon interaction has interesting properties in the
generalized LLA. If one introduces a fictitious time in the $t-$channel,
then at the same moment of ``time'' the interaction occurs only between
two Reggeons (pair--wise interaction) \ci{CW,Bar,KP}.
It conserves the number of
Reggeons (elastic scattering) but changes color and two-dimensional
transverse momenta of Reggeons. As a result, the diagrams contributing
to the amplitude $A_n(s,t)$ have a conserved number $n$ of Reggeons in the
$t-$channel and have a form of generalized ladder diagrams \ci{CW,Bar,KP}.
These diagrams can be effectively resumed using the Bethe--Salpeter
approach \ci{Bar,KP} and the resulting expression for the scattering amplitude
$A_n(s,t)$ can be represented as a sum of contributions of the
$n$ Reggeon compound states propagating in the $t-$channel
\be
A_n(s,t) = i s\sum_{\{q\}} s^{E_{n,\{q\}}}\
        \beta_{n\to h_1}^{\{q\}}(t) \beta_{n\to h_2}^{\{q\}}(t)\,.
\lab{An}
\ee
Here, $E_{n,\{q\}}$ is the energy of the $n$ Reggeon compound state
$\ket{\chi_{n,\{q\}}}$, which satisfies the
(2+1)--dimensional Schrodinger equation, the so--called
Bartels--Kwiecinski--Praszalowicz equation \ci{Bar,KP}
\be
{\cal H}_n\ket{\chi_{n,\{q\}}}
= E_{n,\{q\}}\ket{\chi_{n,\{q\}}}
\lab{BKP}
\ee
with $\{q\}$ being the set of quantum numbers parameterizing all possible
solutions.
The hamiltonian ${\cal H}_n$ corresponds to the elastic pair--wise
interaction of $n$ Reggeons. It acts on 2--dimensional transverse momenta
of Reggeons and describes the evolution of the $n$ Reggeon compound state
$\ket{\chi_{n,\{q\}}}$ in the $t-$channel.
The residue factors in \re{An} measure the coupling of the Reggeon compound 
states to the hadronic state $\ket{\Phi_h}$ and they are defined as 
\be
\beta_{n\to h}^{\{q\}}(t)
= \vev{\Phi_h(t)|\chi_{n,\{q\}}}\,.
\lab{beta}
\ee
For short--distance hadronic states like onium $\beta_{n\to h}^{\{q\}}(t)$ 
can be calculated in perturbative QCD \ci{bfkl,Bar,onium}. Expression \re{An}
for the scattering amplitude is consistent with the Regge
theory predictions provided that we interpret the Regge poles as the
$n$ Reggeon compound states and identify the intercept as their maximal
energy
\be
\alpha(t) - 1 = {\rm max}_{_{\{q\}}} E_{n,\{q\}}\,.
\lab{max}
\ee
Combining relations \re{A} and \re{An} together we
obtain the Regge asymptotics of the scattering amplitude in the generalized
LLA as an infinite sum over all possible $n$ Reggeon compound states
propagating in the $t-$channel. For $n=2$ the first term in the sum \re{A}
corresponds to the BFKL Pomeron.
The next term, $n=3$, is associated with three Reggeon compound states
which belong to the Odderon family of the Regge poles \ci{odd2}. Although
they have a zero color charge, in contrast with the BFKL Pomeron
their charge conjugation is negative, $C=-1$. As a result, they cannot
couple to the hadronic states with $C=1$ like virtual photon in the deep
inelastic scattering, but for the hadronic states with $C=-1$ like proton
they contribute to the total cross sections $\sigma^{\rm tot}_{pp}$
and $\sigma^{\rm tot}_{p\bar p}$. Moreover, their contribution is
responsible for the increasing with the energy of the difference \ci{odd2},
$\Delta\sigma=
\sigma^{\rm tot}_{p\bar p}-\sigma^{\rm tot}_{pp}$. The Regge behaviour
of $\Delta\sigma$
is controlled by the intercept of the Odderon \ci{odd,odd1}, which
despite of a lot of efforts is unknown yet.

As it follows from \re{A}, the contributions of the $n$ Reggeon states
to the scattering amplitude is suppressed by powers of
$\as$ with respect to that of the BFKL Pomeron. Nevertheless, they have
to be taken into account in order to preserve unitarity of the $S-$matrix
of QCD \ci{CW,Bar}.
To derive the higher Reggeon compound states we have to solve the
(2+1)--Schrodinger equation \re{BKP} for
$n$ interacting Reggeons. For $n=2$ it was done
in \ci{bfkl}, but for $n\ge 3$ the problem becomes extremely complicated
partially
due to interaction between color charges of Reggeons. The latter interaction
can be drastically simplified by taking the multi--color limit \ci{Ven},
$N\to\infty$ and $\as N = {\rm fixed}$, and passing by means of
Fourier transformation from two-dimensional transverse momenta of Reggeons,
$k_\perp=(k_1,k_2)$, to two-dimensional impact parameter space,
$b_\perp=(b_1,b_2)$. After these transformations, the Reggeon
hamiltonian takes the following form \ci{Lip2}
\be
{\cal H}_n = \frac{\as N}{4\pi}\lr{H_n + \bar H_n} + \CO(N^{-2})\,.
\lab{RH}
\ee
The hamiltonians $H_n$ and $\bar H_n$ act on
holomorphic and antiholomorphic coordinates of the Reggeons in the
impact parameter space, $z=b_1+ib_2$ and $\bar z=b_1-ib_2$, respectively,
and describe nearest--neighbour interaction between $n$ Reggeons
$$
H_n = \sum_{k=1}^n H(z_k,z_{k+1})\,,\qqquad
\bar H_n = \sum_{k=1}^n H(\bar z_k,\bar z_{k+1})\,,\qqquad
[H_n, \bar H_n] = 0\,,
$$
where $z_k$ and $\bar z_k$ are holomorphic and antiholomorphic coordinates
of the $k-$th Reggeon and  periodic boundary conditions $z_{n+1}=z_1$
and $\bar z_{n+1}=\bar z_1$ are imposed. Here,  $H(z_1,z_2)+
H(\bar z_1,\bar z_2)$ is the interaction hamiltonian of two Reggeons
with coordinates $(z_1,\bar z_1)$ and $(z_2,\bar z_2)$
in the impact parameter space, the so--called BFKL kernel \ci{Lip1,Lip2},
\be
H(z_1,z_2)=-\psi(-J_{12})-\psi(1+J_{12})+2\psi(1)
\lab{psi}
\ee
where $\psi(x)=\frac{d}{dx}\ln\Gamma(x)$ and the operator $J_{12}$ is
defined as a solution of the equation
$$
J_{12} (1+J_{12}) = -(z_1-z_2)^2\partial_1\partial_2
$$
with $\partial_k=\partial/\partial z_k$. The expression
for $H(\bar z_1,\bar z_2)$ is similar to \re{psi}.
Thus, in the multi--color limit, the 2--dimensional Reggeon
hamiltonian ${\cal H}_n$ describing the pair--wise interaction of
$n$ Reggeons turns out to be equivalent to the sum \re{RH}
of two 1--dimensional
mutually commuting hamiltonians $H_n$ and $\bar H_n$. This allows us to
reduce the original (2+1)--dimensional problem \re{BKP}
to the system of two
(1+1)--dimensional Schrodinger equations \ci{Lip2}
\be
H_n \ket{\varphi_{n,\{q\}}} =
\varepsilon_{n,\{q\}} \ket{\varphi_{n,\{q\}}}\,,\qqquad
\bar H_n \ket{\bar\varphi_{n,\{q\}}} = \bar\varepsilon_{n,\{q\}}
\ket{\bar\varphi_{n,\{q\}}}
\lab{H}
\ee
where the wave functions ${\varphi_{n,\{q\}}}$ and ${\bar\varphi_{n,\{q\}}}$
depend only on holomorphic and antiholomorphic coordinates of the Reggeons,
respectively. Then, in the multi--color limit the spectrum of the $n$ Reggeon
states can be found as
$$
E_{n,\{q\}}=\frac{\as N}{4\pi}
\lr{\varepsilon_{n,\{q\}}+\bar\varepsilon_{n,\{q\}}}\,,
\qqqquad
\ket{\chi_{n,\{q\}}} = \ket{\varphi_{n,\{q\}}} \ket{\bar\varphi_{n,\{q\}}}
\,.
$$
Although equations \re{H} are much simpler than the original Schrodinger
equation \re{BKP} it is not obvious that they can be solved exactly for an
arbitrary number of Reggeons. A significant progress has been achieved
recently \ci{FK,Lip}
after it was realized that the system \re{H} of $(1+1)-$dimensional
Schrodinger equations has interesting interpretation in terms
of one--dimensional lattice models \ci{Q}.

Let us consider the one-dimensional lattice with periodic boundary
conditions and the number of sites, $n$, equal to the number of Reggeons.
We parameterize $n$ sites by holomorphic coordinates $z_k$
$(k=1,...,n)$ and introduce the interaction between nearest neighbors
on the lattice with the holomorphic Reggeon hamiltonian $H_n$, \re{psi}. Thus
defined ``holomorphic'' lattice model is described by the same Schrodinger
equation as in \re{H} and it obeys the following remarkable property
\ci{FK,Lip}.
It turns out to be equivalent to the celebrated XXX Heisenberg magnet of
spin $s=0$ corresponding to the principal series representation of the
noncompact conformal $SL(2,\IC)$ group.
The same is true, of course, for
the antiholomorphic Schrodinger equation in \re{H}.
Therefore, the $n$ Reggeon compound states in multi--color QCD share all
their properties with the eigenstates of the XXX Heisenberg magnet defined
on the periodic one--dimensional lattice with $n$ sites and their intercept
\re{max} is closely related to the ground state energy of the magnet.%
\footnote{Indeed, from
point of view of lattice models it is more natural to change a sign of
the Reggeon hamiltonian, ${\cal H}_n$, and interpret $\lr{-{\rm max}\ E_n}$
in \re{max} as a ground state energy of the XXX magnet.}
This result opens a possibility to apply the powerful
methods of exactly solvable models \ci{BA1,BA2,BA3,BA4} to the solution of the
Regge problem in QCD. The first step has been undertaken in \ci{FK,K}
where the generalized Bethe ansatz has been developed for diagonalization
of the Reggeon hamiltonians. The expressions for the $n$ Reggeon wave
functions and their corresponding energies were found in terms of
the fundamental $Q-$function which satisfies the Baxter equation for the
XXX Heisenberg magnet of spin $s=-1$. The solution of the Baxter equation
was found in the simplest case of $n=2$ Reggeon state, or BFKL Pomeron.
For higher Reggeon states the problem becomes much more complicated and
it is still open \ci{MW}. In the present paper we continue the study of the
Baxter equation initiated in \ci{FK,K} and develop a method which allows us
to find its general solutions in the form of asymptotic expansion similar to
the well--known quasiclassical approximation in quantum mechanics.

The paper is organized as follows. In Section 2 we summarize
the Bethe ansatz solution for $n$ Reggeon compound states and introduce
the Baxter equation. We interpret the Baxter equation
as a discrete one--dimensional Schrodinger equation and identify
inverse conformal weight of the Reggeon states as
a small parameter which plays a role of the Planck constant.
The quasiclassical expansion of the solution of the Baxter equation in
powers of this parameter is performed in Section 3. Similar
to the situation in quantum mechanics it leads to the asymptotic series
for the energy of Reggeon states whose properties are studied in Section 4.
The asymptotic expansions for the Reggeon quantum numbers are
derived in Section 5. In Section 6 we use
the obtained results to estimate the energy of higher Reggeon states.
Summary and concluding remarks are given in Section 7.
The analytical properties of the energy of the Reggeon states are 
considered in Appendix A.
Relation between Reggeon states and conformal operators is discussed in
Appendix B. Some useful properties of the asymptotic series are described 
in Appendix C.

\sect{Generalized Bethe Ansatz}

The fact that the system of Schrodinger equations \re{H}
is completely integrable
implies that there exists a family of ``hidden'' holomorphic and
antiholomorphic conserved charges, $\{q\}$ and $\{\bar q\}$, which commute
with the Reggeon hamiltonian and among themselves. Their explicit form can
be found using the quantum inverse scattering method as \ci{Lip,FK}
\be
q_k=\sum_{n\ge i_1>i_2>\cdots >i_k\ge 1}
i^k z_{i_1i_2} z_{i_2i_3} \ldots z_{i_ki_1}
\partial_{i_1}\partial_{i_2} \ldots \partial_{i_k}
\lab{opqn}
\ee
with $z_{ij}\equiv z_i-z_j$
and the expression for $\bar q_k$ is similar. The appearance of these
operators is closely related to the invariance of the Reggeon hamiltonian
\re{RH} under conformal $SL(2,\IC)$ transformations \ci{Lip1}
\be
z \to \frac{az+b}{cz+d}\,,\qquad
\bar z \to \frac{\bar a\bar z+ \bar b}{\bar c\bar z+\bar d}
\lab{ct}
\ee
where $ac-bd=\bar a\bar c-\bar b\bar d=1$. Indeed, we recognize $q_2$ and
$\bar q_2$ as the quadratic Casimir operators of the $SL(2,\IC)$ group while
the remaining conserved charges $\{q_k,\bar q_k\}$, $k=3,...,n$ can be
interpreted as higher Casimir operators. The Reggeon compound states
belong to the principal series representation of the $SL(2,\IC)$ group
and under the conformal transformations \re{ct} they are transformed as
quasiprimary fields with conformal weights $(h,\bar h)$ \ci{BPZ}.

The $n$ Reggeon states diagonalize the operators $\{q,\bar q\}$ and the
eigenvalues of the
conserved charges $q_2,$ $q_3$, $\ldots$, $q_n$ play a role of their
additional quantum numbers.
In particular, the eigenvalues of the quadratic Casimir operators
are related to the conformal weights of the $n$ Reggeon state as
$$
q_2=-h(h-1) \,,\qqquad
\bar q_2 =-\bar h (\bar h-1)\,,\qqquad
\bar q_2=q_2^*\,.
$$
where the possible values of $h$ and $\bar h$
can be parameterized by integer $m$ and
real $\nu$
\be
h=\frac{1+m}2 + i\nu\,,\qquad
\bar h=\frac{1-m}2 + i\nu\,,\qquad
m=\IZ\,,\quad \nu = \IR\,.
\lab{h}
\ee
As to remaining charges, their possible values also become quantized.
The explicit form of the corresponding quantization conditions
is more complicated and will be discussed in Sect.~5.

To find the explicit form of the eigenstates and eigenvalues of the
$n$ Reggeon compound states corresponding to a given set of quantum
numbers $\{q,\bar q\}$ we apply the generalized Bethe ansatz
developed in \ci{FK,K}.
The Bethe ansatz for Reggeon states in multi--color QCD is based on
the solution of the Baxter equation
\be
\Lambda(\lambda)
Q(\lambda)=(\lambda+i)^n Q(\lambda+i) + (\lambda-i)^n Q(\lambda-i)\,.
\lab{Bax}
\ee
Here, $Q(\lambda)$ is a real function of the spectral parameter
$\lambda$, $\Lambda(\lambda)$ is the eigenvalue of the so--called
auxiliary transfer matrix for the XXX Heisenberg magnet of spin $s=0$
\be
\Lambda(\lambda)=2\lambda^n+q_2 \lambda^{n-2} + \ldots + q_n
\lab{Lam}
\ee
and $n$ is the number of Reggeized gluons or, equivalently,
the number of sites of the one-dimensional spin chain.
For a fixed $n$ it is convenient to introduce the function
\be
\widetilde Q(\lambda)=\lambda^n Q(\lambda)
\lab{tilde}
\ee
and rewrite the Baxter equation \re{Bax} as
\be
\widetilde Q(\lambda+i)+\widetilde Q(\lambda-i) - 2\widetilde Q(\lambda)
= \lr{
-\frac{h(h-1)}{\lambda^2}
+ \frac{q_3}{\lambda^3} + \ldots + \frac{q_n}{\lambda^n}}
\widetilde Q(\lambda)\,.
\lab{tildeBax}
\ee
Once we know the function $\widetilde Q(\lambda)$, the energy $E_n$ of
the $n-$Reggeon compound state can be evaluated using the relation
\be
E_n=\frac{\as N}{2\pi} \Re \varepsilon_n(h,q_3,\ldots,q_n)
\lab{En}
\ee
where the holomorphic energy $\varepsilon_n$ is defined as
\be
\varepsilon_n(h,q_3,\ldots,q_n)=i\frac{d}{d\lambda}
\log\frac{\widetilde Q(\lambda-i)}
         {\widetilde Q(\lambda+i)}\Bigg|_{\lambda=0}\,.
\lab{en}
\ee
The expression for the wave function of the $n$ Reggeon states in terms
of the function $\widetilde Q$ can be found in \ci{FK,K} and it will not
be discussed in the present paper.

The Baxter equation \re{tildeBax}
has the following properties \ci{K}. We notice that for
$q_n=0$ it is effectively reduced to a similar equation for the
states with $n-1$ Reggeized
gluons. The corresponding solution, $Q(\lambda)$, gives rise to the degenerate
unnormalizable $n$ Reggeon states with the energy
\be
\varepsilon_n(h,q_3,...,q_{n-1},0)=
\varepsilon_{n-1}(h,q_3,...,q_{n-1})\,.
\lab{dege}
\ee
These states should be excluded from the spectrum of the $n$ Reggeon
hamiltonian and solving the Baxter equation for $n$ Reggeized gluons we
have to satisfy the condition
\be
q_n \neq 0\,.
\lab{neq}
\ee
As a function of the quantum numbers, the holomorphic energy obeys
the relations
\be
\varepsilon_n(h,q_3,...,q_n)
=\varepsilon_n(1-h,q_3,...,q_n)
=\varepsilon_n(h,-q_3,...,(-)^n q_n)\,,
\lab{q3-q3}
\ee
which follow from the symmetry of the Baxter equation \re{tildeBax} under the
replacement $h\to 1-h$ or $\lambda\to-\lambda$ and $q_k\to (-)^k q_k$.
This relation means that the spectrum of the Reggeon hamiltonian
is degenerate with respect to quantum numbers $h$, $q_3$, $...$, $q_n$.
Then, assuming that the ground state of the XXX Heisenberg magnet of
spin $s=0$
is not degenerate we can identify the quantum numbers corresponding to
the maximal value of the Reggeon energy as \ci{K}
\be
h\bigg|_{\rm max} = \frac 12
\lab{1/2}
\ee
and for the states with only even number of Reggeons, $n=2k$,
$$
q_3\bigg|_{\rm max}=q_5\bigg|_{\rm max}=...=q_{2k-1}\bigg|_{\rm max}=0\,.
$$
For the states with odd number of Reggeons the latter condition is not
consistent with \re{neq}.

We notice that the conformal weight $h$ enters as a parameter into the
Baxter equation \re{Bax} and, in general, one is interesting to find the
solutions of \re{Bax} only for its special values \re{h}.
Moreover, since the equation \re{tildeBax} is invariant under the
replacement $h\to 1-h$ we may restrict ourselves to the region
\be
\Re h \ge  1/2\,,
\lab{integer}
\ee
or equivalently $m\ge 0$ in \re{h}.

Our strategy in solving the Baxter equation will be the following \ci{FK,K}. 
We will first try to solve \re{Bax} for integer positive values of the 
conformal weight $h=\IZ_+$ and then analytically continue the resulting 
expression for the energy \re{en} to all possible values \re{h} including 
the most interesting one \re{1/2}. It is clear that this procedure is
ambiguous since for any integer $Z$ one could multiply the energy by a factor 
$\exp(i\pi Z h)$. However, according to the Carlson's theorem \ci{Col}, the 
holomorphic energy is uniquely defined in the
region \re{h} by its values at the integer positive values of the conformal 
weight $h$ provided that the function $\varepsilon_n$ is regular in \re{h} 
and its asymptotics at infinity is $\CO\lr{\e^{c |h|}}$ with $c < \pi$. As 
we show in Appendix A, these two conditions are indeed satisfied.

The Baxter equation \re{Bax} has two linear independent solutions and in
order to select only one of them we impose the additional condition on the
function $Q(\lambda)$ for $h=\IZ_+$
\be
Q(\lambda)\stackrel{\lambda\to\infty}{\sim} \lambda^{h-n}\,.
\lab{as}
\ee
For integer positive conformal weight, $h\ge n$, the solution,
$Q(\lambda)$, of the Baxter equation \re{Bax} under the additional condition
\re{as} is given by a polynomial of degree $h-n$ in the spectral
parameter $\lambda$. This implies in turn that $\widetilde Q(\lambda)$
is a polynomial of degree $n$ in $\lambda$, which can be expressed
in terms of its roots $\lambda_1$, $...$,
$\lambda_n$ as follows \ci{FK,K}
\be
\widetilde Q(\lambda) = \prod_{i=1}^h (\lambda-\lambda_i)
                      = \exp\lr{\sum_{i=1}^h \ln (\lambda-\lambda_i)}
                      \stackrel{\lambda\to\infty}{\sim} \lambda^h\,.
\lab{roots}
\ee
Substituting \re{roots} into \re{tildeBax} and putting $\lambda=\lambda_k$
we obtain that the roots satisfy the Bethe equation for the XXX spin chain
of spin $s=0$
\be
1 =\prod_{j=1 \atop j\neq k}^n
\frac{\lambda_k-\lambda_j+i}{\lambda_k-\lambda_j-i}\,,
\qqquad k=1,...,n\,.
\lab{Be}
\ee
There is, however, one additional important condition on the roots, which
follows from the definition \re{tilde}. Comparing \re{roots} with \re{tilde}
we find that $\lambda_1$, $...$, $\lambda_{h-n}$ are roots of
$Q(\lambda)$ and $\lambda_{h-n+1}=\ldots=\lambda_h=0$.
This means that among all the solutions of the Bethe equation \re{Be}
we have to select only those which have the $n-$time degenerate root
$\lambda_k=0$.

The explicit solution of the Baxter equation \re{Bax}
is known only for $n=2$
\ci{FK,K}
\be
Q_{n=2}(\lambda;h)= i^h h(1-h)\
{_3F_2}\left({1+h,2-h,1-i\lambda \atop 2, 2}; 1
\right)
\lab{3F2}
\ee
and it can be identified as a continuous symmetric Hahn orthogonal
polynomial \ci{Sl}. Substituting the solution \re{3F2}
into \re{en} we obtain the holomorphic energy of
$n=2$ Reggeon states as \ci{K}
\be
\varepsilon_2(h)=-4\left[\psi(h)-\psi(1)\right]
\lab{en-n=2}
\ee
where $\psi-$function was defined in \re{psi}. We substitute
$\varepsilon_2(h)$ into \re{En} and analytically continue the
result from integer $h$ to all possible complex values \re{h}
$$
E_2(h)=-2\frac{\as N}{\pi}
\Re\left[\psi\lr{\frac{1+|m|}2+i\nu}-\psi(1)\right]\,.
$$
This relation coincides with the well--known expression \ci{bfkl}
for the energy of
the $n=2$ Reggeon compound state, the BFKL Pomeron. The maximum
value of the energy,
$E_2^{\rm max}=\frac{\as N}{\pi}4\ln 2$, is achieved at $h=1/2$
and it is in agreement with \re{bfkl}, \re{max} and \re{1/2}.

Solving the Baxter equation \re{Bax}
for $n\ge 3$ one may try to look for the solution
as a linear combination of the $n=2$ solutions \ci{K},
$Q(\lambda)=\sum_k C_k(h,q_3,...,q_n) Q_{n=2}(\lambda;k)$.
Using the properties of $Q_{n=2}$ as orthogonal polynomials we obtain
that the coefficients $C_k(h,q_3,...,q_n)$ satisfy the multi-term
recurrent relations. Most importantly, the quantum numbers
$q_3$, $\ldots$, $q_n$ corresponding to the polynomial solutions of
the Baxter equation turned out to be quantized and their possible
values as well as the values of the roots $\lambda_k$ were found to
be real \ci{K}
\be
\Im \lambda_k = \Im q_k =0\,.
\lab{real}
\ee
Although the explicit form of the recurrent relations for $\{C_k\}$
is known \ci{K} and it is not difficult to solve them numerically
for lowest values of the conformal weight $h$ and then find the quantized
charges $\{q_k\}$ and the energy $\varepsilon_n$,
the analytical expression for the solution $Q(\lambda)$ similar
to \re{3F2} is still missing. At the same time, the results of numerical
solution of the Baxter equation for $n=3$ and $n=4$ presented in
\ci{K} indicate that the analytical solution should exist.
In this paper we will find such kind of solution for an arbitrary $n$ in the
special limit, $h\gg 1$. We will show that it is possible to
develop an asymptotic expansion for the solutions of the Baxter equation, as
well as for the energy of the $n$ Reggeon states, in powers of $1/h$.

\subsection{The origin of the quasiclassical approximation}

Let us rescale in \re{tildeBax} the spectral parameter as
$\lambda\to\lambda h$ in order to get rid of a large factor $h(h-1)$ in the
r.h.s.\ of the Baxter equation \re{tildeBax}
as $h\to\infty$. Then, we introduce new
functions $f(\lambda)$ and $\Phi(\lambda)$
\be
f(\lambda)=\exp\lr{h \Phi(\lambda)}
=\e^{-h\ln h} \ \widetilde Q(h\lambda)\,,
\lab{f}
\ee
where according to the definition \re{roots}
\be
\Phi(\lambda)=\frac1{h}\sum_{i=1}^h \ln\lr{\lambda-\frac{\lambda_i}{h}}\,.
\lab{Phi}
\ee
Substituting \re{f} into \re{en}
we find the following expression for the holomorphic energy of the
$n$ Reggeon state in terms of the function $\Phi(\lambda)$
\be
\varepsilon_n=-i\left[\Phi'\lr{\frac{i}{h}}-\Phi'\lr{-\frac{i}{h}}\right]
\,,
\lab{ener}
\ee
where prime denotes a derivative with respect to the spectral parameter
$\lambda$.

The main reason for considering the function $\Phi(\lambda)$
is that in the limit $h\to\infty$ we expect the scaling
$\lambda_i=\CO(h)$ and $\Phi(\lambda) \sim h^0$, which allows us
to expand $\Phi(\lambda)$ in powers of $1/h$ as
\be
\Phi(\lambda)= \Phi_0(\lambda)
             + \frac1{h} \Phi_1(\lambda)
             + \frac1{h^2} \Phi_2(\lambda)
             + ...
\lab{decPhi}
\ee
with all functions $\Phi_i(\lambda)$ being $h-$independent.

The substitution of the relation \re{f} into \re{tildeBax}
yields the following equation for $f(\lambda)$
\be
f\lr{\lambda+\frac{i}{h}}+f\lr{\lambda-\frac{i}{h}}-2f(\lambda)
=-V(\lambda) f(\lambda)\,,
\lab{Sch}
\ee
where the notation was introduced for the ``potential''
\be
V(\lambda)=\frac1{\lambda^2}\lr{1-\frac1{h}}-\frac1{\lambda^3}\frac{q_3}{h^3}
-\ldots-\frac1{\lambda^n}\frac{q_n}{h^n}\,.
\lab{pot}
\ee
Taking a naive limit $h\to\infty$ in \re{Sch} we find that $f(\lambda)$
satisfies the one--dimensional Schrodinger equation,
$-\fra1{h^2}\frac{d^2}{d\lambda^2}f=2m(E-V(\lambda))f$,
for the wave function
of a particle with mass $m=1/2$ and energy $E=0$ moving in the singular
potential $V(\lambda)$. In this equation $1/h$ plays a role of the Planck
constant $\hbar$ and the relation \re{f} has a form of the
WKB ansatz,
$f(\lambda)=\exp\lr{\frac1{\hbar} \Phi(\lambda)}$.
This suggests that $h\to\infty$ limit of the Baxter equation \re{Sch}
should be closely related to the quasiclassical approximation $\hbar\to 0$ in
quantum mechanics. 


We would like to notice that there is an intriguing relation between the 
Bethe Ansatz solution of our model and that of the one--dimensional Toda 
chain \ci{GP}. In both cases after the separation of variables one has to 
solve the Baxter equation which has the form of one--dimensional discretized 
Schr\"odinger equation \re{Sch} for a particle in an external potential. 
The only difference between models is in the form of the potential and in 
different boundary conditions which one imposes on the solution of the Baxter
equation.

The condition for all terms in the expression for the potential \re{pot}
to have the same behaviour at large $h$ implies the following scaling
of the holomorphic quantum numbers
\be
\frac{q_3}{h^3}=\hat q_3
               + \frac1{h} q_3^{_{(1)}}
               + \frac1{h^2} q_3^{_{(2)}}
               + \CO\lr{\frac1{h^3}}\,,\quad ...\ \,, \quad
\frac{q_n}{h^n}=\hat q_n
               + \frac1{h} q_n^{_{(1)}}
               + \frac1{h^2} q_n^{_{(2)}}
               + \CO\lr{\frac1{h^3}}\,.
\lab{q3q4}
\ee
with $\hat q_3\equiv q_3^{_{(0)}}$, $...$,
$\hat q_n\equiv q_n^{_{(0)}}$ and all coefficients $q_k^{_{(j)}}$
being $h-$independent. This leads to
the decomposition of the potential $V(\lambda)$ similar to that in
\re{decPhi}
\be
V(\lambda) = V_0(\lambda)
             + \frac1{h} V_1(\lambda)
             + \frac1{h^2} V_2(\lambda)
             + \ldots
\lab{decV}
\ee
where
\ba
V_0(\lambda) &=& \lambda^{-2}-{\hat q_3}\lambda^{-3}
-\ldots-{\hat q_n}\lambda^{-n}
\,,
\lab{V0}
\\[3mm]
V_1(\lambda) &=& - \lambda^{-2} - q_3^{_{\rm (1)}}\lambda^{-3}
-\ldots-q_n^{_{\rm (1)}}\lambda^{-n}
\,,\quad
\nonumber
\\[3mm]
V_2(\lambda) &=& - q_3^{_{\rm (2)}}\lambda^{-3}
-\ldots-q_n^{_{\rm (2)}}\lambda^{-n}
\,,\qquad \ldots
\nonumber
\ea
Finally, to solve \re{Sch} and find the functions $\Phi_i(\lambda)$
we have to substitute \re{decPhi} into \re{f} and \re{Sch}, take into
account the decomposition \re{decV} and equate the coefficients in front of
different powers of $1/h$ in the both sides of the discrete
Schrodinger equation \re{Sch}.

\sect{Leading large$-h$ approximation}

As the conformal weight $h$ increases the number of roots
of the solution of the Baxter equation \re{roots} and \re{Phi}
also increases.
According to \re{real}, the roots are real and in the limit $h\to\infty$
it is convenient to introduce their distribution density on the real
axis as
\be
\rho(\sigma)=\frac1{h}\sum_{i=1}^h \delta\lr{\sigma-\frac{\lambda_i}{h}}
\lab{rho}
\ee
with the normalization condition
\be
\int_{-\infty}^\infty d\sigma\ \rho(\sigma) = 1\,.
\lab{norm}
\ee
In the large$-h$ limit $\rho(\sigma)$ becomes a smooth positive definite
function of a real parameter $\sigma$. It vanishes outside some (not
necessary connected) finite interval ${\cal S}$ on the real axis
\be
\rho(\sigma) \neq 0\,, \qqqquad \mbox{for}\quad\sigma\in
{\cal S}=[\sigma_1,\sigma_2]\cup [\sigma_3,\sigma_4]\cup \ldots \cup
[\sigma_{2k-1},\sigma_{2k}]\,.
\lab{S}
\ee
with the number $k$ depending on the ``potential'' $V(\lambda)$, or
equivalently on the quantum numbers $q_3$, $...$, $q_n$.
The explicit form of $\rho(\sigma)$ can be obtained from the Baxter
equation \re{Sch}
in the limit $h\to\infty$ as follows. Using \re{Phi} and \re{rho}
we get
$$
\Phi(\lambda)=\int_{{\cal S}} d\sigma\ \rho(\sigma) \ln(\lambda-\sigma)
$$
and after differentiation of the both sides with respect to $\lambda$
\be
\Phi'(\lambda)=\int_{{\cal S}} d\sigma\ \frac{\rho(\sigma)}{\lambda-\sigma}
\stackrel{\lambda\to\infty}{\sim} \frac1{\lambda}\,,
\lab{Phi'}
\ee
where integration is performed along the finite interval \re{S}.
The last relation implies that $\Phi'(\lambda)$ is an analytical function of
the spectral parameter $\lambda$ in the complex $\lambda-$plane
with the cut along the finite interval ${\cal S}$ and singularity at
$\lambda=\infty$. Taking discontinuity of $\Phi'(\lambda)$
across the cut we can reconstruct the distribution density of roots as
\be
\rho(\sigma)=\frac{i}{2\pi}\left[
\Phi'(\sigma+i0)-\Phi'(\sigma-i0)
\right]\,.
\lab{disc}
\ee
Similar to \re{decPhi}, the distribution density
can be expanded in powers of $1/h$
\be
\rho(\sigma)=\rho_0(\sigma)+\frac1{h}\rho_1(\sigma)
+\frac1{h^2}\rho_2(\sigma)+ ... \,.
\lab{decrho}
\ee
Here, the functions $\rho_k(\sigma)$ are defined using \re{disc} as a
discontinuity of the functions $\Phi_k(\lambda)$ and they satisfy
the normalization conditions,
$\int_{\cal S} d\sigma \rho_0(\sigma)=1$ and
$\int_{\cal S} d\sigma \rho_k(\sigma)=0$ for $k\ge 1$, which
follow from \re{norm}.

To find the solution of the Baxter equation \re{Sch}
in the leading large$-h$ limit we use the identity
\be
\frac{f(\lambda\pm i/h)}{f(\lambda)} \equiv
\exp(h\left[\Phi(\lambda\pm i/h)-\Phi(\lambda)\right])
=\exp(\pm i \Phi_0'(\lambda)+\CO(1/h))
\lab{exp}
\ee
and keep the leading $1/h-$terms in the both side of \re{Sch} to get
\be
4\sin^2\lr{\frac12\Phi_0'(\lambda)} = V_0(\lambda)\,.
\lab{LLA}
\ee
The equation has two different solutions for $\Phi_0'(\lambda)$ but only
one of them satisfies the additional condition \re{Phi'} as
$\lambda\to\infty$,
\be
\Phi_0'(\lambda)=-2i\ln\left[
\sqrt{1-\frac14 V_0(\lambda)} + i\sqrt{\frac14 V_0(\lambda)}
\right]\,.
\lab{Phi-LLA}
\ee
Notice that $\Phi_0'(\lambda)$ can be defined from \re{LLA} up to a constant
$4\pi\IZ$ which does not contribute, however, neither to the energy \re{ener},
nor to the distribution density \re{disc}.

Let us consider the properties of $\Phi_0'(\lambda)$ in the complex
$\lambda-$plane. Taking into account that the potential $V_0(\lambda)$ is a
real function of $\lambda$, we find from \re{Phi-LLA} that $\Phi_0'(\lambda)$
is not analytical in the regions of $\lambda$, in which the arguments of the
square roots change their sign, that is for
$V_0(\lambda) > 4$ or $V_0(\lambda) < 0$, and in compact form
\be
|V_0(\lambda)-2| > 2\,,\qquad \lambda \in {\cal S}\,.
\lab{crit}
\ee
Solving this inequality we find the interval ${\cal S}$ on the real
axis across which the function $\Phi_0'(\lambda)$ has a discontinuity
and, as a consequence of \re{disc}, the distribution density of roots \re{S}
is different from zero. The same result can be easily deduced from
\re{LLA} as follows. In order to have roots on ${\cal S}$, the function
$f(\lambda)$ should be oscillating on this interval or equivalently
$\Phi_0(\lambda)$ should have an imaginary part for $\lambda\in {\cal S}$.
If $\lambda$ is outside ${\cal S}$, then
the function $\Phi_0(\lambda)$ is real and it follows from \re{LLA} that
$0 \le V_0(\lambda) \le 4$.

Once we solved \re{crit} and obtained the explicit form of ${\cal S}$, we
substitute \re{Phi-LLA} into \re{disc}, evaluate the discontinuity of
$\Phi_0'(\lambda)$ across the interval ${\cal S}$ and get the distribution
density as
\be
\rho_0(\sigma)=\frac2{\pi}\ln
\lr{\sqrt{1-\frac14 V_0(\sigma)}+\sqrt{-\frac14 V_0(\sigma)}}
\,,\qquad
\mbox{for $\qquad V_0(\sigma) <0$}
\lab{rho2}
\ee
and
\be
\rho_0(\sigma)=\frac2{\pi}\ln
\lr{\sqrt{\frac14 V_0(\sigma)-1}+\sqrt{\frac14 V_0(\sigma)}}
\,,\qquad
\mbox{for $\qquad V_0(\sigma) > 4$}\,.
\lab{rho1}
\ee
These expressions define the distribution density of roots in the leading
large$-h$ limit for an arbitrary number of Reggeons,
$n\ge 2$.

\subsection{Critical values of quantum numbers}

\begin{figure}[ht]
   \vspace*{-1cm}
   \epsfysize=9cm
   \epsfxsize=10cm
   \centerline{\epsffile{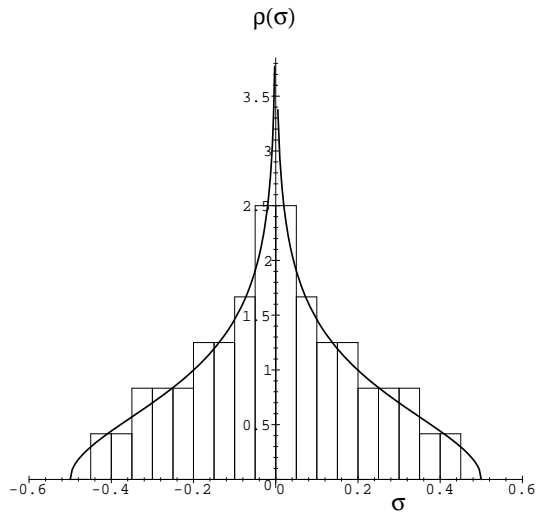}}
   \vspace*{-1.5cm}
\caption{\label{root-n=2} The histogram of the normalized roots
of the solution of the $n=2$ Baxter equation for $h=50$
versus the distribution density $\rho(\sigma)$ in the leading
large$-h$ limit.}
\end{figure}

Let us consider the special case $n=2$, in which we can compare
\re{f} and \re{Phi-LLA} with the exact solution \re{3F2} of the
Baxter equation. For $n=2$ we substitute $V_0(\lambda)=\lambda^{-2}$
into \re{crit} and find the explicit form of ${\cal S}$ as
$$
\rho_0(\sigma) \neq 0\,,\qquad \mbox{for}\quad
{\cal S}=[-\fra12,\fra12]\,.
$$
Then, it follows from the definition \re{rho}
that the zeros of $Q_{n=2}(\lambda;h)$ are located on the interval
$-h/2 < \lambda_i < h/2$. Substituting $V_0(\lambda)$ into \re{rho1} we
obtain the distribution density of zeros of the $n=2$ solutions of the
Baxter equation in the leading large$-h$ limit as
\be
\rho_0(\sigma)=\frac1{\pi}
\ln\frac{1+\sqrt{1-4\sigma^2}}{1-\sqrt{1-4\sigma^2}}
\,,
\qquad
\sigma \in {\cal S}=[-\fra12,\fra12]\,.
\lab{rho-n=2}
\ee
One can check that this expression satisfies the normalization condition
\re{norm}. To test \re{rho-n=2} we use the explicit expression
\re{3F2} for $Q_{n=2}(\lambda;h)$ at $h=50$, find numerical values of
all 48 normalized roots, $\{\lambda_k/h\}$, and compare
on fig.~\ref{root-n=2} the histogram of
their distribution with \re{rho-n=2}.%
\phantom{\ref{root-n=2}}
\vspace*{-4mm}
\par
\begin{figure}[htb]
   \vspace*{-1cm}
   \centerline{
   \epsfysize=9cm\epsfxsize=10cm\epsffile{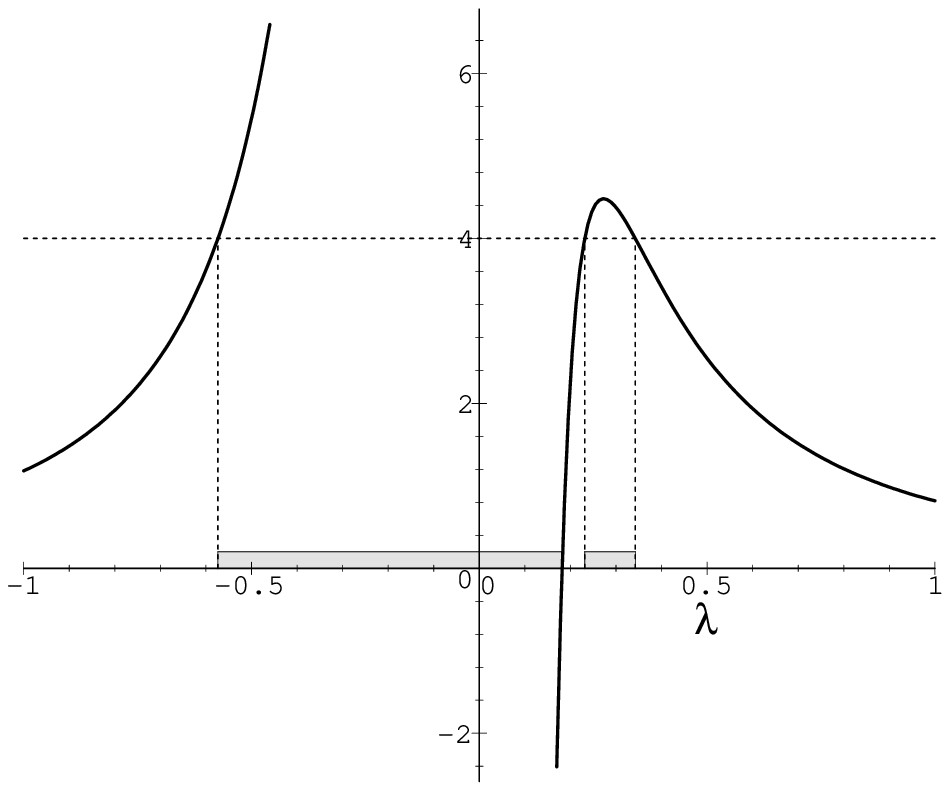}
   \hspace*{-2cm}
   \epsfysize=9cm\epsfxsize=10cm\epsffile{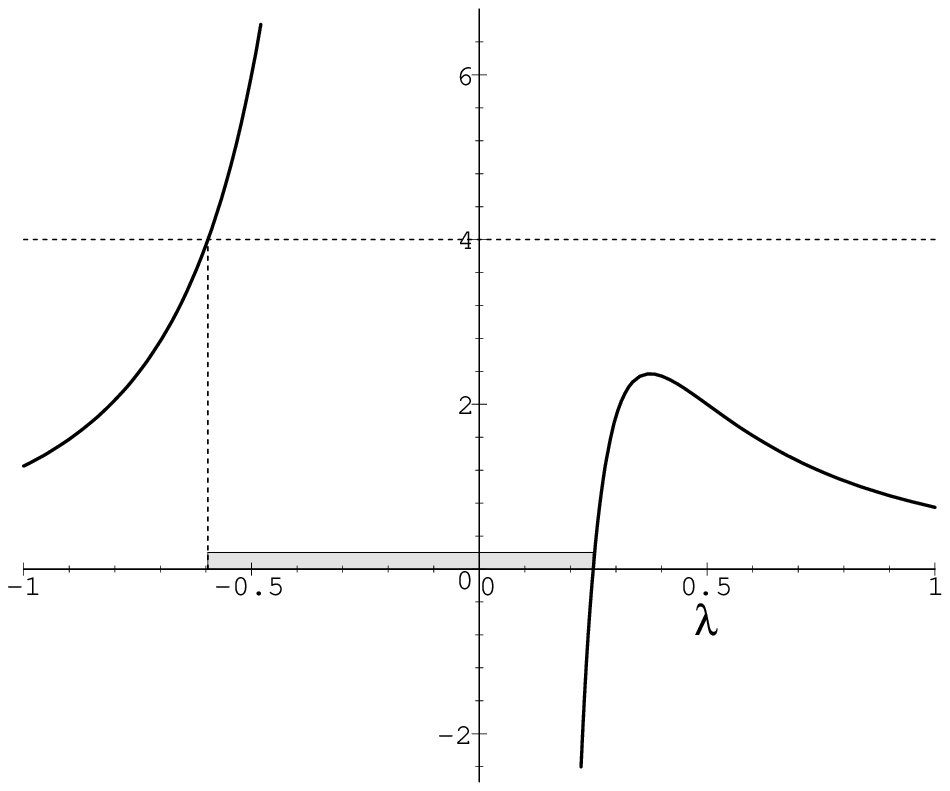}}
   \vspace*{-1cm}
   \centerline{ (a) \hspace*{7.3cm} (b)}
\caption{\label{poten} Possible forms of the potential $V_0(\lambda)$
entering into $n=3$ Baxter equation which lead to two different solutions
of \re{crit}: (a) $\hat q_3 = 0.18 < q_3^*$
and 
(b) $\hat q_3 = 0.25 > q_3^*$.
The shadow area represents the support ${\cal S}$.
}
\unitlength=1mm
\begin{picture}(0,0)(0,0)
\put(51.5,90){\makebox(0,0)[cc]{${\scriptstyle V_0(\lambda)}$}}
\put(132,90){\makebox(0,0)[cc]{${\scriptstyle V_0(\lambda)}$}}
\end{picture}
\vspace*{-0.5cm}
\end{figure}
\begin{figure}[hbt]
   \vspace*{-1cm}
   \centerline{
   \epsfysize=8cm\epsfxsize=10cm\epsffile{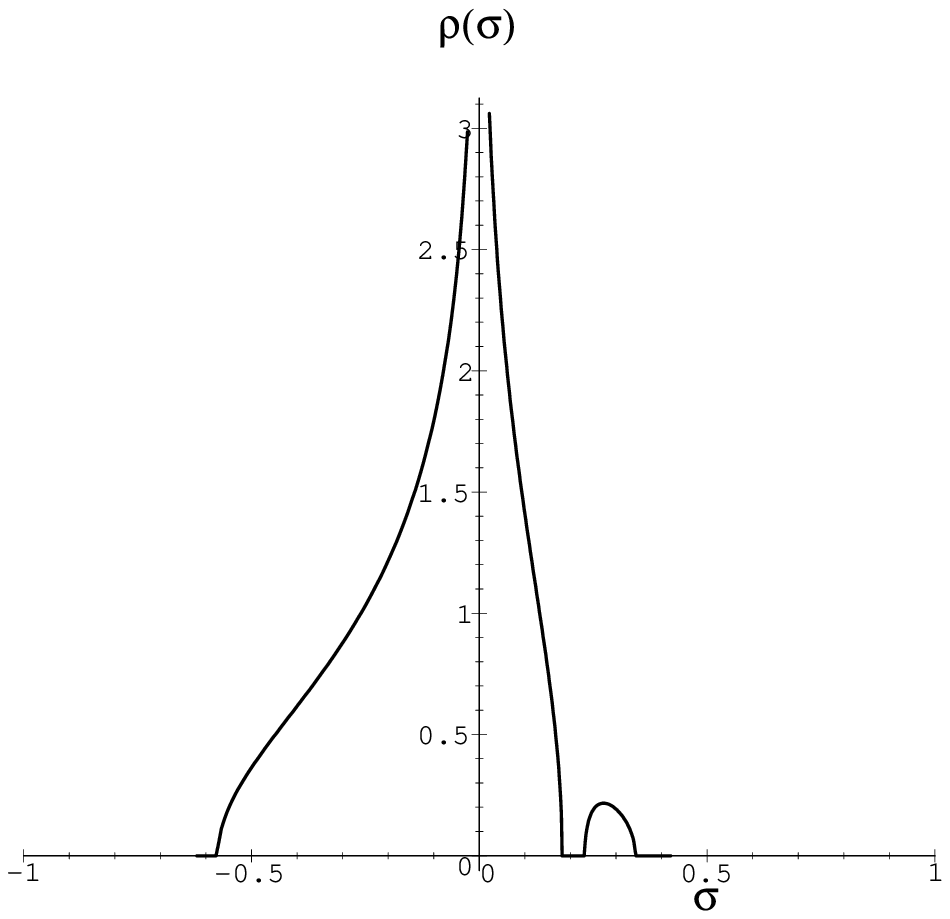}
   \hspace*{-2cm}
   \epsfysize=8cm\epsfxsize=10cm\epsffile{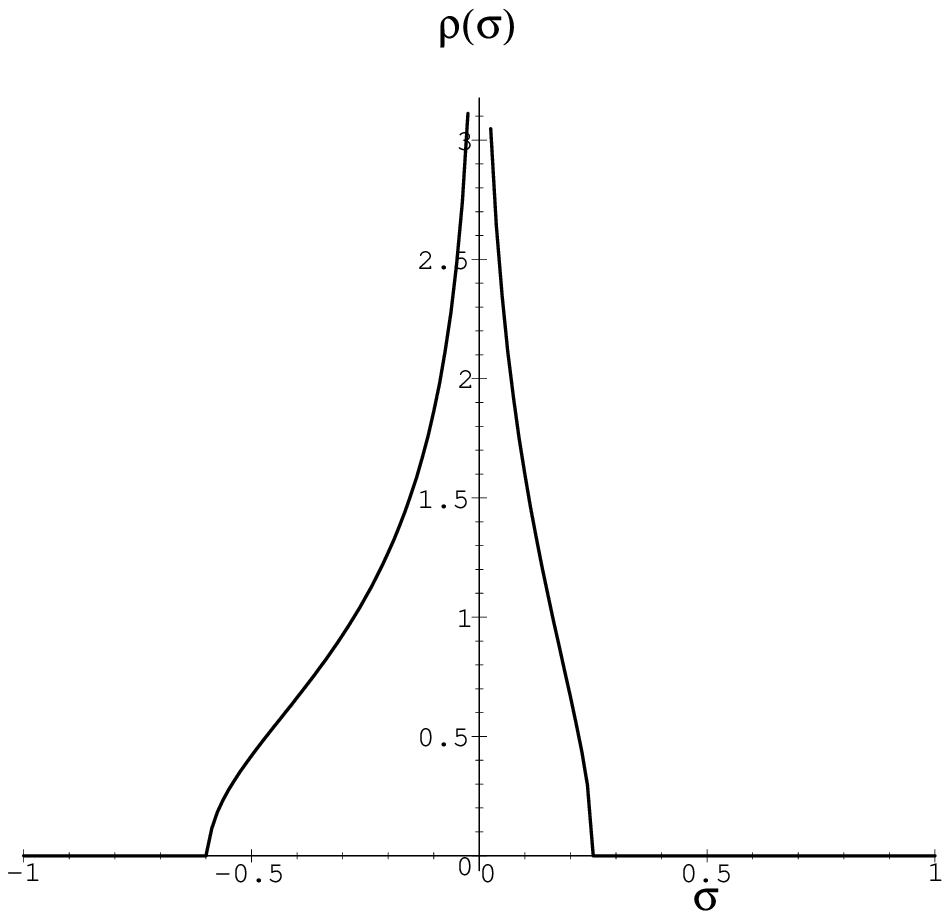}}
   \vspace*{-1cm}
   \centerline{ (a) \hspace*{7.3cm} (b)}
\caption{\label{root-dis} Two different forms of the distribution density
for the $n=3$ Baxter equation: (a) $\hat q_3 < q_3^*$ and
(b) $\hat q_3 \ge q_3^*$, corresponding to the potentials of
fig.~\ref{poten} (a) and (b), respectively.}
\end{figure}

Let us generalize our analysis and consider the possible solutions of
\re{crit} for $n=3$ Reggeon states. For $n=3$ the potential \re{V0} is
given by $V_0(\lambda)=\lambda^{-2}-\hat q_3\,\lambda^{-3}$ and, depending
on the value of the rescaled quantum number $\hat q_3$, there are two
different solutions of \re{crit} shown on fig.~\ref{poten}.%
\phantom{\ref{poten}}
For $\hat q_3^2 < (q_3^*)^2$ the equation $V_0(\lambda)=4$ has
three real roots $\lambda=\sigma_1,\sigma_3,\sigma_4$, while for
$\hat q_3^2 \geq (q_3^*)^2$ there is only one real root,
$\lambda=\sigma_1$. The ``critical'' value of the quantum number
$$
(q_3^*)^2=\frac1{27}
$$
can be found as a solution of the system of equations,
$V_0'(\lambda^*)=0$ and $V_0(\lambda^*)=4$. Notice that in both
cases, $\hat q_3^2 < (q_3^*)^2$ and $\hat q_3^2 \geq (q_3^*)^2$,
the equation $V_0(\lambda)=0$ has only one solution
$\sigma_2=\hat q_3$. Then, taking for simplicity $\hat q_3 >0$ we
identify the finite support ${\cal S}$ for the distribution
density of roots at $n=3$ as
\be
{\cal S}=[\sigma_1,\sigma_2]\cup [\sigma_3,\sigma_4] \,,\qquad
\mbox{for $\quad 0 < \hat q_3 \le \frac1{\sqrt{27}}$}
\lab{gap}
\ee
with
$
\sigma_1 < 0 < \sigma_2=\hat q_3 < \sigma_3 < \frac32 \hat q_3 < \sigma_4
$
and
\be
{\cal S}=[\sigma_1,\sigma_2] \,,\qquad
\mbox{for $\quad\hat q_3 > \frac1{\sqrt{27}}$}
\lab{nogap}
\ee
with $\sigma_1 < 0 < \sigma_2 = \hat q_3$. The generalization of these
relations to the case $\hat q_3 < 0$ is obvious. The distribution densities
corresponding to \re{gap} and \re{nogap} are shown on fig.~\ref{root-dis}(a)
and (b), respectively.
\phantom{\ref{root-dis}}

\begin{figure}[hbt]
   \vspace*{-1.5cm}
   \centerline{
   \epsfysize=9cm\epsfxsize=10cm\epsffile{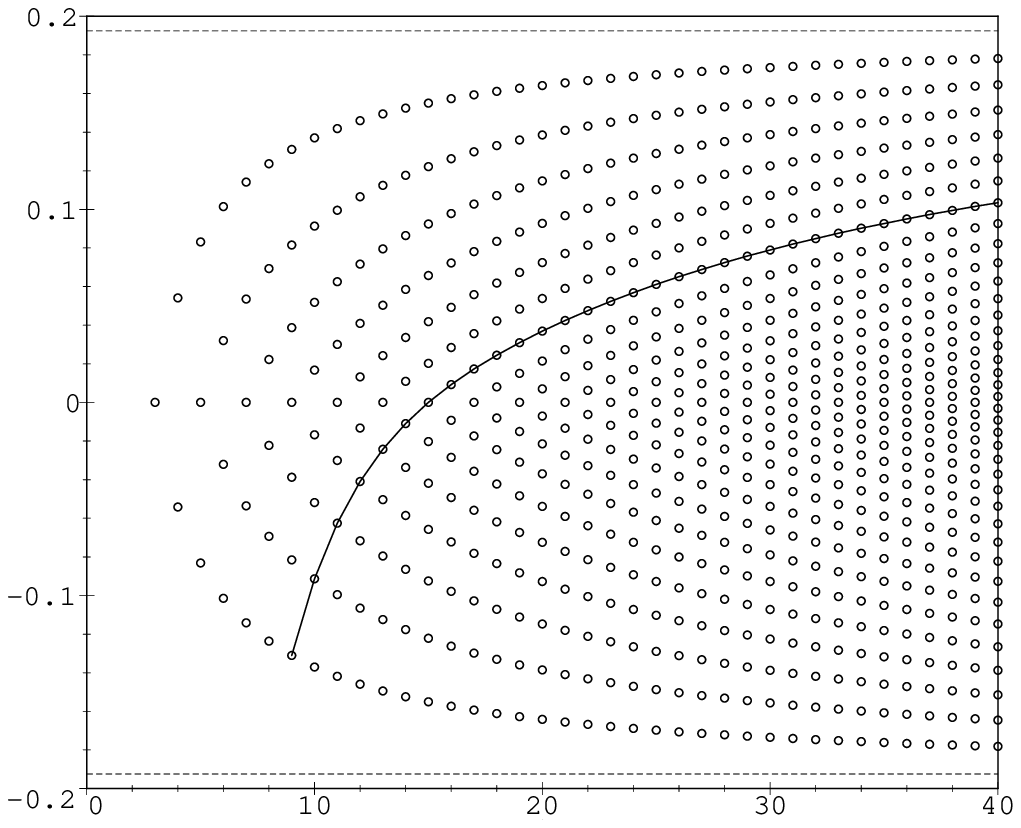}
   \hspace*{-1.5cm}
   \epsfysize=9cm\epsfxsize=10cm\epsffile{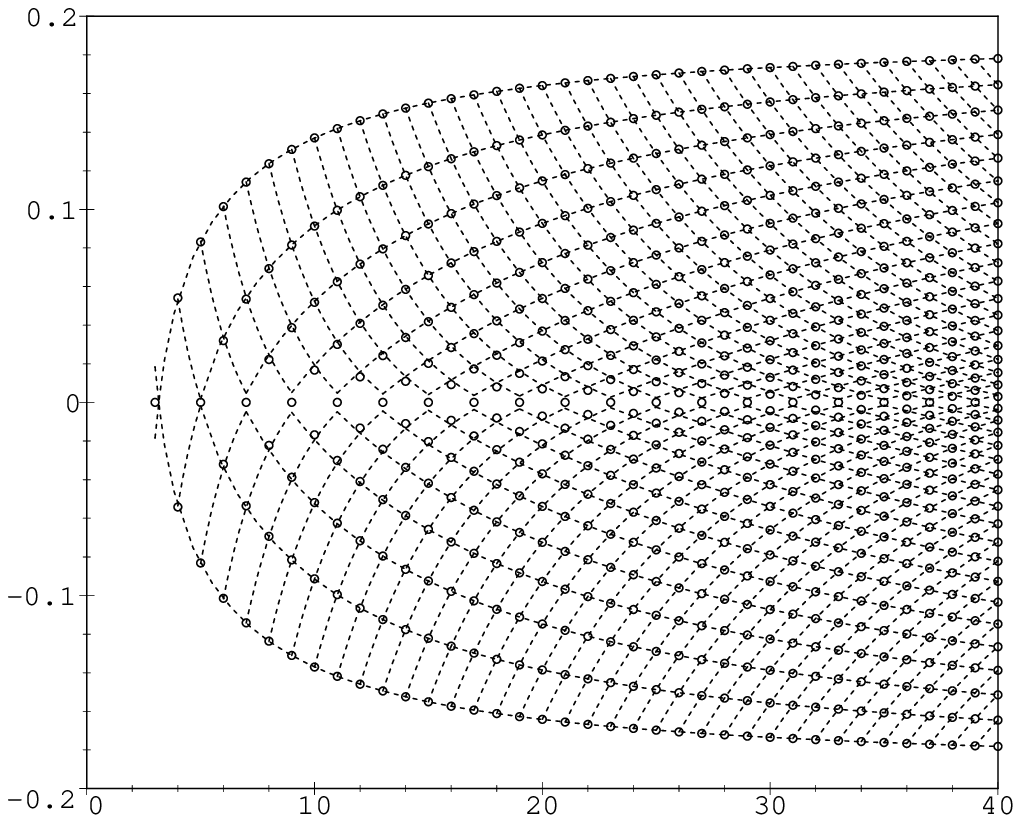}}
   \vspace*{-1cm}
   \centerline{ (a) \hspace*{7.8cm} (b)}
\caption{\label{q3hat} The distribution of quantized
$q_3$ for different values of the conformal weight
$3 \le h \le 40$: (a) The results
of the numerical solution of the $n=3$ Baxter equation. The dashed line
represents the upper and lower limits for $q_3$ in \re{crit-q3}. The solid
line indicates one of the curves to which quantized $q_3$ belong.
(b) One--parametric families of curves defined in \re{Apm} and \re{Bpm}.}
\unitlength=1mm
\begin{picture}(0,0)(0,0)
\put(65,35){\makebox(0,0)[cc]{$h$}}
\put(3,88){\makebox(0,0)[cc]{${q_3}/{h^3}$}}
\put(151,35){\makebox(0,0)[cc]{$h$}}
\put(89,88){\makebox(0,0)[cc]{${q_3}/{h^3}$}}
\end{picture}
\vspace*{-5mm}
\end{figure}

It is important to note that, in contrast with the previous case,
for $n=3$ there exists a ``critical''
value of the quantum number $\hat q_3 = \pm \frac1{\sqrt{27}}$,
at which the gap in the distribution of the roots vanishes and \re{gap} is
replaced by \re{nogap}, or equivalently fig.~\ref{root-dis}(a)
is replaced by fig.~\ref{root-dis}(b). One might expect that
this value separates two different ``phases'' of the Baxter equation
and its solution should have different properties for
$\hat q_3^2 \le \frac1{27}$ and $\hat q_3^2 > \frac1{27}$.
To show that this is really the case we recall
that we are interesting in polynomial solutions of the Baxter equation
\re{roots} under the additional condition that they should have $n-$time
degenerate root $\lambda_k=0$. Till now we did not satisfy this condition
and ``good'' solutions could appear in our analysis together with
``unwanted'' ones.
In order to get rid of them we have to impose additional constraints on
the solutions of the Baxter equation in the large $h$
limit. Our assumption is that for $n=3$ this can be achieved by
restricting the possible values of the quantum number $q_3$ to lie in the
region
\be
-\frac1{\sqrt{27}} \le \hat q_3 \le \frac1{\sqrt{27}}\,.
\lab{crit-q3}
\ee
This condition is equivalent to the statement that for the
$n=3$ Reggeon compound states the distribution density of
roots, $\rho(\sigma)$,
should have the support ${\cal S}$ given by \re{gap} and
consisting of two connected intervals (see fig.~\ref{root-dis}(a)).
To confirm our assumption \re{crit-q3} we present on fig.~\ref{q3hat}
the results (see eq.~(6.46) and fig.~8 in \ci{K}) of the numerical
solution of the quantization conditions for $q_3/h^3$, corresponding
to $n=3$ and integer conformal weight $3 \le h \le 40$. We find that
all numerically found quantized values of $q_3/h^3$ are in perfect
agreement with \re{crit-q3}. They lie inside the strip defined by
\re{crit-q3} and for large $h$ their maximal/minimal values
asymptotically approach the critical values
$q_3/h^3=\pm\frac1{\sqrt{27}}$. Due to the symmetry $q_3\to -q_3$ of
the Baxter equation, \re{q3-q3},
the quantized $q_3$ are distributed symmetrically
with respect to the axis $q_3=0$. Moreover, looking on fig.~\ref{q3hat} it is
difficult do not notice
that the possible values of $q_3$ for different $h$ try to form
a family of one--parametric curves in the two--dimensional
$(h,q_3)-$plane which can be labeled by integer $k=0,1,...\ $.
An example of such a curve is shown on fig.~\ref{q3hat}(a).
The curves start at the points with $h=k+3$ and $q_3$ taking
the absolute minimal value, cross the axis $q_3=0$
at $h=2k+3$ and for $h\to\infty$ they asymptotically approach the
maximal value inside the strip \re{crit-q3}. This suggests that
all possible values of the quantum number $q_3$ can be parameterized
in the following form
\be
\frac{q_3}{h^3}= a_0(k)+ \frac1{h} a_1(k) + \frac1{h^2} a_2(k) + \CO(h^{-3})
\,, \qquad k=0,\,1, ...
\lab{fam}
\ee
where $a_j(k)$ are some unknown coefficients depending on integer $k$.
Comparing this ansatz with \re{q3q4} we find that $a_0(k)=\hat q_3$
and all remaining coefficients $a_1(k)$, $a_2(k)$, $...$ correspond to
higher $1/h-$terms in the expansion \re{q3q4}. To test \re{fam}
we have to evaluate the nonleading $1/h$ corrections to the
quantum number $q_3/h^3$ in \re{q3q4}. This will be done in Sect.~5.
\phantom{\ref{q3hat}}
\par
Let us generalize the condition \re{crit-q3} for the Baxter equation
at $n>3$.
For an arbitrary $n$ the solutions of \re{crit} have the following
properties. Since $V_0(\lambda) \sim 1/\lambda^2 \to 0_+$ as
$\lambda\to\infty$ the interval ${\cal S}$ does not contain infinity and
has a finite size. Then, the potential is singular at the origin,
$V_0(\lambda)\sim -\hat q_n/\lambda^n \to \infty$ as $\lambda\to 0$, and,
according to \re{crit}, the point $\sigma=0$ is always inside ${\cal S}$.
In general, depending on the value of the quantum numbers $\hat q_3$,
$...$, $\hat q_n$, the interval ${\cal S}$ may consist of up to $n-1$
connected intervals. We require that the number of connected intervals
inside ${\cal S}$ should take a maximal possible value, $n-1$, for the
polynomial solutions of the Baxter equation.%
\footnote{This condition has its analog in the Toda chain \ci{Toda,Guz}. 
To separate the variables and reduce the Bethe Ansatz to the solution
of the Baxter equation, one performs the canonical transformation and 
introduces the set of $n-1$ generalized coordinates $x_1$, $...$, $x_{n-1}$
\ci{Skl}. Then, the $n-1$ closed intervals inside ${\cal S}$ correspond to the 
regions of the classical motion of $x_1$, $...$, $x_{n-1}$.
} 
The necessary
condition for this is that the equations $V_0(\lambda)=0$ and
$V_0(\lambda)=4$ should have $n-2$ and $n$ {\it real\/} roots, respectively.
The solution of these constraints leads to the quantization
conditions on $\hat q_3$, $\ldots$, $\hat q_n$
similar to \re{crit-q3}. To satisfy them
the quantized values of the conserved charges should lie inside the domain
in the $n-2$ dimensional space of all possible
$\hat q_3$, $\ldots$, $\hat q_n$.
The boundary of the domain defines the ``critical'' $n-3$ dimensional
hypersurface
\be
\Sigma_n(q_3^*,\ldots,q_n^*)=0
\lab{si}
\ee
which separates different phases of the Baxter equation.
For $n=3$ we have $\Sigma_3(q_3^*)=(q_3^*)^2-27$ and the quantized
$q_3$ belong to the interval \re{crit-q3}. Let us find the explicit form
of the quantization conditions for $n=4$.

\begin{figure}[hbt]
   \vspace*{-1cm}
   \centerline{
   \epsfysize=9cm\epsfxsize=10cm\epsffile{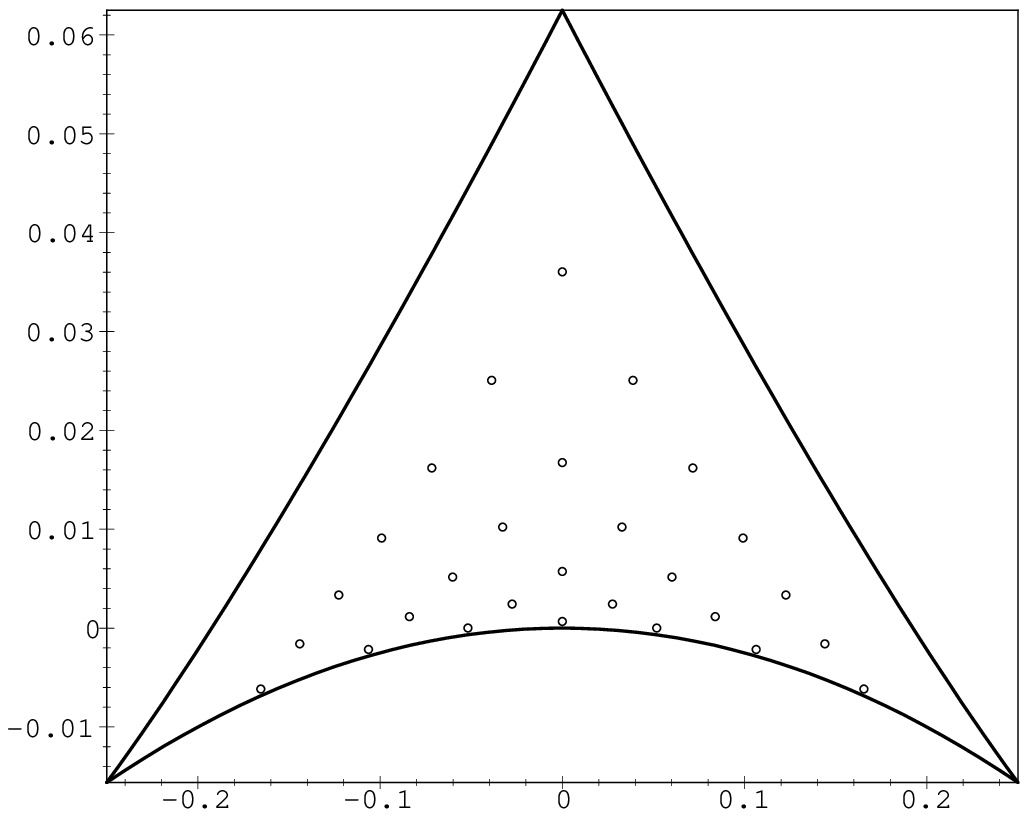}}
   \vspace*{-1.2cm}
\caption{\label{domain} The distribution of the quantized
$q_3$ and $q_4$ corresponding to the polynomial solutions of
the $n=4$ Baxter equation. The quantum numbers lie inside the
generalized triangle defined in \re{crit-q4}.}
\unitlength=1mm
\begin{picture}(0,0)(0,0)
\put(87.5,22){\makebox(0,0)[cc]{${q_3/h^3}$}}
\put(42,57){\makebox(0,0)[cc]{${q_4/h^4}$}}
\end{picture}
\end{figure}

For $n=4$ the potential is given by $V_0(\lambda)=\lambda^{-2}-\hat q_3\,
\lambda^{-3}-\hat q_4\,\lambda^{-4}$ and in order for the equation
$V_0(\lambda)=0$ to have two real roots we have to require that
$\hat q_4 \ge -\frac14 \hat q_3^2$. Under this condition the function
$V_0(\lambda)$ has two extremums -- two local maximums for $\hat q_4>0$ and
one maximum and one minimum for $\hat q_4 <0$. Their positions,
$\lambda=\lambda_{\rm max}$ or $\lambda=\lambda_{\rm min}$, can be found as
solutions of the equation $\frac{d V_0}{d\lambda}=0$. Then, one can easily
check that the second condition on $V_0(\lambda)$, namely that
$V_0(\lambda)=4$ has four real solutions, becomes equivalent to the
requirement $V_0(\lambda_{\rm max}) \ge 4$. After some algebra we find
that the
second constraint on $\hat q_3$ and $\hat q_4$ can be represented
together with the first one in the following form
\be
-4 \hat q_4\ \le\ \hat q_3^2\ \le\ \frac 89
\frac{\hat q_4\lr{\sqrt{48\hat q_4+1}-2}^2}{\sqrt{48\hat q_4+1}-1}\,.
\lab{crit-q4}
\ee
These inequalities define the two-dimensional domain on the
$(\hat q_3,\hat q_4)-$plane shown on fig.~\ref{domain}. Using the
results of numerical solutions of the $n=4$ Baxter equation \ci{K}
we plot on fig.~\ref{domain} the numerical values of quantized
$(q_3/h^3,q_4/h^4)$ corresponding to the conformal weight $h=10$
(see fig.~10 in \ci{K}). We observe a complete agreement of the numerical
results with \re{crit-q4}. Moreover, the quantized values of $q_3$ and
$q_4$ are distributed on fig.~\ref{domain} in a regular way, but
similar to the previous case $n=3$, in order to describe their ``fine
structure'' we need to know nonleading $1/h$ corrections to both quantum
numbers.
\phantom{\ref{domain}}
\par
The one--dimensional boundary of the domain \re{crit-q4} defines the
``critical'' values of the quantum numbers. The range of quantized
charges $\hat q_3$ and $\hat q_4$ can be found from \re{crit-q4} as
$$
-\frac14 \le \hat q_3 \le \frac14\,, \qqqquad
-\frac1{64} \le \hat q_4 \le \frac1{16}\,.
$$
We notice that taking $\hat q_4=0$ in \re{crit-q4} we obtain the relation
$\hat q_3^2 \le 1/{27}$, which we identify as the condition \re{crit-q3}
on $\hat q_3$ for $n=3$. In the same manner, the point $\hat q_3=\hat q_4=0$
on fig.~\ref{domain} corresponds to the possible range of the quantum numbers
for the $n=2$ Baxter equation. Continuing this analogy, we may conclude that
the $n-2$ dimensional domain of the quantized charges
$\hat q_3$, $\ldots$, $\hat q_n$ corresponding to the $n$ Reggeon states
lies on an intersection of the hyperplane $\hat q_{n+1}=0$ with
the $n-1$ dimensional domain of $\hat q_3$, $\ldots$, $\hat q_n$,
$\hat q_{n+1}$ corresponding to the $(n+1)$ Reggeon states.
The same property can be expressed in terms of \re{si} as
$
\Sigma_n(q_3^*,\ldots,q_n^*)=\Sigma_{n+1}(q_3^*,\ldots,q_n^*,q_{n+1}^*=0)\,.
$

\subsection{Divergences of quasiclassical approximation}

Let us use the quasiclassical expansion \re{decPhi} of the solution of
the Baxter equation to evaluate the holomorphic energy \re{ener}.
In order to apply \re{ener} we need to know the
behaviour of $\Phi'(\lambda)$ at $\lambda=\pm i/h$. In the limit
$h\to\infty$, the points $\lambda=\pm i/h$ move toward the origin from
different sides of the imaginary axis and the energy \re{ener} is different
from zero because the point $\lambda=0$ belongs to the interval
${\cal S}$. Using \re{Phi-LLA} and \re{V0} we find the asymptotic behaviour of
the function $\Phi_0'(\lambda)$ for $\lambda\to 0$ at the both sides of
the cut as
\be
\Phi_0'(\lambda\pm i0) \stackrel{\lambda \to 0}{\rightarrow}
\mp i \ln\frac{\hat q_n}{\lambda^n} + ...
\lab{Phi-LLA-as}
\ee
and applying \re{disc} we conclude that the density of roots is
logarithmically divergent at the origin (see figs.~\ref{root-n=2}
and \ref{root-dis})
\be
\rho(\sigma)  \stackrel{\sigma \to 0}{\rightarrow}  \frac1{\pi}
\ln\frac{\hat q_n}{\sigma^n} + {\cal O}\lr{\sigma}\,.
\lab{rho-as}
\ee
Substituting \re{Phi-LLA-as} into \re{ener} we obtain
the following expression for the holomorphic energy
\be
\varepsilon_n = -2 \ln (\hat q_n h^n) \approx -2 \ln q_n \,,
\qquad \mbox{as $h\to \infty$}\,.
\lab{en-LLA}
\ee
Thus, the holomorphic energy of the $n$ Reggeon compound states scales
in the limit $h\to\infty$ as a logarithm of the ``higher'' quantum number
$q_n$.%
\footnote{A somewhat similar result was obtained in \ci{Lip2}.}
We stress that \re{en-LLA} was found in the leading large$-h$ limit
and in order to justify this expression we have to estimate the level at
which next--to--leading $1/h-$corrections to the energy appear.
Naively one might expect that nonleading corrections to $\varepsilon_n$
will arise at the ${\cal O}(1/h)$ level. However, as we will show in a
moment, this is not true due to singular behaviour of the potential
$V_0(\lambda)$ at $\lambda=0$ and, as a consequence, the quasiclassical
expansion of the energy does not work properly. Indeed, beyond the leading 
large$-h$ approximation the energy \re{ener} gets
contribution from $\Phi_1$, $\Phi_2$, $\ldots$, which in the quasiclassical
approximation are proportional to the derivatives of the
function $\Phi_0(\lambda)$ (see eq.~\re{next} below).
In particular, $\frac{d\Phi_k}{d\lambda} \sim
\frac {d^{k+1} \Phi_0}{d\lambda^{k+1}} \sim \lambda^{-k}$ as
$\lambda\to 0$ and
the contribution of $\Phi_k$ to the holomorphic energy can be estimated
using \re{ener} as $\varepsilon_n \sim \lambda^{-k} h^{-k} = \CO(h^0)$
for $\lambda=\pm i/h$ and it is of the same order as the leading order
contribution \re{en-LLA}.
Thus, in order to get an expansion of the holomorphic energy in powers of
$1/h$ it is not enough to restrict ourselves by the leading order result
\re{en-LLA}.

\sect{Beyond the leading order}

To find the solution of the Baxter equation \re{Sch}
beyond the leading $1/h$ order
we have to keep nonleading terms $\Phi_1$, $\Phi_2$, $\ldots$ in
\re{decPhi} and expand further the r.h.s.\ of \re{exp} in
powers of $1/h$. Substituting the expansion for
$f(\lambda\pm \fra{i}{h})/f(\lambda)$
into the Baxter equation \re{Sch} and comparing
the coefficients in front of powers of $1/h$ in the both sides of
the equation we obtain the following system of equations
\ba
\Phi_1'(\lambda) &=&
      -\frac12 \Phi_0'' \cot \Phi_0' + \frac{V_1(\lambda)}{2\sin \Phi_0'}
\nonumber
\\
\Phi_2'(\lambda) &=&
       \frac16 \Phi_0''' +\frac12 \Phi_0'' \Phi_1'
         + \cot \Phi_0' \lr{\frac18 (\Phi_0'')^2-\frac12 (\Phi_1')^2
                             -\frac12 \Phi_1''}
         + \frac{V_2(\lambda)}{2\sin \Phi_0'}\,,\quad ...
\lab{next}
\ea
where prime denotes a derivative with respect to the spectral parameter
$\lambda$, the function $\Phi_0(\lambda)$ was defined in \re{LLA} and
\re{Phi-LLA} and the potentials $V_1$, $V_2$, $...$ were introduced in
\re{decV} and \re{V0}. Integrating these relations with the boundary
conditions $\Phi_k(\infty) = 0$ one can get the functions $\Phi_1(\lambda)$,
$\Phi_2(\lambda)$, $...$ and then reconstruct the solution of the Baxter
equation using \re{f} and \re{decPhi}.

Since we would like to use the solution of \re{next} to find the energy of
the $n$ Reggeon states \re{ener} it is of most interest for us to consider
the behaviour of $\Phi_k(\lambda)$ at small $\lambda=\CO(h^{-1})$.
However, as it follows from \re{next} and \re{Phi-LLA-as}, the expansion
for the function $\Phi(\lambda)$ in powers of $1/h$ is not well defined for
small $\lambda$ due to singular behaviour of the potential at the origin,
$V(\lambda) \sim 1/\lambda^n$ as $\lambda\to 0$. Indeed, as was anticipated
at the end of previous section, the functions $\Phi_k'(\lambda)$ are
proportional to the higher derivatives of $\Phi_0(\lambda)$. For small
$\lambda$
one can use \re{Phi-LLA-as} to find from \re{next} the following general
form of their asymptotic behaviour as $\lambda\to 0$
\ba
\Phi_0'(\lambda) &=& a_{-1}\ln(i\lambda) + a_0 + a_1 i\lambda
                   + a_2 (i\lambda)^2 + \CO(\lambda^3)
\nonumber
\\[2mm]
\Phi_1'(\lambda) &=& b_0/(i\lambda) + b_1 + b_2 i\lambda + \CO(\lambda^2)
\lab{Phi-exp}
\\[2mm]
\Phi_2'(\lambda) &=& c_0/(i\lambda)^2 + c_1/(i\lambda) + c_2 + \CO(\lambda)\,,
\qquad ...
\nonumber
\ea
where the coefficients depend on whether $\lambda$ approaches the origin
from above or below the cut ${\cal S}$. To calculate the holomorphic energy
\re{ener} we substitute $\lambda=\pm i/h$ into \re{Phi-exp} and evaluate
$\Phi'(\lambda)$ using \re{decPhi}. In particular, $\Phi'(-i/h)$ has the
following expansion
\be
\Phi'(-i/h)= -a_{-1} \ln h + (a_0+b_0+c_0+\ldots) + (a_1+b_1+c_1+\ldots)/h
           + (a_2+b_2+c_2+\ldots)/h^2 + \CO(1/h^3)
\lab{expe}
\ee
where dots denote the contributions of the functions $\Phi_k$ with $k \ge 3$.
We conclude that different terms in the large$-h$ expansion \re{decPhi}
give the contributions to $\Phi'(\pm i/h)$ of the same order in $1/h$.
As a consequence, in order to calculate the holomorphic energy $\varepsilon_n$
we have to find an effective way to resum the contributions of
$\Phi_k'(\lambda)$ at small $\lambda=\CO(1/h)$ to all orders $k$.

\subsection{Leading order resummation}

Let us try to simplify the system of equations \re{next} in the limit of
small $\lambda$. As was shown in Sect.~3, the function $\Phi_0'(\lambda)$
has a cut along the interval ${\cal S}$ of the real axis and for
$\lambda\to 0$ we have to consider separately two cases when $\lambda$
approaches the origin from above and below the cut, $\Im \lambda > 0$ and
$\Im \lambda < 0$, respectively. Using \re{Phi-LLA-as} we obtain the
following identities
\be
\exp(i \Phi_0'(\lambda)) \stackrel{\lambda \to 0}{=}
\left\{
{
\frac{\hat q_n}{\lambda^n}
\lr{1+\CO(\lambda)}
\,,\qquad \mbox{for $\Im \lambda > 0$}
\atop
\frac{\lambda^n}{\hat q_n}
\lr{1+\CO(\lambda)}
\,,\qquad \mbox{for $\Im \lambda < 0$}
}
\right.
\lab{small}
\ee
Without lack of generality we may concentrate on the first case,
$\Im \lambda >0$, and apply the relation \re{small} in the form
$\exp(i\Phi_0') \gg \exp(-i\Phi_0')$ to simplify the trigonometric
functions entering into the system \re{next} as
\be
\sin(\Phi_0')=\frac1{2i}\exp(i\Phi_0')(1+\CO(\lambda^{2n}))\,,\qqquad
\cot(\Phi_0')=i + \CO(\lambda^{2n})\,.
\lab{app}
\ee
Then, in the limit $\lambda\to 0$ and $\Im \lambda > 0$
the system of equations \re{next} is replaced by
\ba
\Phi_1'(\lambda) &=& -\frac{i}2 \Phi_0'' + i V_1(\lambda) \e^{-i\Phi_0'}
                + \CO(\lambda^{2n-1})\,,
\lab{next-app}
\\
\Phi_2'(\lambda)&=&\frac16 \Phi_0''' +\frac12 \Phi_0'' \Phi_1'
         +  {\frac{i}8 (\Phi_0'')^2-\frac{i}2 (\Phi_1')^2
                             -\frac{i}2 \Phi_1''}
         + i V_2(\lambda)\e^{-i\Phi_0'}
         + \CO(\lambda^{2n-2})\,,\quad ...
\nonumber
\ea
where $\Phi_0'$ can be defined from \re{LLA} as
$\exp(i\Phi_0')=2-V_0(\lambda)+\CO(\lambda^{2n})$.
One easily checks that performed approximation introduces ambiguities into
the definition of the functions $\Phi_k'(\lambda)$
at the level of $\CO(\lambda^{2n-k})$ corrections,
$\delta \Phi_1' \sim \lambda^{2n-1}$, $\delta \Phi_2' \sim \lambda^{2n-2}$,
$...\ $. Therefore substituting these relations into \re{ener} and
\re{decPhi} and putting
$\lambda=\pm i/h$ one can find the expression for the holomorphic energy with
the following accuracy
$$
\delta \varepsilon_n \sim \delta \Phi_0' + \frac1{h} \delta \Phi_1'
+ \frac1{h^2}  \delta \Phi_2' \sim \frac 1{h^{2n}}\,.
$$
This means that using the approximation \re{app} and \re{next-app}
one can define the holomorphic energy up to $\CO(h^{-2n})$ terms.

The reason why we are interesting in the approximation \re{app}
is that instead of solving the infinite set of equations \re{next-app}
for $\Phi_k'$ one can find a closed expression for the function
$\Phi'(\lambda)$. Our observation is that for small $\lambda$ the
Baxter equation \re{Sch} can be effectively replaced by a new
equation
\be
\frac{f(\lambda+i/h)}{f(\lambda)}=2-V(\lambda)\,,\qquad
\mbox{for $\lambda\to 0$ and $\Im\lambda >0$}
\lab{next-1}
\ee
and
\be
\frac{f(\lambda-i/h)}{f(\lambda)}=2-V(\lambda)\,,\qquad
\mbox{for $\lambda\to 0$ and $\Im\lambda < 0$} \,,
\lab{next-2}
\ee
which reproduces the relations \re{next-app} and which can be
solved exactly. A simplest way to show this is to use \re{small} and \re{exp}
and to take into account that for small $\lambda$ in the Baxter equation
\re{Sch} we have
$f(\lambda-i/h) \ll f(\lambda+i/h)$ for $\Im \lambda > 0$ and
$f(\lambda-i/h) \gg f(\lambda+i/h)$ for $\Im \lambda < 0$.%
\footnote{Since in both cases the points $\lambda\pm i/h$ should lie on the
same side from the real axis, $\lambda$ cannot approach the origin too
close, $|\Im \lambda| \ge 1/h$.}

Let us consider the relation \re{next-1} and use the ansatz \re{f} to rewrite
it as
$$
\exp\left[h\lr{\e^{\frac{i}{h}\partial}-1}\Phi(\lambda)\right]=2-V(\lambda)\,.
$$
where $\partial\equiv\frac{\partial}{\partial\lambda}$.
Its solution has a form
\be
i\Phi'(\lambda)=\frac{\frac{i}{h}\partial}
{\exp(\frac{i}{h}\partial)-1}
\ln\lr{2-V(\lambda)}\,,\qquad
\mbox{for $\Im\lambda >0$}\,.
\lab{Phi-1}
\ee
As a check, we substitute $\Phi=\sum_{k\ge 0} h^{-k} \Phi_k$ and
$V=\sum_{k\ge 0} h^{-k} V_k$ into the both sides of \re{Phi-1},
compare the coefficients in front of powers of $1/h$ and reproduce
\re{next-app}. An analog of \re{Phi-1} for $\Im \lambda < 0$ can be
obtained from \re{next-2} as
\be
i\Phi'(\lambda)=\frac{\frac{i}{h}\partial}
{\exp(-\frac{i}{h}\partial)-1}
\ln\lr{2-V(\lambda)}\,,\qquad
\mbox{for $\Im\lambda < 0$}\,.
\lab{Phi-2}
\ee
Substituting \re{Phi-1} and \re{Phi-2} into \re{ener} and putting
$\lambda=\pm i/h$ we derive after some algebra the following expression
for the holomorphic energy of the $n$ Reggeon state
\be
\varepsilon_n=-\frac{\partial}{\exp(\partial)-1}
\ln \lr{\left[2-V\lr{\frac{i\lambda}{h}}\right]
\left[2-V\lr{-\frac{i\lambda}{h}}\right]}
\Bigg|_{\lambda=1} + \CO(h^{-2n})
\lab{en-next}
\ee
and taking into account that the potential \re{decV} is a real function
of $\lambda$
\be
\varepsilon_n=
-2\Re \frac{\partial}{\exp(\partial)-1}
\ln \lr{2-V\lr{\frac{i\lambda}{h}}}\Bigg|_{\lambda=1} + \CO(h^{-2n})\,.
\lab{en-next-1}
\ee
Here, the last term was added to indicate explicitly that the expression
for the holomorphic energy was found within the approximation \re{app}
and it is valid only up to $\CO(h^{-2n})$ terms. This implies in particular,
that expanding the operator
$$
\frac{\partial}{\exp(\partial)-1} = \sum_{k=0}^\infty \frac1{k!}\, B_k
\ (\partial)^k
$$
with $B_k$ being Bernoulli numbers, we may keep in \re{en-next-1} only $2n-1$
first terms.

The expression for the energy \re{en-next-1} is real. Another interesting
property of \re{en-next-1} is that although it has a rather formal
form it can be explicitly evaluated for an arbitrary potential $V(\lambda)$.
The calculation is based on the relation
\be
\frac{\partial}{\exp(\partial)-1} \ln(a\lambda) =
\ln (a\lambda) -\frac1{2\lambda}-\frac1{12\lambda^2}+\frac1{120\lambda^4}
+ \ldots
=\psi(\lambda)+\ln a
\lab{iden}
\ee
which is valid for an arbitrary $a$ and which
follows from the definition of the $\psi-$function given in \re{psi}
and from identity $\lr{\ln \lambda}'= (\e^{\partial}-1)\psi(\lambda)
=\psi(\lambda+1)-\psi(\lambda)=1/\lambda$.
Let us consider the equation $V(\lambda)=2$ with the potential defined
in \re{pot}. It has $n$ roots which we denote as $\delta_1$, $...$,
$\delta_n$
\be
2-V(\lambda)
=2-\frac{h(h-1)}{(h\lambda)^2}+\frac{q_3}{(h\lambda)^3}
+\ldots+\frac{q_n}{(h\lambda)^n}
=\frac2{\lambda^n}\prod_{i=1}^n (\lambda-\delta_i)\,.
\lab{delta}
\ee
Substituting this relation into \re{en-next} and applying the identity
\re{iden} we find
\be
\varepsilon_n = -2\ln 2 - \sum_{i=1}^n
\left[\psi(1+ih\delta_i)+\psi(1-ih\delta_i)
-2\psi(1)\right] + \CO(h^{-2n})\,.
\lab{en-psi-1}
\ee
Notice that the roots $\delta_i$ depend on $h$ and in the limit
$h\to\infty$ they scale as $\delta\sim h^0$. For example, for $n=2$
one can easily obtain from \re{delta} their explicit form as
$\delta_{1,2}^{-2}=\fra{2h}{h-1}$.

It is interesting to note that the parameters $\delta_i$ have a simple
interpretation in terms of the auxiliary transfer matrix $\Lambda(\lambda)$
defined in \re{Lam}. Comparing the definitions \re{Lam} and \re{pot} we
conclude that $h\delta_1$, $...$, $h\delta_n$ entering into
\re{en-psi-1} can be determined as zeros of the auxiliary transfer matrix
$$
\Lambda(h\delta_k)=(h\delta_k)^n(2-V(\delta_k))=0\,,\qquad k=1, ... , n\,.
$$
Using the well--known asymptotic expansion \re{iden} of the $\psi-$function
we get from \re{en-psi-1} the expansion of the energy in
powers of $1/h$
\be
\varepsilon_n= -\ln q_n^2 + 2n \psi(1) -4 \sum_{k=1}^{n-1}
\frac1{h^{2k}}\frac{(2k-1)!\,\zeta(2k)}{(2\pi)^{2k}}
\sum_{i=1}^n \delta_i^{-2k} + \CO(h^{-2n})
\lab{en-psi-3}
\ee
where $\psi(1)=-\gamma_{_{\rm E}} = -0.577216$ is the Euler constant
and $\zeta(2k)$ is the Riemann zeta function. The comparison of \re{en-psi-3}
and \re{en-LLA} shows that, in accordance with our expectations \re{expe},
nonleading $1/h-$corrections to the solution of the Baxter equation
generate $\CO(h^0)$ contribution to the energy, $2n \psi(1)$,
as well as an asymptotic series in $1/h$. We should stress that applying
\re{en-psi-3} we have to take also into account the $h-$dependence of the
roots $\delta_k$ and expand $\sum_{i=1}^n \delta_i^{-2k}$ in powers of $1/h$.
Although \re{en-psi-3} looks like an even function of $1/h$
the latter expansion will give rise to odd powers of $1/h$.

Let us apply \re{en-psi-3} for $n=2$. Using the explicit form of the roots,
$\delta_i^{-2}=\frac{2h}{h-1}$, together with $q_2=-h(h-1)$
we obtain from \re{en-psi-3}
\be
\varepsilon_2(h)= - 2 \ln (h(h-1)) - 4\gamma_{_{\rm E}} - \frac2{3h(h-1)}
                 + \CO(h^{-4})
                = - 4 \ln h - 4\gamma_{_{\rm E}} + \frac2{h} + \frac1{3h^2}
                 + \CO(h^{-4})\,.
\lab{n=2-next}
\ee
Comparing this expression with the exact result \re{en-n=2} we verify that
up to $\CO(h^{-4})$ corrections both expressions do coincide.

Expression \re{en-psi-3} defines only $2n-1$ first terms of the infinite
asymptotic series for $\varepsilon_n$ and in order to apply \re{en-psi-3} it
is very important to understand how accurately these few terms extrapolate
the exact result for $\varepsilon_n$ for an {\it arbitrary\/} value of the
conformal weight, $1/2 \le h < \infty$. The answer to this question is based
on the properties of the asymptotic series \ci{as2,as3,as1} and
it will be discussed in Sect.~6. To anticipate our conclusions, we substitute
the most interesting value of the conformal weight, \re{1/2},
into \re{n=2-next} and \re{en-n=2} and obtain the following numerical values
\be
\varepsilon_2^{\rm app}(1/2) = 5.797059\,,\qquad
\varepsilon_2(1/2) = 8\ln 2 = 5.545177\,.
\lab{comp}
\ee
Remarkably enough, the approximate result turns out to be only $4\%$ bigger
the exact expression, which determines the intercept of the BFKL Pomeron
\re{bfkl}.

\subsection{Improved approximation}

The approximation performed in the previous section has a strong limitation.
It does not allow us to evaluate the holomorphic energy with the accuracy
better that $\CO(h^{-2n})$. Since we do not know in advance how many terms in
the $1/h-$expansion of the energy we will need in order to approach the
physically interesting value $\varepsilon_n(h=1/2)$ it is desirable to have
an improved scheme, which would allow us to predict higher $1/h$ corrections
to the energy.

There is a simple way how one can improve the approximation. Let us start
with the Baxter equation \re{Sch} and rewrite it in the following form
\be
\frac{f(\lambda+i/h)}{f(\lambda)}=2-V(\lambda)
-\lr{\frac{f(\lambda)}{f(\lambda-i/h)}}^{-1}\,.
\lab{BB}
\ee
As we have seen in Sect.~4.1, in the limit of small $\lambda$ and
$\Im \lambda > 0$ the ratio ${f(\lambda+i/h)}/{f(\lambda)}$ behaves as
$\sim -V(\lambda) \sim q_n\,\lambda^{-n}$. This fact allows us to
neglect the
last term in the r.h.s.\ of the Baxter equation \re{BB} and reproduce
\re{next-1}. The next step will be to consider the same term as a small
perturbation and iterate the Baxter equation \re{BB} as
$$
\frac{f(\lambda+i/h)}{f(\lambda)}=
2-V(\lambda)-\frac1{2-V(\lambda-i/h)-
\lr{\frac{f(\lambda-i/h)}{f(\lambda-2i/h)}}^{-1}}\,.
$$
Repeating this procedure $k-$times we obtain
\ba
&&\frac{f(\lambda+i/h)}{f(\lambda)}=2-V(\lambda)
\lab{f-c}
\\
&& 
-\lrs{2-V(\lambda-\fra{i}{h})
     -\lrs{2-V(\lambda-\fra{2i}{h}) - \ldots -
          \lrs{2-V(\lambda-\fra{ik}{h})-
                \lr{\fra{f(\lambda-ik/{h})}{f(\lambda-i(k+1)/h)}}^{-1}
             }^{-1}
       }^{-1}
   }^{-1}\,.
\nonumber
\ea
Here, the ratio
$\lr{f(\lambda-ik/h)/f(\lambda-i(k+1)/h)}^{-1}$
is much smaller than $V(\lambda-ik/h)$ at small $\lambda\sim 1/h$ provided that
$\Im \lambda > 0$ and the arguments of $f$ still lie in the upper half plane.
Otherwise, for $\Im \lambda < 0$, the same ratio
becomes large and we are not allowed to treat it as a small parameter.
This means, that for fixed number of iterations, we have to replace
$\Im \lambda > 0$ by a stronger condition $\Im \lambda > (k+1)/h$.
In particular, we cannot continue the iterations infinitely and
replace the r.h.s.\ of \re{f-c} by continued fraction since this will
push the allowed region of $\lambda$ away from the origin.

Let us consider \re{f-c} in the case when $\lambda=\CO(1/h)$ and
$\Im \lambda > (k+1)/h$. Then, using the relations
$
V(\lambda-ik/h)=\CO(h^n)
$
and
$
{f(\lambda-ik/h)}/{f(\lambda-i(k+1)/h)}\sim -V(\lambda-i(k+1)/h)=\CO(h^n)
$
we can find that neglecting $\lr{{f(\lambda-ik/h)}/{f(\lambda-i(k+1)/h)}}^{-1}$
in \re{f-c} we modify the r.h.s.\ of the Baxter equation \re{f-c}
at the level of $\CO(h^{-(2k+1)n})$ terms. Performing this transformation
and introducing notation for the ``iterated'' potential
\be
\CV_k(\lambda) = 2-V(\lambda)
-\lrs{2-V(\lambda-\fra{i}{h})
     -\lrs{2-V(\lambda-\fra{2i}{h}) - \ldots -
          \lrs{2-V(\lambda-\fra{ik}{h}}^{-1}
       }^{-1}
   }^{-1}
\lab{CV}
\ee
we can rewrite the Baxter equation \re{f-c} in the following form
$$
\frac{f(\lambda+i/h)}{f(\lambda)}=\CV_k(\lambda) + \CO(h^{-(2k+1)n})\,,
$$
where $\Im \lambda > (k+1)/h$ and $\lambda=\CO(1/h)$. For $k=0$
this equation is equivalent to \re{next-1} and similar to \re{Phi-1} its
solution can be represented for an arbitrary $k$ as
\be
i\Phi'(\lambda)=\frac{\frac{i}{h}\partial}
{\exp(\frac{i}{h}\partial)-1}
\ln\lr{\CV_k(\lambda)+\CO(h^{-(2k+1)n})}
=\frac{\frac{i}{h}\partial}
{\exp(\frac{i}{h}\partial)-1}
\ln \CV_k(\lambda)+\CO(h^{-(2k+2)n})
\lab{Phi-im}
\ee
where we took into account that $\CV_k(\lambda) = \CO(h^n)$ for
$\lambda=\CO(1/h)$. We recall that \re{Phi-im} was derived under
the additional conditions $\Im \lambda > (k+1)/h$ and $\lambda=\CO(1/h)$.
To find the expression for $i\Phi'(\lambda)$ in the lower half--plane,
$\Im\lambda < 0$, we have to iterate the Baxter equation \re{BB}
considering
${f(\lambda+\fra{i}{h})}/{f(\lambda)}$ as a small parameter and follow
the same steps which led to \re{Phi-2}. The final expression
is similar to \re{Phi-2} with $2-V(\lambda)$ replaced by complex conjugated
iterated potential $\CV_{k}^*(\lambda)$ and it is defined
for $\Im \lambda < -(k+1)/h$ and $\lambda=\CO(1/h)$.

Although \re{Phi-im} was found for $\Im \lambda > (k+1)/h$ we may analytically
continue the result to $\lambda=i/h$ and substitute it into \re{ener}
to get the following expression for the holomorphic energy
$$
\varepsilon_n = -\frac{\partial}{\exp(\partial)-1}
\ln \lr{\CV_k\lr{\fra{i\lambda}{h}} \CV_k^*\lr{-\fra{i\lambda}{h}}}
\Bigg|_{\lambda=1} + \CO(h^{-2(k+1)n})
$$
and using the fact that $V(\lambda)$ is a real function of $\lambda$
\be
\varepsilon_n = -2 \Re\ \frac{\partial}{\exp(\partial)-1}
\ln \CV_k\lr{\fra{i\lambda}{h}}
\Bigg|_{\lambda=1} + \CO(h^{-2(k+1)n})\,.
\lab{en-im}
\ee
Here, $k$ is a positive integer which enters into the definition \re{CV}
of the potential $\CV_k$. We conclude from \re{en-im} that the accuracy of
the approximation is
controlled by the number of iterations $k$. Increasing this number
in the definition \re{CV} of the potential $\CV_k$ and using \re{en-im}
we can obtain the expansion of the holomorphic energy $\varepsilon_n$ in
powers of $1/h$ with an arbitrary accuracy. For $k=0$ the expressions
for the energy \re{en-im} and \re{en-next-1} coincide.
Each new iteration, $k\to k+1$, adds $2n$ additional terms to the expansion
of $\varepsilon_n$.

One can apply \re{en-im} in a two different ways. The simplest one is to
expand $\ln\CV_k(i\lambda/h)$ and ${\partial}/({\exp(\partial)-1})$
in powers of $1/h$ and $\partial$, respectively, and keep only
first $2(k+1)n-1$ terms. The second way is based on the identity \re{iden}.
Similar to \re{delta} one has to decompose $\CV_k(\lambda)$ into simple
factors and obtain the expression for $\varepsilon_n$ as a sum of
$\psi-$functions.

\sect{Fine structure of quantum numbers}

The expressions for the energy, \re{en-im} and \re{CV}, depend on the
potential $V(\lambda)$. Applying \re{en-im} one finds that the higher terms
in the asymptotic expansion of the holomorphic energy become sensitive to
the nonleading $1/h$ corrections to the quantum numbers $q_k/h^k$ entering
into the definition \re{decV} and \re{V0} of $V(\lambda)$. The results of
the numerical solution of the Baxter equation indicate that the nonleading
corrections to the quantum numbers are organized in such a way that for
different values of the conformal weight $h$ the quantized values of
$q_3$, $...$, $q_n$ belong to the family of one--parametric curves for $n=3$
(see fig.~\ref{q3hat}), two-parametric surfaces for $n=4$ and, in general,
to $n-2$ parametric $(n-2)-$dimensional hypersurfaces for an arbitrary $n$.
There is a simple way how one can understand this fine structure within the
quasiclassical approximation.

Let us recall that in the large $h$ limit, the distribution density of roots
of the polynomial solutions of the Baxter equation has the support ${\cal S}$
consisting of $n-1$ connected intervals. The total number of roots is equal
to the conformal weight $h$ and, as a consequence, the distribution density
\re{rho} satisfies the normalization condition \re{norm}. All $h$ roots
take real values and they are distributed on the interval
${\cal S}=[\sigma_1,\sigma_2]\cup ... \cup [\sigma_{2n-3},\sigma_{2n-2}]$.
Let us denote by $N_k$ the number of roots belonging to the
interval $[\sigma_{2k-1},\sigma_{2k}]$ with $k=1$, $...$, $n-1$.
We remember that one of the intervals, say $[\sigma_{2n-3},\sigma_{2n-2}]$,
necessary contains the origin $\sigma=0$ and the corresponding root of the
polynomial solution should be $n-$time degenerate. This leads to the
following conditions on $N_k$
$$
\sum_{k=1}^{n-1} N_k =h\,,\qquad
0 \le N_1, N_2, ... , N_{n-2} \le h-n
\,,\qquad
n \le N_{n-1} \le h
\,.
$$
We recognize that for a given $n$ and $h$ among all integer numbers $N_1$, 
$...$, $N_{n-1}$ there are only $n-2$ independent. It is natural to expect 
that these are the sets of the integer numbers $\{N_k\}$, which parameterize
$(n-2)-$dimensional hypersurfaces describing the distribution of the
quantized $q_3$, $...$, $q_n$ for different values of $h$.

To show this we use the definition \re{rho} of the distribution
density and express the numbers $N_k$ as
$$
\int_{\sigma_{2k-1}}^{\sigma_{2k}} d\sigma \ \rho(\sigma) = \frac{N_k}{h}\,.
$$
In the large $h$ limit, we substitute \re{decrho} into this relation
and obtain expansion of its l.h.s.\ in powers of $1/h$. At the same time,
depending on the value of $N_k$, the r.h.s.\ contains only $\sim h^0$ and
$\sim h^{-1}$ terms. Then, from the comparison of the both sides we get
\be
\int_{\sigma_{2k-1}}^{\sigma_{2k}} d\sigma \
\lr{\rho_0(\sigma)+\frac1{h} \rho_1(\sigma)} = \frac{N_k}{h}\,,
\qquad
\int_{\sigma_{2k-1}}^{\sigma_{2k}} d\sigma \ \rho_j(\sigma) = 0\,,
\lab{eq-1}
\ee
where $k=1, ..., n-1$ numerates the intervals inside ${\cal S}$ and
$j=2,\, 3, ... $ refers to the level of nonleading $1/h-$terms in
the expansion of
the distribution density \re{decrho}. We notice that for small values
of $N_k$, such that $N_k=\CO(h^0)$, the first equation in \re{eq-1}
can be further split into two independent relations
\be
\int_{\sigma_{2k-1}}^{\sigma_{2k}} d\sigma \ \rho_0(\sigma) = 0\,,
\qquad
\int_{\sigma_{2k-1}}^{\sigma_{2k}} d\sigma \ \rho_1(\sigma) = N_k\,,
\qquad \mbox{for $N_k=\CO(h^0)\,.$}
\lab{eq-2}
\ee
Finally, using the definition \re{disc} of the distributions $\rho_j$ as
discontinuity of the function $\Phi_j'$ across the cut, we rewrite the
integral of $\rho_j$ over the cut $[\sigma_{2k-1},\sigma_{2k}]$ as a contour
integral of $\Phi_j'$ around the cut
\be
\int_{\sigma_{2k-1}}^{\sigma_{2k}} d\sigma\ \rho_j(\sigma)
=\oint_{C_k} \frac{d\lambda}{2\pi i}\ \Phi_j'(\lambda)
\lab{eq-3}
\ee
where the contour $C_k$ encircles the interval $[\sigma_{2k-1},\sigma_{2k}]$
in the complex $\lambda-$plane in the anti-clockwise direction. Using
the relations \re{next} and replacing the potentials $V_j(\lambda)$ by
their definitions \re{decV} and \re{V0}, we can express the r.h.s.\ of
\re{eq-3} in terms of quantized $q_3^{_{(j)}}$, $...$, $q_n^{_{(j)}}$.
Combining \re{eq-1}, \re{eq-2} and \re{eq-3} together we obtain an infinite
set of equations, in which we treat the numbers $N_1$, $...$, $N_{n-1}$ as
parameters and $q_3^{_{(j)}}$, $...$, $q_n^{_{(j)}}$ as unknown variables.
For a given set of integers $N_1$, $...$, $N_{n-1}$ their solution
will give us the expressions for $q_3^{_{(j)}}$, $...$, $q_n^{_{(j)}}$
which we will substitute into \re{q3q4} to find the asymptotic expansion
of the quantum numbers $q_3$, $...$, $q_n$. This result can be expressed
in the following form
\be
h =  \sum_{j=1}^{n-1}N_j\,,\qquad
q_k = q_k (h;N_1, N_2, ..., N_{n-2})\,, \qquad
N_1, ..., N_{n-2} \ge 0\,,\quad N_{n-1} \ge n
\lab{para}
\ee
where $k=3$, $...$, $n$. If we relax the condition for $N_k$ to be integer,
then these parametric relations define the $n-2$ dimensional domain in the
space of quantum numbers $h,q_3,...,q_n$. The interpretation of \re{para}
within the framework of two--dimensional conformal field theories is
proposed in Appendix B.

\subsection{Nonleading corrections at $n=3$}

Let us find the explicit form of the function $q_k=q_k(h;N_1,...,N_{n-2})$
for the $n=3$ Reggeon states. For $n=3$ the distribution density of roots
on ${\cal S}=[\sigma_1,\sigma_2]\cup[\sigma_3,\sigma_4]$ is described by
two integer numbers $0 \le N_1\le h-3$ and $N_2=h-N_1\ge 3$. According to our
notations, $N_2\ge 3$ counts the number of roots on the interval
$[\sigma_3,\sigma_4]$ including the 3--time degenerate root $\lambda_k=0$.
For integer conformal weight $h=N_1+N_2$ one can form only $h-2$ possible
sets of integers $(N_1,N_2)=(N_1,h-N_1)$ with $N_1=0,...,h-3$. This means
that in agreement with the numerical results shown on fig.~\ref{q3hat},
for fixed conformal weight $h\ge 3$ there are only $h-2$ possible quantized
values of $q_3$.

Let us analyze the system of equations \re{eq-1} in the special limit
$N_1=\CO(h^0)$, in which \re{eq-2} holds. The first equation in \re{eq-2}
involves the leading distribution density $\rho_0(\sigma)$, which
according to \re{rho1} and \re{rho2} is a positive definite smooth function
on ${\cal S}$ (see fig.~\ref{root-dis}(a)). Therefore, in order to satisfy
$\int_{\sigma_1}^{\sigma_2} d\sigma\, \rho_0(\sigma)=0$ we have to require
that the interval of the integration is shrinking to a point,
$\sigma_{1}=\sigma_{2}$. As we have shown in Sect.~3.1, this is the same
condition which defines the critical value of the quantum numbers.
We conclude that the solution of the first equation in \re{eq-2} is that
the leading quantum numbers $\hat q_3$, $...$, $\hat q_n$, should
take their critical values $\hat q_k=q_k^*$, or equivalently belong to the
critical hypersurface $\Sigma_n$, \re{si}. For $n=3$ this amounts to say
that
\be
\hat q_3\equiv q_3^{(0)}=\pm\frac1{\sqrt{27}}\,.
\lab{f-1}
\ee
Let us consider the second equation in \re{eq-2}. It contains the integral
of the first non-leading correction to the distribution density,
$\rho_1(\sigma)$, over the vanishing interval. In order for the
integral to be different from zero, the function $\rho_1(\sigma)$ should have
singularities on $[\sigma_1,\sigma_2]$. To show this we introduce a
small parameter $\varepsilon$,
\be
\sigma_2-\sigma_1=\varepsilon \,,\qqquad
V_0(\sigma_1)=V_0(\sigma_2)=4
\lab{appr}
\ee
and approach the limit $\sigma_{2}-\sigma_{1}\to 0$ as $\varepsilon\to 0$.
Here, the second relation follows from \re{crit}
(see also fig.~\ref{poten}(a))
and the potential $V_0$ for $n=3$ was defined in \re{V0}.
Choosing for simplicity $\hat q_3 >0$, we use \re{appr} to obtain the
following expansions at small $\varepsilon$
\be
\sigma_{1,2}=\frac1{\sqrt{12}}\pm \frac{\varepsilon}2\,,\qquad
\hat q_3=\frac1{\sqrt{27}}-\frac{\sqrt{3}}{2}\varepsilon^2\,,\qquad
V_0(\sigma)=4+144 x(1-x) \varepsilon^2\,,
\lab{simple}
\ee
where $\sigma=\sigma_1 + \varepsilon x$ and $0 \le x \le 1$. Substituting
these expressions into \re{LLA} and \re{next} we find the asymptotic behaviour
of $\Phi_0'(\lambda)$ and $\Phi_1'(\lambda)$ at the vicinity of
the infinitesimal interval $[\sigma_1,\sigma_2]$ as
\be
\Phi_0'(\lambda)=12\varepsilon\sqrt{z(z-1)} + \CO(\varepsilon^2)\,,
\qquad
\Phi_1'(\lambda)=-\frac1{4\varepsilon}\lr{\frac1{z}+\frac1{z-1}}
+\frac{V_1(\fra{1}{\sqrt{12}})}{24\varepsilon\sqrt{z(z-1)}}
+\CO(\varepsilon^0)
\lab{rh2}
\ee
where $\lambda=\sigma_1+z\varepsilon$, $z$ has an infinitesimal
imaginary part and $0 \le \Re z \le 1$. Taking discontinuity \re{disc}
across the cut $0 \le z \le 1$ we obtain the first nonleading
correction to the distribution density
\be
\rho_1(\sigma_1+x\varepsilon)=\frac1{\varepsilon}
\left[
-\frac14\lr{\delta(x)+\delta(x-1)}
+\frac{1}{24\pi}\frac{V_1(\fra{1}{\sqrt{12}})}{\sqrt{x(1-x)}}
\right]
+\CO(\varepsilon^{0})\,.
\lab{rh1}
\ee
In accordance with our expectations it is divergent for $\varepsilon\to 0$.
Applying the identity
\be
\int_{\sigma_1}^{\sigma_2} d\sigma\ \rho_j(\sigma) =
\varepsilon \int_0^1 dx \ \rho_j(\sigma_1+x\varepsilon) =
\varepsilon \oint_{C_1} \frac{dz}{2\pi i}  \Phi_j'(\sigma_1+\varepsilon z)
\lab{co}
\ee
with $C_1$ encircling the interval $[0,1]$ in complex $z-$plane, we
may use \re{rh1} to evaluate the integral in \re{co}.
However, a more effective way
to obtain the same result is to consider the contour integral of
$\Phi_1'(\sigma_1 + z\varepsilon)$ in \re{co}
and realize, using \re{rh2},
that the contour $C_1$ can be deformed away from the interval $[0,1]$.
Then, the contour integral in \re{co} is given by a residue of
$\Phi_1'(\sigma_1 + z\varepsilon)$ at $z=\infty$ (with
$z\varepsilon=\,$fixed!) and its substitution
into the second equation \re{eq-2} yields
$$
V_1\lr{\fra1{\sqrt{12}}}=24 \lr{N_1+\fra12}\,.
$$
Finally, using the definition \re{V0} of the potential $V_1(\lambda)$
we obtain the first nonleading correction to the quantum number $q_3$ as
\be
q_3^{_{(1)}} = -\frac{N_1+1}{\sqrt 3}
\lab{q3[1]}
\ee
where $N_1=\CO(h^0)$ is a positive integer parameterizing all possible
solutions of the quantization conditions \re{eq-1} and \re{eq-2}.

Let us repeat similar calculation and find the next correction,
$q_3^{_{(2)}}$. We start with the equation \re{next} for $\Phi_2'$ and use
the asymptotics \re{rh2} to estimate the leading singularity of
$\Phi_2'(\sigma_1+z\varepsilon)$ in the limit $\varepsilon\to 0$
as $\sim \varepsilon^{-3}$. This seems to imply that the integral
over $\rho_2(\sigma)$ in \re{co} should be divergent at $\varepsilon\to 0$.
However, despite of the fact that all these singular terms potentially
contribute to
$\rho_2(\sigma)$, the integral $\int_{\sigma_1}^{\sigma_2}d\sigma
\rho_2(\sigma)$ gets a nonzero contribution only from
$\CO(\varepsilon^{-1})$ term and is finite. To understand
this property we consider the contour integral of $\Phi_2'(\sigma_1+
\varepsilon z)$ in \re{co} and take into account, using \re{rh2},
that to any {\it finite\/} order of the expansion of the functions
$\Phi_j'(\sigma_1+\varepsilon z)$ $(j=0,\,1,..)$ in powers of $\varepsilon$
their singularities are located on the interval $[0,1]$
and at $z=\infty$ in the complex $z-$plane.
As a result, the contour integral in \re{co} is given by the
residue of the function $\Phi_j'(\sigma_1+\varepsilon z)$ at infinity
$z=\infty$ but with $\varepsilon z=\,$fixed. One can find from \re{rh2}
that in the limit $z=\infty$ and $\varepsilon z=\,$fixed the functions
$\Phi_j'(\sigma_1+\varepsilon z)$ have the following  asymptotic behaviour
$\Phi_j' \sim \sum_k (\varepsilon z)^{-k}\, f_k$ with some coefficients $f_k$.
All $(\varepsilon z)^{-k}-$terms with $k \neq 1$ in the expansion of
$\Phi_2'(\sigma_1+\varepsilon z)$ have a zero residue at infinity and
the contour integral \re{co} gets a nonzero contribution only from
$\CO((\varepsilon z)^{-1})$ term. Carefully expanding \re{next} in
powers of $\varepsilon$ we identify after some algebra the proper terms in
$\Phi_2'(\sigma_1+\varepsilon z)$ as
\be
\Phi_2'(\sigma_1+\varepsilon z) \stackrel{z\to\infty}{\sim}
(\varepsilon z)^{-1}
\lr{
2 (q_3^{_{(1)}})^2 -\frac1{\sqrt{3}} q_3^{_{(1)}}-\frac 59
+
\frac1{24}V_2\lr{\fra1{\sqrt{12}}}} +
...
\lab{rh3}
\ee
where dots denote terms with another (smaller or larger) powers of
$1/(\varepsilon z)$. Here, the potential
$V_2(\lambda)=-\lambda^{-3} q_3^{_{(2)}}$ was defined in \re{V0}
and $q_3^{_{(2)}}$ is the next nonleading correction to the quantum number
$q_3$. The quantization condition \re{eq-1} for $\rho_2$ has a form
$\int_{\sigma_1}^{\sigma_2} d\sigma\, \rho_2(\sigma)
= \oint dz\, \Phi_2' = 0$ and it requires that
the residue of \re{rh3} at $z=\infty$ should be equal to zero.
As a result, the second correction to the quantum number $q_3$ can be
found from \re{rh3} as
$$
q_3^{_{(2)}}=\frac1{\sqrt 3}
\lr{2 (q_3^{_{(1)}})^2 -\frac1{\sqrt{3}} q_3^{_{(1)}}-\frac 59}\,,
$$
and after substitution of \re{q3[1]} the explicit dependence of
$q_3^{_{(2)}}$ on integer number $N_1$ is
\be
q_3^{_{(2)}}=\frac{\sqrt 3}{27}\lr{6 N_1^2 + 15N_1 +4}\,.
\lab{q3[2]}
\ee
Continuing the same procedure it is straightforward to obtain the next
nonleading corrections to $q_3$ from the quantization conditions \re{eq-1}.
Here, we present the results of our computations using
{\it Maple~V\/}  Symbolic Computation System
\ba
q_3^{_{(3)}}&=&
-{\frac{\sqrt {3}}{81}}\left (2\,N_1^{3}+12\,N_1^{2}+32\,N_1+8\right
)
\nonumber
\\
q_3^{_{(4)}}&=&
{\frac {\sqrt {3}}{2187}}\left (30\,N_1^{4}+60\,N_1^{3}-942\,N_1^{2}-
972\,N_1-688\right )
\nonumber
\\
q_3^{_{(5)}}&=&
{\frac {\sqrt {3}}{6561
}}\left (114\,N_1^{5}+330\,N_1^{4}-5478\,N_1^{3
}-10050\,N_1^{2}-17012\,N_1-7360\right )
\lab{q3s}
\\
q_3^{_{(6)}}&=&
{\frac {\sqrt {3}}{59049}}\left (1344\,N_1^{6}+5058\,N_1^{5}-96210\,N_1
^{4}-254382\,N_1^{3}-709062\,N_1^{2} \right.
\nonumber
\\
& &
\hspace*{20mm} \left. -668508\,N_1-261472 \right )
\nonumber
\\
q_3^{_{(7)}}&=&
{\frac {\sqrt {3}}{
531441}}\left (16560\,N_1^{7}+76104\,N_1^{6}-1708338
\,N_1^{5}-5887110\,N_1^{4}-23283058\,N_1^{3}
\right.
\nonumber
\\
& &  \left. \hspace*{20mm}
-34667562
\,N_1^{2}-30609348\,N_1-10272032\right )
\nonumber
\ea
Finally, combining \re{f-1}, \re{q3[1]}, \re{q3[2]} and \re{q3s}
together we obtain the first eight terms of the asymptotic expansion
\re{q3q4} of the quantized $q_3$ in powers of $h$. We notice that
\re{q3[1]}, \re{q3[2]} and \re{q3s} were found for positive
$q_3^{_{(0)}}$. To get the corresponding expressions for negative
$q_3^{_{(0)}}$ we may use the symmetry of the Baxter equation \re{q3-q3} under
$q_3\to -q_3$ to change a sign of all nonleading coefficients
$q_3^{_{(j)}}$ in \re{q3[1]}, \re{q3[2]} and \re{q3s}.

\subsection{Comparison with numerical calculations}

Substituting \re{q3s} into \re{q3q4} we find that the resulting expression
for quantized $q_3=q_3(h;N_1)$ has a parametric form \re{para} with
$h=N_1+N_2$, $N_1\ge 0$ and $N_2\ge 3$. Choosing different values of the
integers $N_1$ and $N_2$ we get the spectrum of quantized $q_3$ and $h$ which
should be compared with the numerical results shown on fig.~\ref{q3hat}(a).
To perform the comparison we remove the condition for $N_1$ and $N_2=h-N_1$
to be integer and consider the dependence of $q_3$ on $N_1$ and $h$ in two
different cases. In the first case, to which we will refer as to $A$, we
take $N_1$ to be zero or positive integer and $N_2\ge 3$ to be positive real.
This gives us two families of curves
\be
A_{\pm}:
\qqqquad
q_3=\pm q_3(h;N_1)\,,\qquad
N_1=\IZ_+\,,\ h \ge 3
\lab{Apm}
\ee
where $\pm$ refers to the symmetry $q_3\to -q_3$ and $h$ is a continuous.
In the second case, $B$, we choose $N_2$ to be positive integer and
$N_1=h-N_2$ to be positive real,
\be
B_{\pm}:
\qqqquad
q_3=\pm q_3(h;h-N_2)\,,\qquad
N_2=3+\IZ_+\,,\ h\ge N_2\,.
\lab{Bpm}
\ee
As a result, we obtain the families $A_\pm$ and $B_\pm$ of the
one--parametric curves in the $(h,q_3/h^3)-$plane shown on fig.~\ref{q3hat}(b).
The curves from $A_+$ and $B_+$ and from $A_-$ and $B_-$ lie above and below
the axis $q_3=0$, respectively. The points in which the curves from
different families cross each other correspond to
integer values of both $N_1$ and $N_2$ and their coordinates define the
quantized values of $h$ and $q_3$. Trying to compare the expressions
\re{f-1}, \re{q3[1]}, \re{q3[2]} and \re{q3s} for
$q_3$ with the numerical results we should keep in mind
that we calculated only first few terms of the asymptotic expansion of the
functions $q_3=q_3(h;N_1)$ for small values of integer $N_1=\CO(h^0)$.
Nevertheless, a careful examination of fig.~\ref{q3hat}(b)
shows that for all quantized values of $q_3$ except of those close
to the degenerate value $q_3=0$ agreement is very good.

Let us summarize the properties of the different families of the curves on
fig.~\ref{q3hat}(b). For fixed $N_1$ the function $q_3(h,N_1)$ from $A_+$
increases as the conformal weight $h$ grows and for large $h$ it approaches
the maximal value. For different $N_1$ the following hierarchy holds
\be
\frac{h^3}{\sqrt{27}} > q_3(h;0) > q_3(h;1) > ... > 0  \,.
\lab{hie1}
\ee
As conformal weight $h$ decreases, the nonleading terms in the asymptotic
expansion of $q_3(h;N_1)$ become important. Using the first 8 terms in the
expansion of $q_3$ we find that the function $q_3(h;N_1)$ vanishes as $h$
approaches the value $h=2N_1+3$ (see fig.~\ref{q3hat}(b))
\be
q_3(h;N_1) \approx 0 \,,\qquad \mbox{for $h=2N_1+3=\IZ_+$}\,.
\lab{zero}
\ee
We recall, however, that the quantum number $q_3=0$ has a special meaning
as corresponding to the degenerate solutions of the Baxter equation \re{neq}.
Notice that equality in \re{zero} is approximate because the integer
$N_1$ takes the value, $N_1=\CO(h)$, beyond the approximation
\re{eq-2}, $N_1=\CO(h^0)$, under which the corrections to quantized $q_3$
have been
found in \re{q3s}. Therefore, using our expressions for $q_3(h;N_1)$
we are not allowed to approach $q_3=0$ very closely. That is the reason
why all curves from $A_+$ and $B_+$ should be terminated
at the vicinity of the axis $q_3=0$. To describe the region near
$q_3=0$ we have to improve the asymptotic approximation for $q_3=q_3(h;N_1).$
In contrast with $A_+$, the curves $q_3=q_3(h;h-N_2)$ from $B_+$ are
decreasing functions of $h$ for fixed $N_2$. They take maximal value
$q_3(N_2;0)$ at $h=N_2$ and for $N_2=3,\,4,...$ all these points belong
to the curve $q_3(h;0)$ from $A_+$. As $h$ grows the function $q_3(h;h-N_2)$
decreases and it crosses subsequently another curves from $A_+$ with
$N_1=1,\,2,...,N_2-3$.  At $h=2N_2-3=2N_1+3$ it takes a zero value in
accordance with \re{zero}.

Although we cannot approach small values of the quantized $q_3$ within the
approximation \re{eq-2}, a careful examination of fig.~\ref{q3hat}(b)
allows us to speculate about possible behaviour of the function $q_3(h;N_1)$
below the axis $q_3=0$. The comparison of figs.~\ref{q3hat}(a) and (b)
suggests that close to the axis $q_3=0$ the different curves from $A_+$
can be considered as a continuation of the curves from $B_-$
to the upper half--plane $q_3>0$. One of such possible curves is
shown on fig.~\ref{q3hat}(a). Together with \re{Apm} and
\re{Bpm} this assumption corresponds to the following property of the
function $q_3(h;N_1)$
\be
q_3(h;N_1) \approx -q_3(h;h-3-N_1)\,,\qquad 0 \le N_1 \le h-3\,.
\lab{N1N2}
\ee
This relation is consistent with \re{zero}. It allows us now to
identify the curve on fig.~\ref{q3hat}(a) as $q_3=q_3(h;6)$.
The value of the integer, $N_1=6$, can be fixed using \re{zero}
from the position of zero $q_3=0$.

\subsection{Quantization conditions for higher $n$}

Let us generalize the analysis of the quantization conditions
\re{eq-1} to the higher
$n$ Reggeon states. As we have shown in Sect.~5.1, the possible values of
quantized $q_3$, $...$, $q_n$ are parameterized by integers
$N_1$, $...$, $N_{n-1}$. According to the definition \re{para}, their
values satisfy either $N_j =\CO(h^0)$, or $N_j=\CO(h)$. Let us solve the
quantization conditions \re{eq-1} in the special case
\be
N_1\,, N_2\,, ...\,, N_{n-2} = \CO(h^0) \,,\quad N_{n-1}=\CO(h) \,,
\lab{N-small}
\ee
when almost all roots of the solution of the Baxter equation belong
to a single interval. As we will see in a moment, the advantage of
\re{N-small} is that for small $N_1$, $...$, $N_{n-2}$ one is able
to find analytical expressions for quantized $q_3$, $...$, $q_n$.
This does not mean however that the quantization conditions \re{eq-1}
can not be solved for the values of integers different from \re{N-small}.
In the latter case the equations for quantized $q_3$, $...$, $q_n$
become more complicated and their solution is more involved.

Using \re{eq-3} we represent the quantization conditions
\re{eq-2} in the following form
\be
\oint_{C_j} \frac{d\lambda}{2\pi i} \Phi_0'(\lambda) = 0\,,
\qquad
\oint_{C_j} \frac{d\lambda}{2\pi i} \Phi_1'(\lambda) = N_j\,,
\qquad
j=1,\, ...,\,n-2\,,
\lab{quac}
\ee
where the functions $\Phi_0'$ and $\Phi_1'$ were defined in \re{LLA} and
\re{next}. As was explained in Sect.~5.1, for each $j=1\,,...,\,n-2$
the solution of the first condition \re{quac} is that the quantum numbers
$q_3^{_{(0)}}$, $...$, $q_{n}^{_{(0)}}$ take their critical values, that is
they belong to the critical $n-2$ dimensional hypersurface \re{si}.
The system of $n-2$ equations, \re{quac}, fixes their position on $\Sigma_n$
``almost'' uniquely. Namely, the solution of \re{quac} defines
some points on $\Sigma_n$, which we identify as quantized values of
$q_3^{_{(0)}}$, $...$, $q_{n}^{_{(0)}}$.
For $n=3$ we found $q_3^{_{(0)}}=\pm 1/\sqrt{27}$,
but for $n=4$ these points can be identified as ``corners'' of the
generalized triangle on fig.~\ref{domain}:
\be
\lr{q_3^{_{(0)}}=0\,,\quad q_4^{_{(0)}}=\frac1{16}}
\qquad {\rm and} \qquad
\lr{q_3^{_{(0)}}=\pm\frac14\,,\quad q_4^{_{(0)}}=-\frac1{64}}  \,.
\lab{ver}
\ee
To solve the second equation in \re{quac} and find the next corrections,
$q_3^{_{(1)}}$, $...$, $q_{n}^{_{(1)}}$, we have to define the
limit in which $q_3^{_{(0)}}$, $...$, $q_{n}^{_{(0)}}$ approach
their quantized values. Similar to the situation for $n=3$,
we introduce small parameters
$\varepsilon_1$, $...$, $\varepsilon_{n-2}$, which measure the
size of the infinitesimal intervals, $\varepsilon_j=\sigma_{2j}-
\sigma_{2j-1}$. None of these intervals contains the
origin $\sigma=0$ and their boundaries satisfy the relations
$V_0(\sigma_{2j})=V_0(\sigma_{2j-1})=0$ or
$V_0(\sigma_{2j})=V_0(\sigma_{2j-1})=4$. Let us concentrate on
the latter case. Then, the potential $V_0(\lambda)$
has a local maximum on the interval $[\sigma_{2j},\sigma_{2j-1}]$
at $\sigma=\sigma_j^*(\varepsilon_j)$ and
$$
\sigma_{2j,\,2j-1}=\sigma_j^*(\varepsilon_j)\pm \frac{\varepsilon_j}2
                + \CO(\varepsilon_j^2)\,.
$$
The expansion of $V_0(\lambda)$ around $\lambda=\sigma_j^*$
looks like
$$
V_0(\lambda)=V_0(\sigma_j^*(\varepsilon_j))
            +\frac12 V_0''(\sigma_j^*(\varepsilon_j))(\lambda-\sigma_j^*)^2
            + ...   \,.
$$
Substituting $\lambda=\sigma_{2j-1}+\varepsilon_j x$ and taking into
account that $V_0(\lambda)=4$ for $x=0$ and $x=1$, we obtain the expansion
of the potential in the limit $\varepsilon_j\to 0$ as
\be
V_0(\sigma_{2j-1}+\varepsilon_j x)-V_0(\sigma_j^*)=\frac{\varepsilon_j^2}2
V_0''(\sigma_j^*)\, x(x-1) + \CO(\varepsilon_j^2)\,.
\lab{V0-exp}
\ee
Here, $\sigma_j^*=\sigma_j^*(0)$ is the position of the maximum of
$V_0(\lambda)$ in the limit $\varepsilon_j=\sigma_{2j}-\sigma_{2j-1}\to 0$.
It can be found as a solution of the equations
\be
V_0(\sigma_j^*)=4\,,\qquad
V_0'(\sigma_j^*)=0 \qquad
{\rm and}
\qquad
V_0''(\sigma_j^*) < 0\,.
\lab{si1}
\ee
Repeating analysis for
$V_0(\sigma_{2j})=V_0(\sigma_{2j-1})=0$ one can show that the same
expansion \re{V0-exp}
holds provided that we define $\sigma_j^*$ as a local minimum
of the potential,
\be
V_0(\sigma_j^*)=0\,,\qquad
V_0'(\sigma_j^*)=0 \qquad
{\rm and}
\qquad
V_0''(\sigma_j^*) > 0\,.
\lab{si2}
\ee
One can check that for $n=3$ the relation \re{V0-exp} coincides with
\re{simple}.
Substitution of the expansion \re{V0-exp} into \re{LLA} and \re{next} yields
\baa
\Phi_0'(\sigma_{2j-1}+\varepsilon_j z) &=&
\frac{\varepsilon_j}2\sqrt{2 |V_0''(\sigma^*_j)|}\sqrt{z(z-1)}
+\CO(\varepsilon_j^2)
\\
\Phi_1'(\sigma_{2j-1}+\varepsilon_j z) &=&
\frac1{\varepsilon_j}
\left[
-\frac14\lr{\frac1{z}+\frac1{z-1}}+\frac1{\sqrt{z(z-1)}}
\frac{V_1(\sigma^*_j)}{\sqrt{2|V_0''(\sigma^*_j)|}}
\right]
+\CO(\varepsilon_j^0)\,,
\eaa
where the potential $V_1(\lambda)$ was defined in \re{V0} and
$j=1,...,\,n-2$ numerates the intervals inside ${\cal S}$. The evaluation
of the contour integral of $\Phi_1'(\lambda)$ in \re{quac} is similar
to that in \re{co}. It is given by the residue of
$\Phi_1'(\sigma_{2j-1}+\varepsilon_j z)$ at
$z\to\infty$ and $\varepsilon_j z =\, $fixed. Finally, the solution of
the second equation \re{quac} has the form
$$
V_1(\sigma^*_j)=\sqrt{2|V_0''(\sigma^*_j)|} \lr{N_j+\frac12}\,,
\qquad j=1,...,n-2\,.
$$
Using the definition \re{V0} of $V_1(\lambda)$ we obtain from this
relation the system of $(n-2)$ equations for quantized
$q_3^{_{(1)}}$, $...$, $q_{n}^{_{(1)}}$
\be
\frac{q_3^{_{(1)}}}{\sigma^*_j}+\frac{q_4^{_{(1)}}}{(\sigma^*_j)^2}+
... +\frac{q_n^{_{(1)}}}{(\sigma^*_j)^n} = -1 -
\sqrt{2(\sigma^*_j)^4|V_0''(\sigma^*_j)|} \lr{N_j+\frac12}\,.
\lab{high}
\ee
Here, $\sigma^*_j$ is defined as a solution of \re{si1} or \re{si2}.
As an example, we consider the system of equation \re{high} for the
$n=4$ Reggeon compound states. Using two different sets \re{ver}
of the quantum numbers $(q_3^{_{(0)}},q_4^{_{(0)}})$ we find the
solution of \re{high} in the following form
$$
q_3^{_{(1)}}=-\frac12(N_1-N_2)\,,\qquad
q_4^{_{(1)}}=-\frac18-\frac{\sqrt 2}{8}(N_1+N_2+1)
$$
for the first set in \re{ver} and
$$
q_3^{_{(1)}}=\pm\frac14\lr{N_1-2\sqrt 2 N_2-1-\sqrt 2} \,,\qquad
q_4^{_{(1)}}=\frac{\sqrt 2}{32}-\frac{1}{16}\lr{N_1- \sqrt 2 N_2}
$$
for the second set in \re{ver}. Here, $0\le N_1\,,N_2 \le h-4$ are integers
parameterizing all possible solutions for $q_3$ and $q_4$.

\sect{Asymptotic expansions for the energy of the Reggeon states}

In previous sections we developed the method for evaluation of the
energy and conserved charges of the $n$ Reggeon states in the quasiclassical
approximation, $h \to \infty$. The resulting expressions have a form of the
asymptotic expansion in powers of $1/h$. Although they were found
for large $h$, it is of most importance to calculate their values
for the conformal weight $h=1/2$, which corresponds to the maximal energy of
the Reggeon state, \re{1/2}, and which defines the intercept of the QCD
Pomerons and Odderons \re{max}.
The natural question appears whether it is meaningful to substitute
$h=1/2$ into the large $h$ asymptotic expansion and estimate the approximate
value of the energy. To test this idea we start with $n=2$ Reggeon states
and apply \re{en-psi-3} and \re{en-im} to find the first few terms of the
expansion of $\varepsilon_2(h)$ in $1/h$
\be
\varepsilon_2(h)=
-4\ln (h\e^{\gamma_{_{\rm E}}})+{\frac {2}{h}}+{\frac {1}{3\,{h}^
{2}}}-{\frac {1}{30\,{h}^{4}}}+{\frac {1}{63\,{h}^{6}}}
-{\frac {1}{60\,{h}^{8}}}+{\frac {1}{33\,{h}^{10}}}-{
\frac {691}{8190\,{h}^{12}}}+{\frac {1}{3\,{h}^{14}}}
+\CO\lr{{h}^{-16}}\,.
\lab{div}
\ee
One can check that this series coincides with the asymptotic expansion of
the exact result \re{en-n=2}. The series \re{div} is a sign--alternating and
its coefficients rapidly grow in higher orders. It is obvious that for
$h=1/2$ the series \re{div} becomes divergent while the correct
answer for the energy $\varepsilon_2(1/2)$ is known to be finite \re{comp}.
On the other hand, as follows from \re{n=2-next} and \re{comp}, the sum of
the first four terms of the divergent series \re{div} gives for $h=1/2$ the
result which is close to the exact answer. These two properties are remarkable
features of the asymptotic series \ci{as2,as3,as1}, which we will explore
below to find the energy of the Reggeon states. Our goal will be to identify
the partial sums
which give the best asymptotic approximation to $\varepsilon_n$. However,
any asymptotic approximation should be accompanied by bounds for the
corresponding error terms and in the case of the asymptotic series their
determination becomes extremely nontrivial \ci{as2,as3,as1}.
In what follows, instead of giving the rigorous proof we apply a
``practical''
version of the asymptotic approximations \ci{as1} based on the well--known
examples and briefly described in Appendix C.

\subsection{Euler transformation for $n=2$ Reggeon states}

Let us apply the ``practical'' asymptotic approximation to
the asymptotic series \re{en-psi-3} and \re{en-im} for the energy
of the $n=2$ Reggeon state and then compare the approximant
$\varepsilon_2^{\rm app}(h)$ with the exact result
\re{en-n=2} for $\varepsilon_2(h)$. To start the analysis, we
substitute $h=1/2$ into \re{div} and examine the first few terms of the series
\be
\varepsilon_2(1/2)=0.46+4+1.33-0.53+1.02-4.27+31.03
                  -345.58+5461.33 + ...\ \,.
\lab{as1/2}
\ee
We find that apart from the first term their absolute value decreases until
the $n=4$ term is reached and then it rapidly increases. This suggests to
truncate the series \re{as1/2} and approximate $\varepsilon_2(1/2)$
by a partial sum of the first 4 terms. For our purposes, however, it is
convenient to exclude the 4--th term from the partial sum because after
this the approximant has a simple interpretation. We notice that the 4--th
term in \re{div} and \re{as1/2} has a sign opposite to all preceding terms
and its inclusion diminishes the partial sum. Therefore, the sum of
the first 3 terms corresponds to the {\it local\/} maximum, $S_{\rm max}$,
of the partial sum $S_n$ as a function of a discrete variable $n$,
\be
\varepsilon_2^{\rm app}(h) = S_{\rm max} = {\rm max}_{_{n=1,2,...}} S_n \,.
\lab{best2}
\ee
It is this interpretation of the approximant which will be used below.

Thus defined asymptotic approximation $\varepsilon_2^{\rm app}(h)$
to \re{div} coincides
with \re{n=2-next} and it is in a good agreement with the exact result
\re{comp}. We notice that \re{n=2-next} provides us asymptotic approximation
to $\varepsilon_2(h)$ not only for $h=1/2$ but for larger values of the
conformal weight $h$ as well. The accuracy of the approximation \re{best2} and
\re{n=2-next} is not clear yet. However, the fact that the asymptotic
series \re{div} and \re{as1/2} are sign--alternating allows us to estimate
the optimum remainder term \ci{as3} (see Appendix C for details).
If we assume that the first
three terms in \re{div} and \re{as1/2} correctly describe the asymptotics of
$\varepsilon_2(h)$, then the absolute value of the optimum remainder term,
$\delta \varepsilon_2(h)=\varepsilon_2(h)-\varepsilon_2^{\rm app}(h)$,
is smaller than the last term included into the approximant and the first term
excluded from the approximant. In application to the series \re{as1/2}
this property allows us to estimate the remainder as
$
-0.53 < \delta \varepsilon_2(1/2) < 0 \,,
$
which is in agreement with the numerical results \re{comp}.

\begin{figure}[htb]
   \vspace*{-1cm}
   \centerline{
   \epsfysize=9cm\epsfxsize=10cm\epsffile{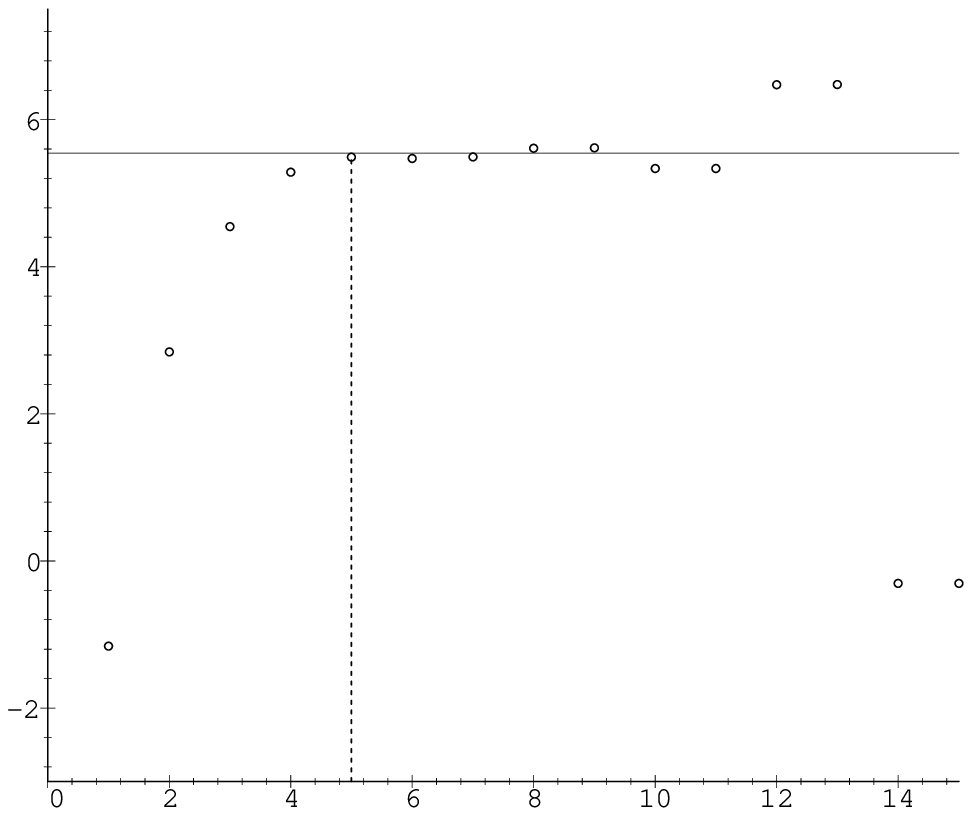}
   \hspace*{-2cm}
   \epsfysize=9cm\epsfxsize=10cm\epsffile{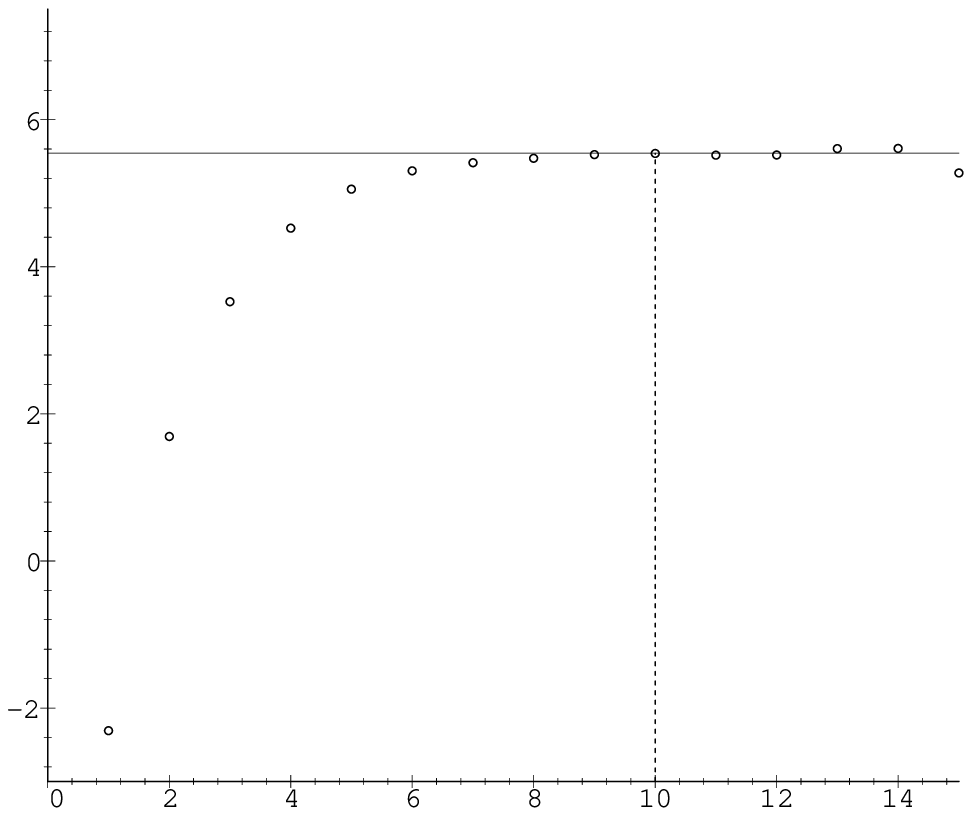}}
   \vspace*{-1cm}
   \centerline{ (a) \hspace*{7.3cm} (b)}
\caption{\label{Euler} The partial sums $S_n$ of the Euler
transformed asymptotic series for the energy, \re{div},
corresponding to $h=1/2$ and two different values of the parameter
of transformation:
(a) $x_0=0.25$ and (b) $x_0=0.5$. The solid line denotes the exact
expression for the energy $\varepsilon_2(1/2)=8\ln 2$, the dotted line
indicates the position of the local maximum $S_{\rm max}$.}
\unitlength=1mm
\begin{picture}(0,0)(0,0)
\put(8,94){\makebox(0,0)[cc]{${S_n}$}}
\put(89,94){\makebox(0,0)[cc]{${S_n}$}}
\put(70,35){\makebox(0,0)[cc]{$n$}}
\put(151,35){\makebox(0,0)[cc]{$n$}}
\end{picture}
\vspace*{-0.5cm}
\end{figure}

\begin{figure}[htb]
   \vspace*{-1cm}
   \centerline{
   \epsfysize=9cm\epsfxsize=10cm\epsffile{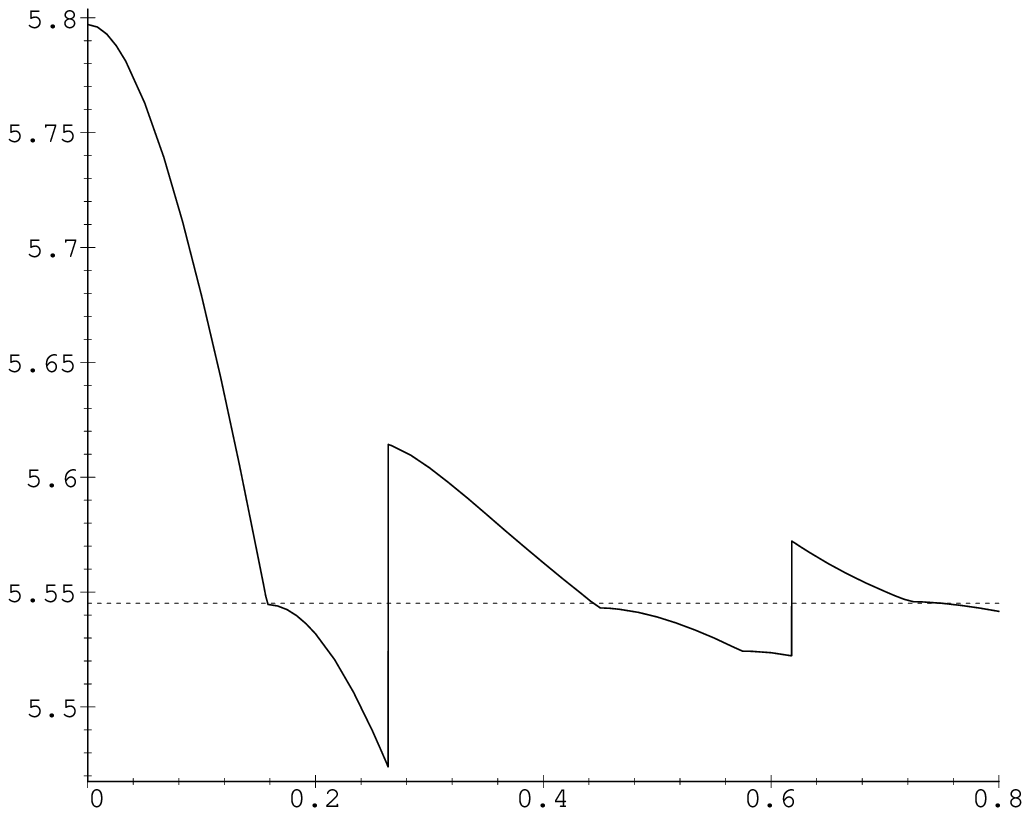}
   \hspace*{-2cm}
   \epsfysize=9cm\epsfxsize=10cm\epsffile{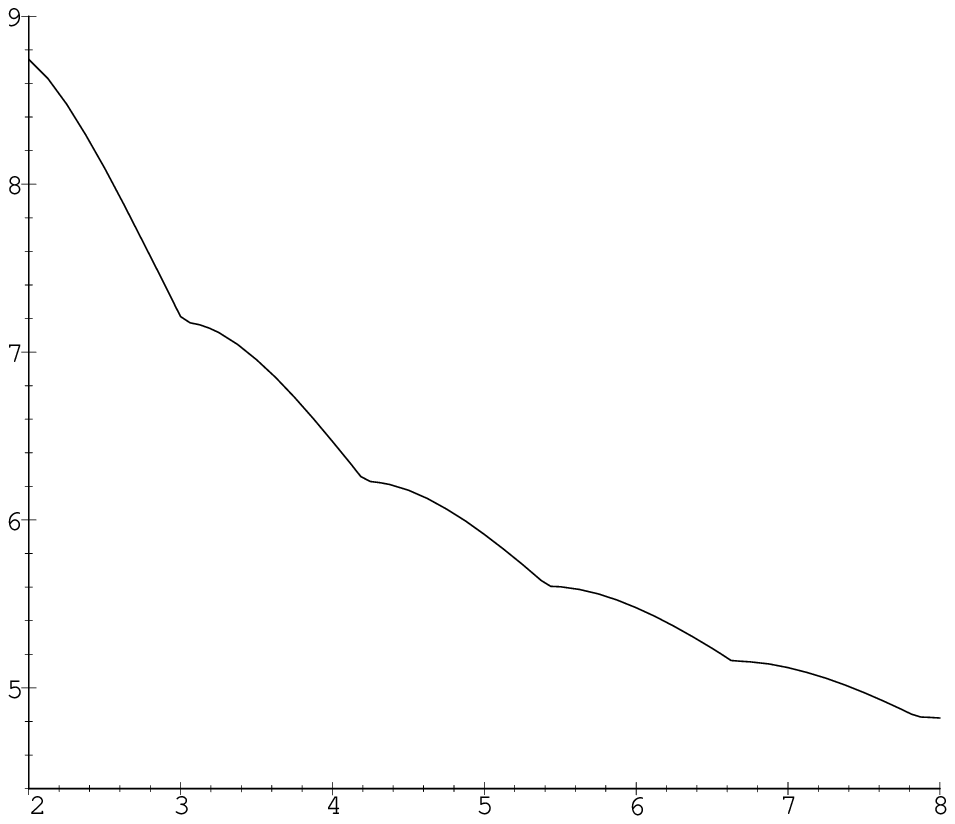}}
   \vspace*{-1cm}
   \centerline{ (a) \hspace*{7.3cm} (b)}
\caption{\label{f2} The dependence of the asymptotic approximation
to the energy, $S_{\rm max}(h=1/2,x_0)$, on the parameter of the Euler
transformation $x_0$ for (a) $n=2$ and (b) $n=3$ Reggeon compound state.
The dotted line represents the exact value of the energy
$\varepsilon_2(h=1/2)=8\ln 2$.}
\unitlength=1mm
\begin{picture}(0,0)(0,0)
\put(5,89){\makebox(0,0)[cc]{${S_{\rm max}}$}}
\put(86,89){\makebox(0,0)[cc]{${S_{\rm max}}$}}
\put(70,30){\makebox(0,0)[cc]{$x_0$}}
\put(151,30){\makebox(0,0)[cc]{$x_0$}}
\end{picture}
\vspace*{-0.5cm}
\end{figure}

Let us apply the Euler transformation \ci{as2,as1} (for definition see
Appendix C) to find better approximation to
the maximal energy $\varepsilon_2(1/2)$. To this end, we take the
asymptotic series \re{div} and reexpand it around the point $h+x_0$ up to
$\CO((h+x_0)^{-16})$ terms. Although the original series \re{div} contains
only even powers of $1/h$ in higher orders, the transformed series has all
powers of $1/(h+x_0)$. Let us denote the partial
sum of the first $n$ terms of the transformed series as $S_n(h,x_0)$.
According to the ``practical'' asymptotical approximation \re{best2},
to find the best approximation to
$\varepsilon_2(h)$ we have to examine the behaviour of $S_n(h,x_0)$ as
a function of $n$, identify the lowest value of $n=n_0$ at which $S_n$
has a maximum and estimate the best approximation to $\varepsilon_2(h)$
as $S_{\rm max}=S_{n_0}(h,x_0)$.
To understand the role of the parameter $x_0$ in this procedure we
fix the conformal weight as $h=1/2$ and evaluate $S_n(1/2,x_0)$ for
$n=1,...,15$ and two different values of $x_0$: $x_0=0.25$ and $x_0=0.5$.
The results of our calculations are shown on fig.~\ref{Euler}.
\phantom{\ref{Euler}}
We notice that in both cases the partial sum $S_n$ grows with $n$ until it
approaches the local maximum, $S_{\rm max}$.
Then, the region of stability follows after which
$S_n$ starts to oscillate around $S_{\rm max}$ and the size of the
fluctuations rapidly grows with $n$.
It is important to realize that the position, $n$, of the maximum
$S_n=S_{\rm max}$ depends on the parameter $x_0$. For $x_0=0.25$ we
have $S_{\rm max}=5.489602$ at $n=5$ and for $x_0=0.5$ it is
$S_{\rm max}=5.539171$ at $n=10$. The dependence of $S_{\rm max}(h,x_0)$
on $x_0$ for $h=1/2$ is shown on fig.~\ref{f2}(a). The cusps correspond
to the ``critical'' values of $x_0$, at which the number of terms contributing
to $S_{\rm max}$ increases by 1.
\phantom{\ref{f2}}
As parameter $x_0$ increases the number of terms contributing to
$S_{\rm max}$ also increases and, hence, the approximation with
which one can evaluate $S_{\rm max}$ becomes better.
However, for large $x_0$ the position of the
local maximum of $S_n$ is shifted toward larger values of $n$ and in order to
apply the Euler transformation we have to add higher $1/h-$terms to the
asymptotic expansion of the energy \re{div}. Namely, the expression \re{div}
defines the asymptotic expansion of the energy up to $\CO(h^{-16})$ terms
and under the Euler transformation the maximal value of the $n$ corresponding
to $S_{\rm max}=S_n$ cannot exceed $n=15$. One can show that this
condition is still satisfied for $x_0=0.75$ and the corresponding value of the
partial sum gives the approximate value of the energy
$$
\varepsilon_2^{\rm app}(1/2)=5.545150
$$
which is remarkably close to the exact answer \re{comp}. One might
expect that the accuracy of the asymptotic approximation should
be better for larger values of the conformal weight, $h> 1/2$.
Indeed, taking the parameter of the Euler transformation as $x_0=0.75$
and repeating the calculation of $S_{\rm max}$ for different values
of $h$ we obtain the results for $\varepsilon_2^{{\rm app}}(h)$
and $\delta \varepsilon_2(h)$ summarized in Table 1.
\begin{table}[htb]
\centerline{
\begin{tabular}{|l|c|c|c|c|c|c|c|c|}
\hline
$h$ & $0.5$ & $1$ & $1.5$ & $2$ & $2.5$ & $3$ & $3.5$ & $4$
\\
\hline
$\varepsilon_2^{\rm app}$ &
$5.5452$ & $\sim 10^{-5}$ & $-2.4548$ & $-4.0000$ & $-5.1215$ & $-6.0000$
& $-6.7215$ & $-7.3333$  \\
\hline
$\delta \varepsilon_2$ &
$\sim 10^{-5}$ & $\sim 10^{-5}$ & $\sim 10^{-6}$ & $\sim 10^{-7}$
& $\sim 10^{-8}$ & $\sim 10^{-9}$ & $\sim 10^{-9}$ &
$\sim 10^{-9}$ \\
\hline
\end{tabular}}
 \centerline{\parbox{12cm}{\caption{\label{tab1}
The asymptotic approximation to $\varepsilon_2(h)$ based on the Euler
transformation for $x_0=0.75$.
   }}}
\end{table}
We conclude that the
``practical'' version of the asymptotic approximation combined with the
Euler transformation works perfectly well for the energy of the $n=2$
Reggeon states. Although we are not able to find the exact value of
the energy we have the method which allows us to find the approximation,
$\varepsilon_2^{\rm app}(h)$, and improve its accuracy by varying the
value of the parameter of the Euler transformation, $x_0$.
This gives us a hope that the same procedure can be applied to the
evaluation of the energy of the higher Reggeon states.

\subsection{Analytical continuation}

To apply a similar consideration to the higher Reggeon states we need
to know the asymptotic expansion of the energy $\varepsilon_n$
similar to \re{div}. In general, it can be found from
\re{en-im}, but examining the definitions \re{CV} and \re{decV} we
immediately realize an important difference between $n=2$ and $n\ge 3$ cases.
For higher Reggeon states the energy $\varepsilon_n$
depends on the quantum numbers
$q_3$, $...$, $q_n$. Their values are quantized and
depend on the conformal weight $h$ as well as on the set of integer
numbers defined in \re{para}. The latter dependence was studied in Sect.~5
and the expressions for the quantum numbers were obtained
in the form of asymptotic series in $1/h$.
To find the asymptotic expansion of the energy of the
higher Reggeon states we have to supplement the relation \re{en-im}
by analogous expressions \re{para}
for the quantum numbers $q_3$, $...$, $q_n$.
Moreover, as we have shown in Sect.~4, the quasiclassical expansion
of the energy in powers of $1/h$ becomes divergent. The lowest
order terms in the expansion of $\varepsilon_n$ in powers of $1/h$
may depend, in general, on an arbitrary higher order terms in the
expansion of the quantum numbers $q_3$, $...$, $q_n$ and this makes the
definition of the asymptotic expansion of the energy very problematic.

Luckily enough, a close examination of the expression \re{en-im}
shows that the $\CO(h^{-k})-$term in the expansion of the
energy $\varepsilon_n$ depends on the finite number of nonleading terms
in the expansion \re{q3q4} of the quantum numbers:
$q_n^{_{(k)}}$, $q_{n-1}^{_{(k-2)}}$, $...$, $q_3^{_{(k-2(n-3))}}$.
Therefore, having an expression for the quantum numbers up to
$\CO(h^{-k})$ terms we will able to derive from \re{en-im}
the expansion of the energy with the same accuracy. To find the
asymptotic series for $q_3$, $...$, $q_n$ we have to solve the
quantization conditions \re{eq-1} and \re{eq-2} as was described in Sect.~5.
Their solution is well--defined for integer
values of the conformal weight $h \ge n$ and it was parameterized
in \re{para} by the set of integer numbers $N_1$, $...$, $N_{n-1}$.
As a result, the energy of the $n$ Reggeon states
also depends on the same numbers
\be
\varepsilon_n
\equiv \varepsilon_n\lr{h,q_3,...,q_n}
 = \varepsilon_n(h;N_1,...,N_{n-2})
\,.
\lab{epsn}
\ee
This implies, in particular, that similar to the distribution of the
quantized $q_3$, $...$, $q_n$, all possible values of the energy
belong to the families of curves in the $(h,\varepsilon_n)-$plane labelled
by integers $N_1$, $...$, $N_{n-2}$.

\begin{figure}[htb]
   \vspace*{-1cm}
   \centerline{
   \epsfysize=10cm\epsfxsize=10cm\epsffile{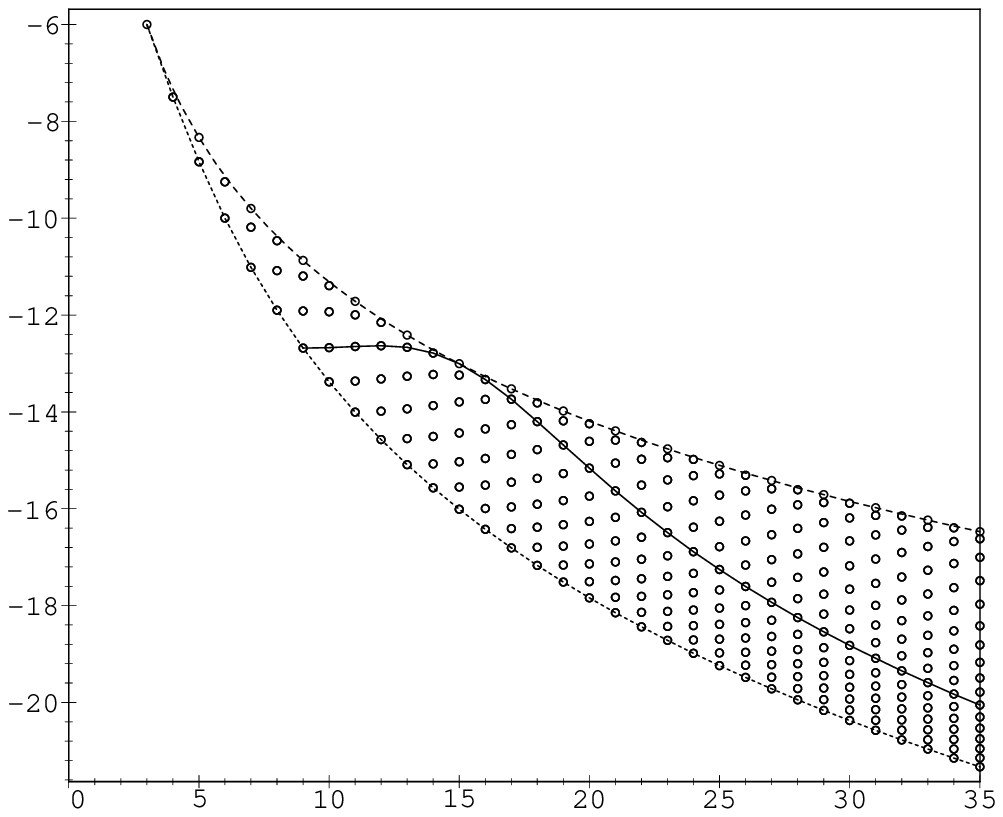}
   \hspace*{-2cm}
   \epsfysize=10cm\epsfxsize=10cm\epsffile{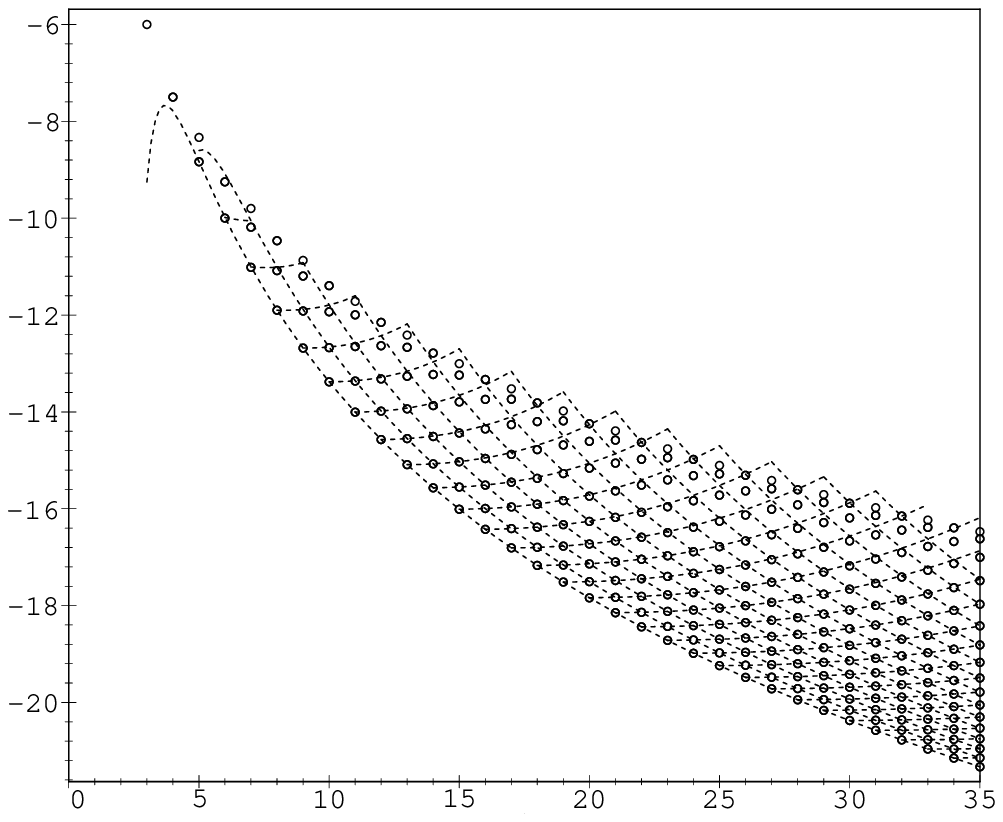}}
   \vspace*{-1cm}
   \centerline{ (a) \hspace*{7.3cm} (b)}
\caption{\label{ener-n=3} Holomorphic energy of the $n=3$ Reggeon compound
states: (a) the results of the numerical solution; (b) two different
families of curves defined in \re{AB}. The solid line on (a)
corresponds to $\varepsilon_3=\varepsilon_3(h,q_3)$ with the function
$q_3=q_3(h;6)$ shown on fig.~\ref{q3hat}(a). The dashed line on (a)
represents the energy of the $n=2$ Reggeon states \re{en-n=2} and the
dotted line on (a) is described by the function $\varepsilon_3(h;0)$.
}
\unitlength=1mm
\begin{picture}(0,0)(0,0)
\put(8,106){\makebox(0,0)[cc]{${\varepsilon_3}$}}
\put(89,106){\makebox(0,0)[cc]{${\varepsilon_3}$}}
\put(67,43){\makebox(0,0)[cc]{$h$}}
\put(148,43){\makebox(0,0)[cc]{$h$}}
\end{picture}
\vspace*{-0.5cm}
\end{figure}

Let us find the explicit form of the function
$\varepsilon_3=\varepsilon_3(h;N_1)$ for the $n=3$ Reggeon
states. Substituting the asymptotic expansion \re{q3q4} and \re{q3s}
of the quantized
$q_3=q_3(h;N_1)$ into \re{en-im} and \re{CV} we find the first 8 terms
of the expansion of $\varepsilon_3$ in powers of $1/h$ as
\be
\varepsilon_3(h;N_1)=-6\,\ln \lr{\frac{h\,\e^{\gamma_{_{\rm E}}}}{\sqrt 3}}
+\frac1{h}\varepsilon_3^{_{(1)}}
+ ... + \frac1{h^7}\varepsilon_3^{_{(7)}}
+\CO(h^{-8})\,,
\lab{en-hN1}
\ee
where the coefficients are given by
\baa
\varepsilon_3^{_{(1)}}&=&6\,N_1+6
\\
\varepsilon_3^{_{(2)}}&=&5\,N_1^{2}+8\,N_1+{\frac {11}{6}}
\\
\varepsilon_3^{_{(3)}}&=&{\frac {58}{9}}N_1^3+{\frac {44}{3}}N_1^2
-{\frac{35}{9}}N_1-{\frac{56}{9}}
\\
\varepsilon_3^{_{(4)}}&=&{\frac{1553}{162}}N_1^4+{\frac{2336}
{81}}N_1^3-{\frac{5359}{162}}N_1^2-
72\,N_1-{\frac{190883}{4860}}
\\
\varepsilon_3^{_{(5)}}&=&{\frac{18538}{1215}}N_1^5+{\frac{13928}{243}}
N_1^4-{\frac{32402}{243}}N_1^3
-{\frac{96188}{243}}N_1^2-{\frac{1735909}{3645}}N_1-{\frac{135976}{729}}
\\
\varepsilon_3^{_{(6)}}&=&{\frac{55345}{2187}}N_1^6+{\frac{83152}{729}}N_1^5
-{\frac{625865}{1458}}N_1^4-{\frac{3545360}{2187}}N_1^3-{\frac{
13670593}{4374}}N_1^2
\\
&&\hspace*{20mm}
-{\frac{5748704}{2187}}N_1-{\frac{80780333}{91854}}
\\
\varepsilon_3^{_{(7)}}&=&{\frac{1983742}{45927}}N_1^7+{\frac{
1489976}{6561}}N_1^6-{\frac{24244915}{19683}}N_1^5
-{\frac{111180580}{19683}}N_1^4-{\frac{902984221}{59049}}N_1^3
\\
&&
\hspace*{20mm}
-{\frac{398272432}{19683}}N_1^2-{\frac{
2037528641}{137781}}N_1-{\frac{259171832}{59049
}}
\eaa
The expression \re{en-hN1} should be compared with the numerical results
for the energy of the $n=3$ Reggeon states presented on fig.~\ref{ener-n=3}(a)
(see also fig.~3 in \ci{K}). Let us perform the comparison following
the same procedure as we have used in Sect.~5.2 for quantized
$q_3=q_3(h;N_1)$. Similar to \re{Apm} and \re{Bpm} the function
\re{en-hN1} defines two different families of the curves in the
$(h,\varepsilon_3)-$plane shown on fig.~\ref{ener-n=3}(b),
\be
A: \quad \varepsilon_3=\varepsilon_3(h;N_1)\,,
\quad h \ge 2N_1+3\,;
\qqquad
B: \quad \varepsilon_3=\varepsilon_3(h;h-N_1-3)\,,
\quad 2N_1+3 \ge h \ge N_1+3\,,
\lab{AB}
\ee
with $N_1\ge 0$ being an integer. Notice
that due to the invariance \re{q3-q3}
under replacement $q_3\to -q_3$ the energy
$\varepsilon_3(h;N_1)$ takes the same value for the curves from $A_+$
and $A_-$. The quantized values of the energy correspond to the points
on fig.~\ref{ener-n=3}(b), at which the curves from $A$ and $B$ cross
each other. There is one--to--one correspondence between different
curves on figs.~\ref{q3hat}(b) and \ref{ener-n=3}(b), such that
each function $q_3=q_3(h;N_1)$ is mapped into the corresponding
function $\varepsilon_3=\varepsilon_3(h;N_1)$. As an example,
the solid line on fig.~\ref{q3hat}(a) corresponding to
$q_3=q_3(h;6)$ induces the curve $\varepsilon_3=
\varepsilon_3(h;6)$ shown on fig.~\ref{ener-n=3}(a).
In the same manner, the axis $q_3=0$ on fig.~\ref{q3hat}(a)
is mapped into the dashed line on fig.~\ref{ener-n=3}(a)
describing the degenerate values of the energy $\varepsilon_3(h,q_3=0)=
\varepsilon_2(h)$. Similar to the distribution of the quantized $q_3$ on
fig.~\ref{q3hat}, we observe a good agreement on fig.~\ref{ener-n=3}
between numerical results
for the energy and analytical expression \re{en-hN1}
everywhere except of the points close to the ``degenerate'' curve
$\varepsilon_3=\varepsilon_3(h,q_3=0)$. Nevertheless, as one can see from
fig.~\ref{ener-n=3}(a), the function $\varepsilon_3(h,q_3)$ is
a smooth function of $h$ at the vicinity of $q_3=0$ and according to \re{N1N2}
it satisfies the following relation
\be
\varepsilon_3(h;N_1)\approx \varepsilon_3(h;h-N_1-3)\,,
\lab{en-sym}
\ee
which allows us to consider \re{AB} as definition of the unique
function $\varepsilon_3=\varepsilon_3(h;N_1)$ for $N_1+3\le h < \infty$.
\phantom{\ref{ener-n=3}}
Solving the Baxter equation, we are interesting to find the maximal
value of the energy of the $n$ Reggeon states, \re{max} and \re{En}.
As was shown in \ci{K} and as it can be easily seen from
fig.~\ref{ener-n=3}(a), for
positive integer conformal weight, $h\ge n$, the energy $\varepsilon_n$
cannot exceed the upper bound
\be
\varepsilon_n(h,q_3,...,q_n) \le -4(\psi(h)-\psi(1))\,,\qqquad
h \ge n
\lab{upper}
\ee
given by the energy of $n=2$ Reggeon state with the conformal
weight $h$. As $h$ decreases the maximal value of the energy
${\rm max}_{_{q_3,...,q_n}}\varepsilon_n(h,q_3,...,q_n)$
grows until $h$
approaches the
value $h=n$, beyond which the Baxter equation does not have polynomial
solutions. On the other hand, we expect from \re{1/2} that
the absolute maximum of energy $\varepsilon_n$ is achieved
for $h=1/2$ and not for $h=n$. This means that in order to approach
physically interesting values of the energy we have to penetrate through
the $h=n$ barrier to the region of smaller values of the conformal weight
$h$. In what follows we restrict consideration to the $n=3$ Reggeon states
while generalization to the higher Reggeon states is straightforward.

To perform analytical continuation of the energy $\varepsilon_3(h,q_3)$
for $h < 3$ we use fig.~\ref{ener-n=3}(a) to observe the following
properties of the function $\varepsilon_3(h;N_1)$.
Let us consider the flow of the quantized values of the energy along
the curve $\varepsilon_3=\varepsilon_3(h;N_1)$ for fixed $N_1$.
The energy increases along $\varepsilon_3(h;N_1)$ as conformal
weight decreases from $h=\infty$ to $h=2N_1+3$. At the point
$h=2N_1+3$ it takes the degenerate value
$\varepsilon_3(h;N_1)=\varepsilon_2(h)$ corresponding to
$q_3(h;N_1)=0$ in \re{zero}. For smaller values of the conformal
weight, $h < 2N_1+3$, the curve $\varepsilon_3=\varepsilon_3(h;N_1)$
is ``reflected'' from the upper limit $\varepsilon_3=\varepsilon_2(h)$
and for $h=N_1+3$ it approaches the end--point,
$\varepsilon_3=\varepsilon_3(h;N_1)=\varepsilon_3(h;0)$, and is
terminated.

We notice from fig.~\ref{ener-n=3} that any two curves
$\varepsilon_3(h;N_1)$
and $\varepsilon_3(h;N_1')$, corresponding to different values of
integers $N_1$ and $N_1'$, cross each other at only one point. Its
position can be easily found from \re{en-sym} as
$$
\varepsilon_3(h;N_1)=\varepsilon_3(h;N_1')\,,
\qquad {\rm for}\quad h=N_1+N_1'+3\,.
$$
It is clear, that if the value of one of
the functions is bigger for $h<N_1+N_1'+3$, then for $h>N_1+N_1'+3$
the relation between them becomes opposite. As a result, the functions
satisfy the following hierarchy
$$
-4(\psi(h)-\psi(1)) \ge \varepsilon_3(h;N_1) > \varepsilon_3(h;N_1')
\,,\qquad {\rm for}\quad h > N_1+N_1'+3\,,\ \ N_1 > N_1'
$$
and
$$
-4(\psi(h)-\psi(1)) \ge \varepsilon_3(h;N_1') > \varepsilon_3(h;N_1)
\,,\qquad {\rm for}\quad h < N_1+N_1'+3\,,\ \ N_1 > N_1'\,.
$$
These inequalities allow us to identify the maximal value of the energy
corresponding to the integer conformal weight $h\ge 3$. We find that for
the values of the conformal weight inside the interval
$2k+2 \le h < 2k+4$, with $k$ positive integer,
among all the functions
$\varepsilon_3=\varepsilon_3(h;N_1)$ the one with
$N_1=k=\left[\fra{h-2}{2}\right]$ takes a maximal
value
\be
{\rm max}_{_{N_1}} \varepsilon_3(h;N_1) =
\varepsilon_3\lr{h;\left[\fra{h-2}{2}\right]} \,.
\lab{up}
\ee
Therefore, as $h$ decreases from $h=\infty$ to $h=4$ the
maximum of the energy ``jumps'' from one curve $\varepsilon_3(h;N_1)$
to the next one, $N_1\to N_1-1$, and finally for $h<4$ it follows the
function $\varepsilon_3(h;0)$.

For $h \ge 3$ the quantized values of the energy are confined to
the region on fig.~\ref{ener-n=3} which is restricted from above
by the function $\varepsilon_3\lr{h;\left[\fra{h-2}{2}\right]}$ and by the
function $\varepsilon_3(h;0)$ from below,
\be
{\rm min}_{_{N_1}} \varepsilon_3(h;N_1) =
\varepsilon_3(h;0) \,.
\lab{down}
\ee
As conformal weight $h$ decreases, the upper and lower bounds
for the energy, \re{up} and \re{down}, respectively, move toward each
other leaving no phase space for quantized values of energy $\varepsilon_3$.
At $h<4$ both boundaries coincide and the energy is given by
\be
\varepsilon_3=\varepsilon_3(h;0)\,,\qqquad {\rm for}\quad h < 4\,.
\lab{the}
\ee
For $h=3$ it takes the value
$\varepsilon_3=\varepsilon_3(3;0)=-6$ corresponding to the $n=3$ degenerate
Reggeon state with $q_3(3;0)=-0$. It is of the most importance now to
understand the behaviour of the function $\varepsilon_3(h;0)$ beyond
the barrier, $h < 3$, where the Baxter equation does not have polynomial
solutions. We conclude from \re{up} and \re{down}
that once the conformal weight approaches the barrier $h=3$,
among all functions $\varepsilon_3(h;N_1)$ describing quantized values
of energy only one, \re{the}, with $N_1=0$ survives.
Therefore, performing analytical continuation to $h<3$ we may restrict
ourselves to the function $\varepsilon_3(h;0)$.

\subsection{Euler transformation for $n=3$}

To define $\varepsilon_3(h;0)$ for $h\le 3$ we
use its asymptotic expansion in powers of $1/h$ and apply the
Euler transformation to evaluate the best asymptotic approximation.
The asymptotic expansion for $q_3(h;0)$ can be found from \re{q3q4}
and \re{q3s} up to $\CO(h^{-10})$ terms as%
\footnote{In comparison with \re{q3s} we calculated and added to
$q_3(h;0)$ two additional nonleading terms.}
\ba
q_3(h;0)&=&{\frac{\sqrt {3}}{9}} h^3
\left(
1-{\frac {3}{h}}+{\frac {4}{3\,{h}^{2}}}-{\frac {8}{9\,
{h}^{3}}}-{\frac {688}{243\,{h}^{4}}}-{\frac {7360}{729
\,{h}^{5}}}
\right.
\lab{q3-h0}
\\[3mm]
&&
\hspace*{20mm}
\left.
-{\frac {261472}{6561\,{h}^{6}}}-{\frac {
10272032}{59049\,{h}^{7}}}
- {\frac{447738400}{531441\,{h}^{8}}}
- {\frac{7244437856}{1594323\,{h}^{9}}}
+\CO\lr{h^{-10}}
\right)
\,.
\nonumber
\ea
As was explained in Sect.~6.2, substituting \re{q3-h0} into
\re{en-im} we can obtain the expansion of the energy with the same
accuracy
\ba
\varepsilon_3(h;0)&=&
-6\,\ln \lr{\frac{h\,\e^{\gamma_{_{\rm E}}}}{\sqrt 3}}
+{\frac {6}h}+{\frac {11}{6\,h^{2}}}
-{\frac {56}{9\,h^{3}}}-{\frac {190883}{4860\,h^{4}}}-{\frac {
135976}{729\,h^{5}}}
\lab{en-h0}
\\[3mm]
&&
\hspace*{20mm}
-{\frac {80780333}{91854\,h^{6}}}
-{\frac {
259171832}{59049\,h^{7}}}-{\frac {506229906961}{21257640\,h^{8}}}-
{\frac {226978281368}{1594323\,h^{9}}}+\CO(h^{-10})\,.
\nonumber
\ea
Let us compare this expression with an analogous expansion \re{div}
of the energy of the $n=2$ Reggeon states. We notice that in both cases
the absolute value of the coefficients in front of powers of $1/h$
rapidly increases in higher orders revealing the asymptotic nature
of the series. However, in contrast with \re{div} the series for
$\varepsilon_3(h;0)$ is not a sign--alternating. Starting from
$\CO(h^{-3})$ all terms in the series \re{en-h0} are negative
and this makes its properties different from \re{div}. As one can see from
fig.~\ref{ener-n=3}(b), the expression \re{en-h0} is in a good agreement
with the numerical results for the energy for large conformal
weight $h\ge 5$. For small $h$ the absolute value of the higher order
terms in \re{en-h0} starts to grow and the series \re{en-h0} becomes
divergent. In particular, substituting $h=1/2$ into \re{en-h0}
we obtain the series
\ba
\varepsilon_3(1/2;0)&=&3.99+12+7.33-49.78-628.42
           -5968.77
\lab{asy}
\\
&& \hspace*{20mm}
-56284.34 -561804.51-6096389.64 -72891678.83 + ... \,,
\nonumber
\ea
which should be compared with an analogous series in \re{as1/2}.

\begin{figure}[htb]
   \vspace*{-1cm}
   \centerline{
   \epsfysize=8cm\epsfxsize=10cm\epsffile{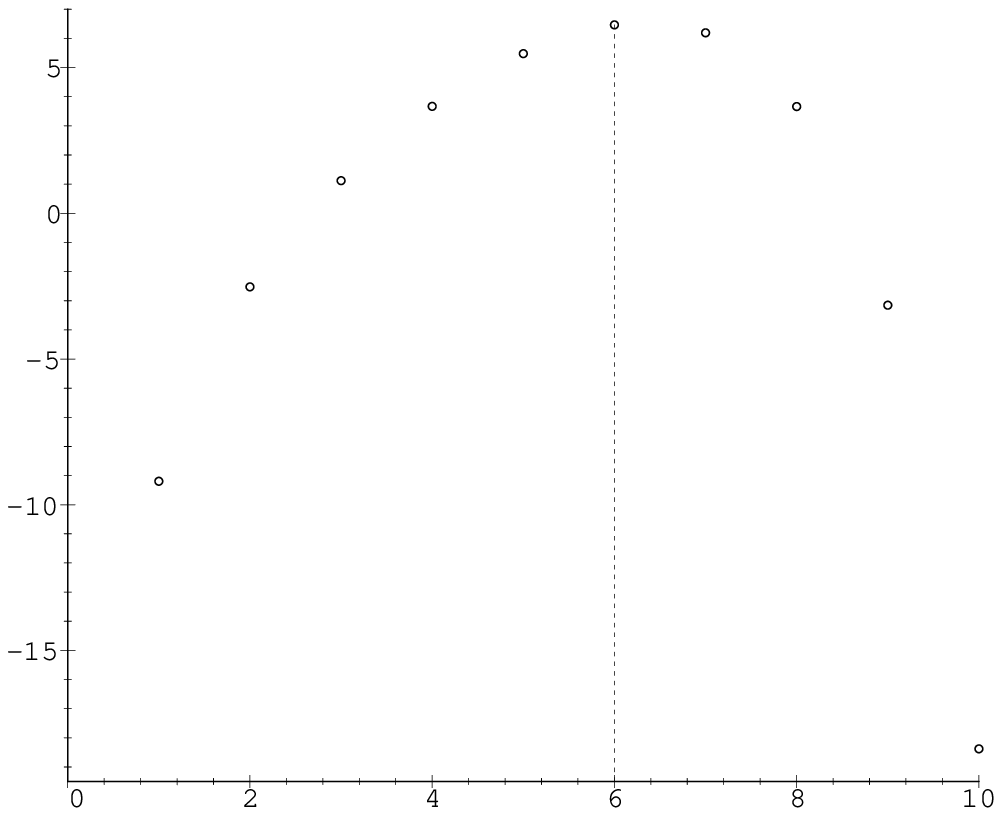}
   \hspace*{-2cm}
   \epsfysize=8cm\epsfxsize=10cm\epsffile{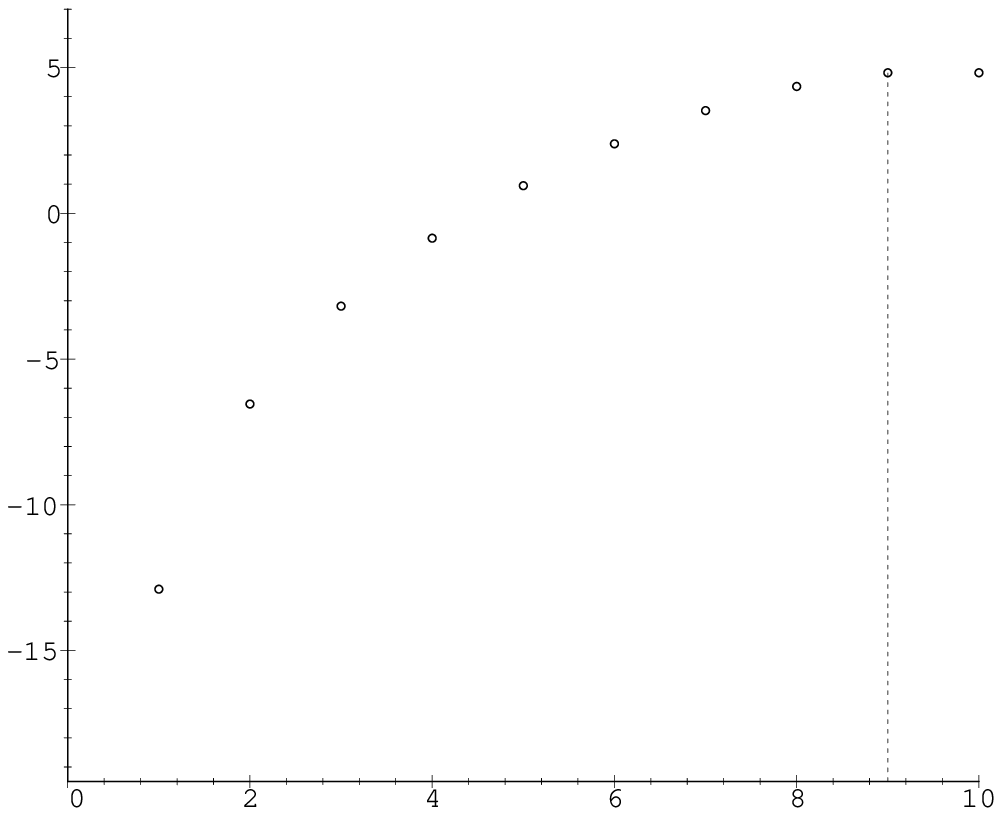}}
   \vspace*{-1cm}
   \centerline{ (a) \hspace*{7.3cm} (b)}
\caption{\label{f1} Partial sums $S_n$ of the Euler transformed series for
the energy of $n=3$ Reggeon states, \re{en-h0}, corresponding to $h=1/2$
and two values of the parameter of transformation:
(a) $x_0=4$ and (b) $x_0=7.85$. The dotted line indicates the position
of the asymptotic approximation $S_{\rm max}$.}
\unitlength=1mm
\begin{picture}(0,0)(0,0)
\put(8,80){\makebox(0,0)[cc]{${S_n}$}}
\put(89,80){\makebox(0,0)[cc]{${S_n}$}}
\put(70,29){\makebox(0,0)[cc]{$n$}}
\put(151,29){\makebox(0,0)[cc]{$n$}}
\end{picture}
\vspace*{-0.5cm}
\end{figure}

To evaluate the energy $\varepsilon_3(h;0)$ for $h < 5$ we
have to define the asymptotic approximation to the series \re{en-h0}.
Let us denote the partial sum of the first $n$ terms of the
series \re{en-h0} as $S_n$. Then,
it follows from \re{en-h0} and \re{asy} that the partial sum $S_n$
grows as $n$ increases, at $n=3$ it takes a maximal value, $S_{\rm max}$,
and for $n\ge 4$ it rapidly decreases. It is clear that $S_{\rm max}$
overestimates the energy while the partial sum $S_n$ for large $n$
underestimates it. This means that the remainder term, as a function
of a discrete $n$, first decreases for small $n$ until it approaches a
minimal negative value corresponding to $S_n=S_{\rm max}$ and then it rapidly
grows to $\infty$. To our practical needs, we identify the asymptotic
approximation to the energy as
\be
\varepsilon_3^{\rm app}(h;0)=S_{\rm max}={\rm max}_{_{n=1,\,2,...}} S_n\,.
\lab{app3}
\ee
Under this choice, the remainder term $\delta\varepsilon_3=\varepsilon_3
-\varepsilon_3^{\rm app}$ is always negative and the approximant \re{app3}
defines an {\it upper\/} limit
for the energy $\varepsilon_3(h;0)$. For the series \re{en-h0} and \re{asy}
the expression for $S_{\rm max}$ is given by the sum of the first three
terms. To increase the number of terms contributing to $S_{\rm max}$ and,
as a consequence, to improve the accuracy of the asymptotic approximation
\re{app3} we apply the Euler transformation to the series \re{en-h0}.
The behaviour of the partial sum, $S_n(h=1/2,x_0)$, as a function of $n$ is
shown on figs.~\ref{f1}(a) and (b) for two different values of the parameter
of transformation, $x_0=4$ and $x_0=7.85$, respectively.
\phantom{\ref{f1}}
Changing $x_0$ we find from fig.~\ref{f1}
that as $x_0$ increases, the growth of $S_n$ with $n$ becomes
less steep and the value of $S_{\rm max}$ slowly
decreases. Since $S_{\rm max}$
defines the upper bound for the energy $\varepsilon_3(h;0)$,
the latter property implies that the accuracy of the asymptotic
approximation \re{app3} becomes better. The dependence of
$S_{\rm max}(h;x_0)$ on $x_0$ for $h=1/2$ is shown on fig.~\ref{f2}(b).
It suggests that
approximant $S_{\rm max}$, being a decreasing function of $x_0$,
asymptotically approaches the energy $\varepsilon_3(h;0)$ from
above for large enough $x_0$.

\begin{figure}[htb]
   \vspace*{-1cm}
   \centerline{
   \epsfysize=10cm\epsfxsize=10cm\epsffile{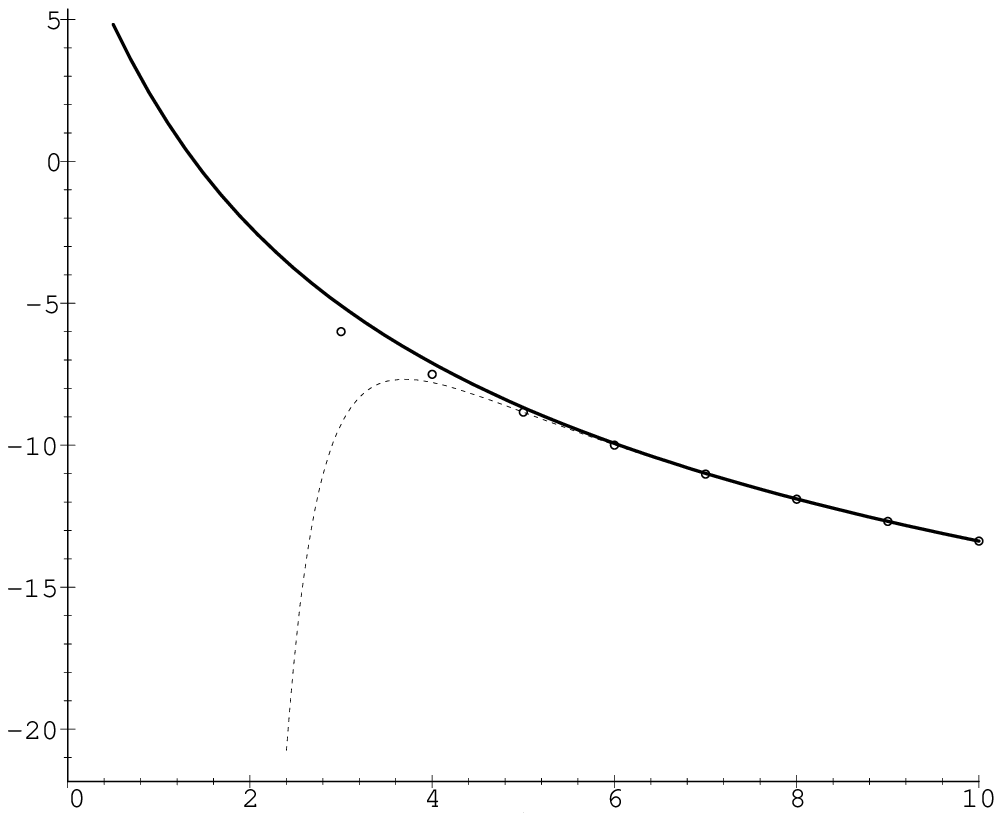}
   \hspace*{-2cm}
   \epsfysize=10cm\epsfxsize=10cm\epsffile{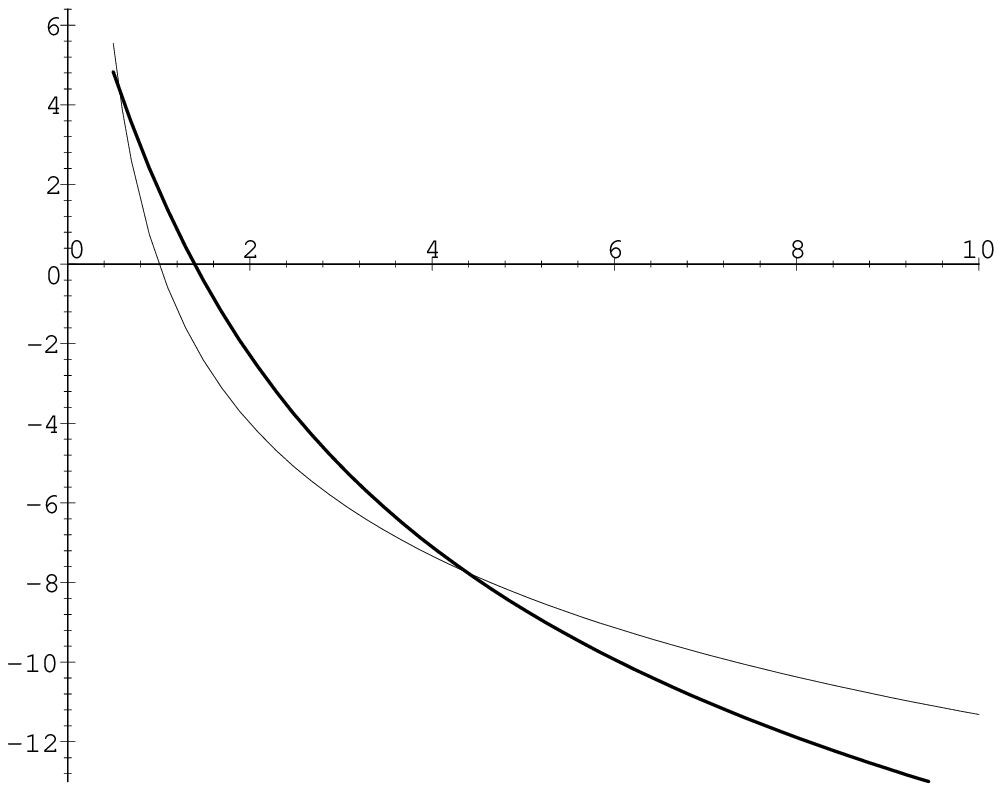}}
   \vspace*{-1cm}
   \centerline{ (a) \hspace*{7.3cm} (b)}
\caption{\label{f3} Holomorphic energy of the $n=3$ Reggeon compound
states: (a) Dots represent numerical results for the energy
$\varepsilon_3=\varepsilon_3(h;0)$. Dotted line corresponds to the
asymptotic series \re{en-h0}. Thick line on (a) and (b),
$\varepsilon_3=\varepsilon_3^{\rm app}(h;0)$, was found after the Euler
transformation of the series \re{en-h0} with the parameter $x_0=7.85$.
Thin line on (b) is the energy of the $n=2$ Reggeon states,
$\varepsilon_2(h)$, defined in \re{en-n=2}.}
\unitlength=1mm
\begin{picture}(0,0)(0,0)
\put(8,104){\makebox(0,0)[cc]{$\varepsilon_3$}}
\put(89,104){\makebox(0,0)[cc]{$\varepsilon_3$}}
\put(70,42){\makebox(0,0)[cc]{$h$}}
\put(151,85){\makebox(0,0)[cc]{$h$}}
\end{picture}
\vspace*{-0.5cm}
\end{figure}

Increasing the value of $x_0$ we have to check that the number of
terms contributing to $S_{\max}$ is less than 10, that is the number of
terms in the expansion \re{en-h0}. Otherwise, additional higher $1/h-$terms
should be taken into account in the asymptotic expansion \re{en-h0}.
We found that the maximal value of $x_0$ for the series \re{en-h0}
which satisfies this condition
is close to $x_0=7.85$. Applying the Euler
transformation to the series \re{en-h0} and taking $x_0=7.85$ we obtain the
asymptotic approximation to the energy of the $n=3$ Reggeon states,
$\varepsilon_3^{\rm app}(h;0)$, shown on fig.~\ref{f3}.
\phantom{\ref{f3}}

Let us summarize the properties of the function
$\varepsilon_3^{\rm app}(h;0)$. According to the definition \re{app3},
this function approximates the energy of the $n=3$ Reggeon
states $\varepsilon_3=\varepsilon_3(h;0)$ for an arbitrary
value of the conformal weight $1/2 \le h < \infty$.
Although its accuracy was not defined yet, we
know in advance that the asymptotic approximation \re{app3}
provides {\it upper\/} bound for the energy. For large values
of the conformal weight, $h>5$, the expression \re{en-h0} is in a good
agreement with the numerical results for the energy and the asymptotic
approximation $\varepsilon_3^{\rm app}(h;0)$ makes it
even better. For smaller values of $h$ the expansion \re{en-h0} becomes
divergent and it is approximated by the function
$\varepsilon_3^{\rm app}(h;0)$. To estimate the accuracy of the
approximation we use the numerical results for the energy,
$\varepsilon_3(h;0)$, and compare them
with the corresponding values of $\varepsilon_3^{\rm app}(h;0)$
to find the remainder terms
$\delta\varepsilon_3=\varepsilon_3-\varepsilon_3^{\rm app}$. The
numerical data are
summarized in Table.~2.
As was expected, the remainder terms are negative.
\begin{table}[htb]
\centerline{
\begin{tabular}{|l|c|c|c|c|c|c|c|c|}
\hline
$h$ & $3$ & $4$ & $5$ & $6$ & $7$ & $8$ & $9$ & $10$
\\ \hline
$\varepsilon_3^{\rm app}$ &
$-5.0809$ & $-7.1050$ & $-8.6696$ & $-9.9333$ & $-10.9880$ & $-11.8907$
& $-12.6783$ & $-13.3763$  \\
\hline
$\delta \varepsilon_3$ &
$-0.9191$ & $-0.3950$ & $-0.1637$ & $-0.0667$
& $-0.0273$ & $-0.0114$ & $-0.0050$ &
$-0.0022$ \\
\hline
\end{tabular}}
\caption{
The asymptotic approximation to $\varepsilon_3(h;0)$ based on the Euler
transformation for $x_0=7.85$.
   }
\end{table}

Performing analytical continuation of the energy to $h < 3$ we
use the symmetry \re{q3-q3} of the Baxter equation under the
replacement $h\to 1-h$ to restrict the value of the conformal
weight to the fundamental domain, $1/2 \le h < \infty$.
We notice from fig.~\ref{f3} that $\varepsilon_3^{\rm app}(h;0)$
is a decreasing function of $h$.
This implies that, in accordance with \re{1/2}, the maximal value of the
energy of the $n=3$ Reggeon state corresponds to $h=1/2$ and
it is given by
\be
{\rm max}_{_{h}}\varepsilon_3^{\rm app}(h;0)=
\varepsilon_3^{\rm app}(1/2;0)=4.826208\,.
\lab{odd-up}
\ee
We stress that this expression should be considered as upper bound
for the energy $\varepsilon_3(1/2;0)$.
The function $\varepsilon_3^{\rm app}(h;0)$ depends on the parameter
of the Euler transformation $x_0$. For $x_0=7.85$ we find from
fig.~\ref{f3} that $\varepsilon_3^{\rm app}(h;0)$ crosses
$\varepsilon_2(h)=-4(\psi(h)-\psi(1))$ at two points,
$h_+=4.38$ and $h_-=0.56$. For fixed $h$, the value of
$\varepsilon_3^{\rm app}(h;0)$ decreases as $x_0$ increases
(see fig.~\ref{f2}(b)) and the distance between
$h_+$ and $h_-$ becomes smaller. One might expect that
the distance $h_+-h_-$ controls the accuracy of the asymptotic
approximation and in the limit $h_+-h_-\to 0$ the function
$\varepsilon_3^{\rm app}(h;0)$ coincides with the
exact expression for the energy, $\varepsilon_3(h;0)$. We recall
that for integer $N_1\ge 1$ the function $\varepsilon_3(h;N_1)$ crosses
$\varepsilon_2(h)$ at only one point $h=2N_1+3$. If one assumes that
the same property holds for $N_1=0$, then $h_+=h_-=3$ and
$\varepsilon_3(h;0) \le -4(\psi(h)-\psi(1))$. For $h=1/2$ this
leads to the relation
\be
\varepsilon_3(1/2;0) < 8\ln 2\,,
\lab{odd-bfkl}
\ee
which is in agreement with \re{odd-up}. Finally, taking into account
\re{odd-up} we find from \re{max} and \re{En} the
asymptotic approximation of the intercept of the Odderon as
\be
\alpha_{_{\rm Odderon}}^{\rm app} = 1+\frac{\as N}{\pi}\, 2.413104\,.
\lab{est}
\ee
This expression is smaller than the intercept of the BFKL Pomeron, \re{bfkl},
and it approximates $\alpha_{_{\rm Odderon}}$ from above. The estimate \re{est}
is also in agreement with the lower bound for the Odderon intercept
proposed in \ci{odd}. 

One may try to repeat the analysis and apply the Euler transformation to
the asymptotic series \re{q3-h0} to find the approximation to
the quantum number $q_3(1/2;0)$ corresponding to the maximal energy of
the $n=3$ Reggeon states. We find however that since all terms in the
power series \re{q3-h0} except of the $\CO(h^{-2})$ term have the same sign,
the Euler transformation does not improve its convergency properties
and we are not able to construct the convergent sequence of the partial
sums similar to \re{app3} for the energy. Moreover,
as we will show in the next section, the coefficients in \re{q3-h0} grow
as factorials and the asymptotic series for $q_3(h;0)$ is not Borel
summable. Unfortunately, there is no regular way for approximating non 
Borel summable series like \re{q3-h0} and in order to calculate
$q_3(1/2;0)$ one has to develop another summation method.

\subsection{Borel summability}

As was found in the previous sections, the asymptotic expansion of the
energy of the $n-$Reggeon states has the following general form
\be
\varepsilon_n(h,\{q\})=-2\ln(h^n \hat q_n\,\e^{n\gamma_{_{\rm E}}}) +
\sum_{k\ge 1} \frac{f_k}{h^k}
\lab{Ser}
\ee
where $\hat q_2=1$, $\hat q_3=\frac1{\sqrt{27}}$, $...$ and
$f_k$ are rapidly growing coefficients. It is important now to
understand whether the asymptotic series \re{Ser} defines uniquely
the energy. To this end we perform the Borel transformation of the
power series entering into \re{Ser}
\be
\varepsilon_n(h,\{q\})=-2\ln(h^n \hat q_n\e^{n\gamma_{_{\rm E}}}) +
\int_0^\infty dt\, B_{\varepsilon_n}(t)\, \e^{-ht}\,.
\lab{Ser-B}
\ee
Here, the function $B_{\varepsilon_n}(t)$ is defined as
\be
B_{\varepsilon_n}(t)=\sum_{k\ge 0}\frac{f_{k+1}}{k!} t^k
                    =\sum_{k\ge 0} A_{k,\{q\}}\, \e^{-kt} - \frac{2n}t
\lab{Bo}
\ee
where in the last relation we used \re{Ak1} and \re{Ak2} 
and the additional term, $-2n/t$, was added to provide regularity of 
$B_{\varepsilon_n}$ as $t\to 0$. The properties of the
function $B_{\varepsilon_n}(t)$ depend on the large order behaviour of 
the coefficients $f_k$. Let us introduce the following ansatz \ci{ZJ}
\be
f_k \sim c\, k!\, a^{-k}\, k^b \left(1+\CO(1/k)\right)
\qquad \mbox{as $k\to \infty$}\,,
\lab{fk}
\ee
which being substituted into \re{Bo} leads to the singularity of
the function $B_{\varepsilon_n}(t)$ at $t=a$
\be
B_{\varepsilon_n}(t)\sim \frac1{(t-a)^{b+2}}\,.
\lab{sin}
\ee
To verify \re{fk} one has to compare $f_k$ with their numerical values
in the asymptotic expansions \re{div} and \re{en-h0} and extract the values
of the parameters $a$, $b$ and $c$.

For the $n=2$ Reggeon states one uses the last three terms in the
expansion \re{div} to find the following values: $a=\pm 6.2746\, i$, 
$b=-1.0201$ and $c=-8.2657$, where the sign of $a$ can not fixed 
since the expansion \re{div} contains only even powers of $h$ in higher 
orders. These values are very close to the exact expressions
\be
\qquad a=\pm 2\pi i\,, \quad b=-1\,, \quad c=-8\,, 
\lab{abc-n=2}
\ee
which follow from the well--known asymptotic expansion of the
$\psi-$function in \re{en-n=2}. Moreover, the function
$B_{\varepsilon_n}(t)$ can be evaluated explicitly using \re{Bo} and 
\re{Ak} as
$$
B_{\varepsilon_2}(t)=\frac4{1-\e^{-t}}-\frac4{t} \,.
$$
We check that, in accordance with \re{sin} and \re{abc-n=2}, the nearest
to the origin singularity of this function is located at $t=\pm 2\pi i$.
At the same time, the function $B_{\varepsilon_2}(t)$ does not have 
singularities on the integration path in \re{Ser-B} and as a 
consequence the asymptotic series \re{Ser-B} for the energy of the $n=2$ 
Reggeon states is Borel summable and it can be approximated using the
well--known methods \ci{ZJ}.

Let us check the large order behaviour of the energy of the $n=3$
Reggeon states \re{en-h0} using the ansatz \re{fk}. We find that for the 
set of parameters
\be
a=1.8098\,,\quad
b=1.8195\,,\quad
c=-1.5000
\lab{abc-n=3}
\ee
the relation \re{fk} reproduces the coefficients $f_4$, ..., $f_9$
with 8\% accuracy. This result indicates that the function 
$B_{\varepsilon_3}(t)$ has singularity \re{sin} at the point $t=1.8098$
which belongs to the integration path in \re{Ser-B}. Therefore the Borel
transformation does not exist and the asymptotic series for the energy
of the $n=3$ Reggeon states is not Borel summable. Different
prescriptions for integrating the singularity of $B_{\varepsilon_3}(t)$ 
lead to the results for the energy which differ in $\sim\exp(-ha)$
terms. Thus, the asymptotic series \re{en-h0} does not define the energy 
uniquely but rather represents one member of an infinite class of
functions having the same asymptotic expansion. To avoid ambiguity and fix a
particular member of the class we have to provide additional information
about analytical properties of the energy. This requires solution of
Baxter equation beyond the large $h$ approximation. 

We notice that the asymptotic expansion of $q_3(h;0)$ defined in \re{q3-h0}
exhibits the large order behaviour similar to \re{en-h0}. Indeed, rewriting
\re{q3-h0} as
\be
q_3(h;0)=\frac{\sqrt 3}9 h^3\left(1+\sum_{k\ge 1} \frac{g_k}{h^k}
\right)
\lab{gk}
\ee
we find that the coefficients $g_k$ $(k=4,...,9)$ satisfy the
asymptotic behaviour \re{fk} with 2\% accuracy provided that
\be
a=1.8174\,,\quad
b=0.8019\,,\quad
c=-0.4650\,.
\lab{abc-q3}
\ee
As a consequence, the asymptotic series for $q_3(h;0)$ suffers from the
same problems as that for $\varepsilon_3(h;0)$. We notice that the
expressions \re{abc-n=3} and \re{abc-q3} suggest the following relation
between asymptotic expansions for the $n=3$ Reggeon states: 
$g_k k / f_k \approx 0.28$.

\sect{Conclusions}

The spectrum of the color singlet $n$ Reggeon compound states --
perturbative Pomerons and Odderons, is expressed by means of the
Bethe Ansatz in terms of the solution of the Baxter equation for the XXX
Heisenberg magnet. In this paper we developed the method which allows us to
find the solution of the Baxter equation for an arbitrary number of
the Reggeized gluons.

The method is based on the observation that in
the limit of large integer conformal weight $h$ of the Reggeon states, the
Baxter equation takes a form
of a discrete one--dimensional Schrodinger equation for a particle in the
external potential $V(\lambda)$, which depends on the conserved charges
$q_3$, $...$, $q_n$ of the Reggeon states and is singular at the
origin $\lambda\to 0$. The inverse conformal weight,
$1/h$, plays a role of the Planck constant and this fact allows us to
apply the well--known quasiclassical expansion and obtain the solution
of the Baxter equation, $\Phi(\lambda)$, as an asymptotic series in $1/h$.
We found that the derivative of the solution, $\Phi'(\lambda)$, has a
discontinuity across a finite disconnected interval ${\cal S}$
on the real axis  in the complex $\lambda-$plane, whose position and the number
of connected intervals inside  ${\cal S}$ depend on the potential or
equivalently on the value of the charges $q_3$, $...$, $q_n$.
The quantization conditions for the charges can be reformulated as certain
constraints on ${\cal S}$. Namely, for the $n$ Reggeon compound states the
interval of nonanaliticity ${\cal S}$ should consist of $n-1$ connected
intervals. The corresponding quantization conditions have solutions
for charges which can be parameterized by the set of integer numbers
$N_1$, $...$, $N_{n-1}$. We have shown that the value of quantized
$q_3$, $...$, $q_n$ can be obtained as a series in $1/h$ and
the expressions for the lowest order coefficients were found. We observed
that for $n=3$ and $n=4$ Reggeon states they are in a complete agreement
with the results of the numerical solutions of the Baxter equation.

The energy of the $n$ Reggeon states depends on the conformal weight $h$
and quantized charges $q_3$, $...$, $q_n$.
Using the quasiclassical solution of the
Baxter equation we obtained the expansion of the energy in powers of
$1/h$ for $h\ge n$ and then analytically continued the result for small
values of conformal weight, $1/2 \le h < n$. For large integer $h$
it agrees with the results of the numerical solutions while for
small $h$ the asymptotic expansion of the energy becomes divergent
and it should be replaced by the asymptotic approximation.
Using the properties of the asymptotic series and applying the Euler
transformation we defined the asymptotic approximation of the energy
for small $h$ and checked its validity in the special case of the
$n=2$ Reggeon states by comparing with the exact expression for the energy.
As first nontrivial application, we derived the asymptotic approximation
of the energy of the $n=3$ Reggeon state, perturbative Odderon, and estimated
its intercept. The generalization of the obtained results to the
higher Reggeon states is straightforward.

\bigskip

This work was supported in part by the National
Science Foundation under grant PHY 9309888.

\appendix{A}{Analytical properties of the energy}

\noindent
Let us show that the energy of the $n-$Reggeon compound states satisfies the
conditions of the Carlson's theorem and therefore it has the unique
analytical continuation in the region \re{integer}. 

We start with the simplest case of the $n=2$ Reggeon states, in which the
holomorphic energy depends only on the conformal weight $h$ and it is given 
by \re{en-n=2}. The function $\varepsilon_2(h)$ has poles at the origin $h=0$ 
and at negative integer $h$
\be
\varepsilon_2(h)\stackrel{h\to -k}{\sim}\frac4{h+k}\,, \qquad
k=0, 1, 2, ...
\lab{poles}
\ee
and its asymptotics at infinity is
\be
\varepsilon_2(h)\stackrel{h\to\infty}{\sim}-4\ln h\,.
\lab{infty}
\ee
Thus, the function $\varepsilon_2(h)$ has the unique continuation from 
integer to complex $h$ defined in \re{h}. 

Let us generalize the relations \re{poles} and \re{infty} to the $n-$Reggeon
states. Their holomorphic energy $\varepsilon_n$ was obtained in Sect.~4 in 
the form of the asymptotic series in $1/h$ whereas the analytical expression 
for $\varepsilon_n$ similar to \re{en-n=2} is not available yet. In the large 
$h$ limit the asymptotic behaviour of $\varepsilon_n$ was found in \re{en-LLA}
as 
\be
\varepsilon_n(h)\stackrel{h\to\infty}{\sim}-2n\ln h\,,
\lab{infty-n}
\ee
and it is in agreement with \re{infty} for $n=2$. Trying to satisfy
the second condition of the Carlson's theorem we find that having only a few
first terms, \re{en-psi-3} and \re{en-im}, of the asymptotic series for 
$\varepsilon_n$ we are not able to identify all singularities of 
$\varepsilon_n$ in the complex $h-$plane. 

There is however another way to study the singularities of the energy of the 
Reggeon compound states based on their relation with the anomalous dimensions 
of composite higher twist operators entering into the operator product 
expansion of the structure function of deep inelastic photon--hadron 
scattering at small values of the Bjorken variable $x$ \ci{Lip1}.
Let us consider the structure function at small $x$ in the generalized leading
logarithmic approximation \ci{K}. We choose for simplicity the hadron to be
a perturbative onium state built from two heavy quarks and created via
the decay of the photon with invariant mass $m$. The structure function 
$F(x,Q^2)$ is defined as an imaginary part of the amplitude of the forward 
photon--onium scattering with the center-of-mass energy $s=Q^2 (1-x)/x$ and 
photon virtuality $-Q^2$. Let us consider the asymptotics of the structure 
function in the double scaling limit of large $Q^2/m^2$, and small 
$x \approx Q^2/s$.

In the limit $Q^2 \gg m^2$, the structure function can be expanded in powers
of $1/Q^2$ using the operator product expansion (OPE). In the leading
$\ln Q-$approximation, $\as \ll 1$ and $\as \ln(Q^2/m^2) \sim 1$, this 
expansion looks like
\be
F_\omega(Q^2)\equiv \int_0^1 dx\, x^{\omega-1} F(x,Q^2)
=\frac1{Q^2}\sum_{k=1}^\infty C_\omega^{(k)}
\lr{\frac{m^2}{Q^2}}^{k-1+\gamma_{\omega}^{(k)}}\,,
\lab{OPE}
\ee
where the $k-$th term is associated with the contribution of twist$-2k$
operators. Their matrix elements have the anomalous dimensions 
$\gamma_\omega^{(k)}=\CO(\as)$ which have a perturbative expansion in 
powers of $\as$.

In the small $x$ limit, or equivalently in the limit of large energy
$s\approx Q^2/x \gg Q^2$, the structure function has the Regge behaviour 
which it is governed by the compound Reggeon states propagating in the 
$t-$channel
$$
F_\omega(Q^2)=\sum_{n=2}^\infty \as^{n-2} F_{\omega,n}(Q^2)\,,
$$
where the contribution of the $n-$Reggeon states can be found using
\re{An} for $s\approx Q^2/x$ and $t=0$ as
\be
F_{\omega,n}(Q^2)=\sum_{\{q\}} \frac{1}{\omega-E_{n,\{q\}}}
\beta_{n\to\gamma^*(Q^2)}^{\{q\}}(0) \lr{\beta_{n\to h(m^2)}^{\{q\}}(0)}^*\,.
\lab{mom}
\ee
Here the summation is performed over all quantum numbers $h$, $q_3$, $...$,
$q_n$ corresponding to the $n$ Reggeon compound state with the energy
$E_{n,\{q\}}$. The residue factors have been defined in \re{beta} and in 
the case of the forward scattering, $t=0$, they depend only on the invariant
masses of scattering particles, $Q^2$ and $m^2$. 
In the generalized leading logarithmic approximation, one may calculate the 
residue factors for perturbative states of virtual photon, $\gamma^*(Q^2)$, 
and onium, $h(m^2)$, in the Born approximation and neglect $\as$ corrections. 
As a result, 
$\beta_{n\to\gamma^*}^{\{q\}}$ and $\beta_{n\to h}^{\{q\}}$ do not have 
anomalous dimension and their scaling dimensions are equal to the sum of 
the scaling dimensions of the $n$ Reggeon state, $h+\bar h=1+2i\nu$, and 
the scaling dimensions of photon and onium states
$$
\beta_{n\to\gamma^*(Q^2)}^{\{q\}}(0)=C_{n\to\gamma^*}^{\{q\}}\ Q^{-1+2i\nu}\,,
\qquad
\beta_{n\to h(m^2)}^{\{q\}}(0)=C_{n\to h}^{\{q\}}\ m^{-1+2i\nu}
$$
where the dimensionless coefficients depend on the quantum numbers of the 
Reggeon states and scattering particles. Substituting these relations into 
\re{mom} we obtain
\be
F_{\omega,n}(Q^2)=\frac1{Q^2}
\sum_{q_3,...,q_n}
\int_{-\infty}^\infty d \nu \sum_{m\ge 0}
\frac{C_{n\to\gamma^*}^{\{q\}}C_{n\to h}^{\{q\}}}
{\omega-\frac{\as N}{4\pi}
\left[\varepsilon_n(\frac{1+m}2+i\nu;\{q\})
   +\varepsilon_n(\frac{1+m}2-i\nu;\{q\})\right]}
\lr{\frac{m}{Q}}^{-1-2i\nu}\,.
\lab{mom-sum}
\ee
Here, we extracted the sum over quantized values \re{h} of the conformal
weight, that is summation over discrete $m$ and integration over continious
$\nu$, from the sum over all quantum numbers in \re{mom}.

Let us consider \re{mom-sum} in the limit $Q^2 \gg m^2$, in which  
one should be able to reproduce $1/Q^2$ expansion \re{OPE}. 
For $Q^2 \gg m^2$ one can enclose the integration contour over $\nu$ into 
the lower half-plane, $\Im \nu <0$, and calculate the integral over $\nu$ 
in \re{mom-sum} by taking the residue at the values of $\nu$ which satisfy 
the relation
\be
\frac{4\pi\omega}{\as N}=\varepsilon_n\lr{\frac{1+m}2+i\nu;\{q\}}
                        +\varepsilon_n\lr{\frac{1+m}2-i\nu;\{q\}}\,.
\lab{nu}
\ee
Solving this equation one can find the values of $i\nu$ which determine the 
power of $m/Q$ in the $1/Q-$expansion of the structure function \re{OPE}, or 
equivalently define the scaling dimensions of the composite operators entering 
into the OPE. The comparison of \re{mom-sum} 
with \re{OPE} requires that the solutions of
\re{nu} should have the following form for $\Im \nu <0$:
\be
i\nu=-\lr{\frac12+Z}+ \CO(\as)\,,\qquad Z=0,\,1\,, ...\,.
\lab{nu-Z}
\ee
Let us now take into account that in order for the contribution of the pole 
in $\nu$ to \re{mom-sum} to be nonvanishing, the residue factors,
or equivalently the coefficients $C_{n\to\gamma^*}^{\{q\}}$ and 
$C_{n\to h}^{\{q\}}$, should be different from zero. This condition imposes
selection rules on the quantum numbers $h$, $q_3$, $...$, $q_n$ of the
Reggeon states. In particular, for the residue factors to be scalar,
the conformal spin of the Reggeon state, $h-\bar h=m$, should be equal
to the conformal spin of the photon and onium states \ci{Lip1}:
$m=0$ or $m=2$. 

Then, substituting \re{nu-Z} into \re{nu} and taking the limit $\as\to 0$ we 
find that the holomorphic energy $\varepsilon_n=\varepsilon_n(h;\{q\})$ has 
simple poles at the origin and at the integer negative values of the conformal
spin
$$
\varepsilon_n(h;\{q\}) \stackrel{h\to -k}{\sim} \frac{A_{k,\{q\}}}{h+k}
\,,\qquad
k=0,\,1,\,2\,,..\ .
$$
Moreover, using the asymptotic behaviour \re{infty-n} and calculating the
discontinuity of the energy at the negative $h$ one can write the
dispersion relation for the function $\varepsilon_n(h;\{q\})$ in the complex
$h-$plane which leads to
\be
\varepsilon_n^{\rm pole}(h;\{q\}) = \sum_{k=0}^\infty \frac{A_{k,\{q\}}}{h+k}
+ C
\lab{Ak1}
\ee
with $C$ some infinite $h-$independent subtraction constant. One can check
that this relation holds for the energy of the $n=2$ Reggeon states \re{en-n=2}
provided that 
\be
A_k=4
\lab{Ak}
\ee
and the constant $C$ can be found using the
condition $\varepsilon_2(1)=0$. The asymptotics of the energy \re{infty-n}
implies the following relation: 
\be
A_{k,\{q\}}{\sim} 2n\,,\qquad
\mbox{as $k \to \infty$.}
\lab{Ak2}
\ee
Thus, the consistency of the small $x$ asymptotics and the large $Q^2$ 
expansion of the structure function allows us to identify singularities of the
holomorphic energy of the $n-$Reggeon states and, as a consequence,
uniquely define $\varepsilon_n$ in the half-plane $\Re h \ge 1/2$ by its
values at the integer positive $h$.

\appendix{B}{Conformal operators}

\noindent
The $n$ Reggeon compound states corresponding to the polynomial
solutions of the Baxter equation have a simple interpretation in
terms of the so--called conformal operators \ci{ER}--\ci{BF}.
Let us consider the
holomorphic wave function of the Reggeon states defined in \re{H}.
As was explained in Sect.~1, the same wave function diagonalizes
the hamiltonian of the XXX Heisenberg magnet of spin $s=0$.
The explicit expression for $\varphi_n(\{z_k\};z_0)$ in terms
of the solution of the Baxter equation was found in \ci{FK,K}
and it is based on the correspondence between eigenstates of the
XXX Heisenberg magnets of spins $s=0$ and $s=-1$
\be
\varphi_n(\{z_k\};z_0) =
(z_{12}z_{23}...z_{n1})\, \varphi_n^{(s=-1)}(\{z_k\};z_0)
\lab{s=-1}
\ee
where $z_k$ are the holomorphic coordinates of the Reggeons and $z_0$
is the holomorophic coordinate of the center--of--mass of the
Reggeon compound state. Invariance of the Reggeon hamiltonian under
the conformal transformations \re{ct} leads to the following constraints
\ba
\lr{\sum_{k=1}^n\partial_k +\partial_0} \varphi_n^{(s=-1)} = 0\,,
\nonumber
\\
\lr{\sum_{k=1}^n z_k\partial_k+z_0\partial_0+h+n} \varphi_n^{(s=-1)} = 0 \,,
\lab{Ward}
\\
\lr{\sum_{k=1}^n z_k^2\partial_k
     +2z_k+z_0^2\partial_0+2hz_0}\varphi_n^{(s=-1)} = 0\,.
\nonumber
\ea
The expression for $\varphi_n(\{z_k\};z_0)$ proposed in \ci{FK,K}
satisfies these relations. However, instead of using the Bethe Ansatz
solution for $\varphi_n(\{z_k\};z_0)$ we would like to interpret
\re{Ward} as conformal Ward identities for $(n+1)-$point correlation
functions in some two--dimensional conformal field theory \ci{Lip1,Lip2}
\be
\varphi_n^{(s=-1)}(\{z_k\};z_0)=\VEV{\phi(z_1) ... \phi(z_n) O_h(z_0)}\,.
\lab{corfun}
\ee
Here, the field $\phi(z_k)$ describes the Reggeon with the holomorphic
coordinate $z_k$ and the operator $O_h(z_0)$ extrapolates the
compound $n$ Reggeon state. To satisfy \re{Ward}, the fields $\phi(z_k)$ and
$O_h(z_0)$ should be quasiprimary operators with conformal weights
$1$ and $h$, respectively \ci{BPZ}. The conformal Ward identities
fix two-- and three--point correlation functions up to a normalization
constant as
$$
\VEV{\phi(z_1)\phi(z_2)} = {\rm const.}\times \, z_{12}^{-2}\,,\qquad
\VEV{\phi(z_1)\phi(z_2)O_h(z_0)} = {\rm const.}\times \,
z_{10}^{-h}z_{20}^{-h}z_{12}^{h-2}\,.
$$
Substituting the last relation into \re{corfun} and \re{s=-1}
one can find the holomorphic wave function of the $n=2$
Reggeon compound state \ci{bfkl}.

For integer $h\ge n$ one can construct $O_h(z_0)$ as a
composite operator built from $n$ Reggeon fields $\phi$ and their
derivatives. The corresponding construction has been developed
in QCD many years ago \ci{M,Ohr,Lam,BF} within the framework of
the so--called ``collinear'' conformal group and it can be trivially
generalized
to the ``transverse'' conformal group which is a symmetry group
of the Reggeon states. The properties of the resulting conformal
operators can be summarized as follows. For positive integer
conformal weight $h\ge n$, one can define the basis of the
conformal operators and then represent all possible quasiprimary
operators $O_h(z_0)$ as their linear combination. For fixed
number of the Reggeons, $n$, the conformal basis consists of the
operators $O_{l_1...l_{n-1}}(z_0)$ labelled by integers
$l_1$, $...$, $l_{n-1}$ satisfying the following conditions
$$
l_1\,,...\,,l_{n-1} \ge 0\,, \qquad \sum_{k=1}^{n-1} l_k = h - n\,.
$$
The explicit form of the operators $O_{l_1,...,l_{n-1}}(z_0)$ for
$z_0=0$ is \ci{Lam,BF}
\be
O_{l_1...l_{n-1}}(0)=\partial^{h-n}
P_{l_1...l_{n-1}}
\lr{\frac{\partial_1}{\partial},...,\frac{\partial_{n-1}}{\partial}}
\phi_1(0)...\phi_{n-1}(0)\phi_n(0)\,,
\lab{O}
\ee
where $\partial=\partial_1+...+\partial_n$, the derivative $\partial_k$
acts on the free field $\phi_k\equiv \phi$ and
$P_{l_1,...,l_{n-1}}(\alpha_1,...,\alpha_{n-1})$ is the so--called
conformal polynomial in $\alpha_1$, $...$, $\alpha_{n-1}$. For different
values of $l_1,...,l_{n-1}$ the functions
$P_{l_1,...,l_{n-1}}(\alpha_1,...,\alpha_{n-1})$ form the system of
biorthogonal polynomials on the simplex
$\alpha_1+...+\alpha_{n-1}+\alpha_n=1$ with the weight function
and the measure
$$
w(\alpha_1,...,\alpha_{n-1},\alpha_n)=(2n-1)!\,
\alpha_1...\alpha_{n-1}\alpha_n\,,\qquad
\int{\cal D}\alpha=\int_0^1 d\alpha_1...d\alpha_n\, \delta
\lr{\sum_{k=1}^n\alpha_k-1}\,.
$$
In the special case of the $n=2$ Reggeon state, thus defined conformal
polynomial coincides with the Gegenbauer polynomial \ci{ER,M,Ohr},
$P_l(\alpha)=
C_l^{3/2}(2\alpha-1)$, and the conformal operator interpolating
the BFKL Pomeron for integer conformal weight $h\ge 2$ has a form
$$
O_h^{_{({\rm BFKL})}}(0)=\lr{\partial_1+\partial_2}^{h-2}\,C^{3/2}_{h-2}
\lr{\frac{\partial_1-\partial_2}{\partial_1+\partial_2}}
\phi_1(0)\phi_2(0)\,.
$$
For $n=3$ Reggeon states, the conformal operators \re{O} are
parameterized by two integers, $l_1,\,l_2 \ge 0$ and $l_1+l_2=h-3$.
The conformal polynomials, $P_{l_1,l_2}$, are given by the
Appell polynomials \ci{Ohr} and the basis of conformal operators looks like
\be
O_{l_1,l_2}(0)=\partial^{h-3}\,
J_{l_1,l_2}\lr{5,2,2;\frac{\partial_1}{\partial},\frac{\partial_2}{\partial}}
\phi_1(0)\phi_2(0)\phi_3(0)
\lab{con-3}
\ee
with $\partial=\partial_1+\partial_2+\partial_3$. We recall that
an arbitrary conformal operator $O_h(0)$ is a linear combination of
the basis operators $O_{l_1,l_2}(0)$. However, for the conformal
operator $O_h(0)$ to interpolate the $n=3$ Reggeon state, the
correlation function \re{corfun} has to satisfy the additional condition
of being the eigenstate of the conserved charge $q_3$ defined in \re{opqn}.
Substituting the operators \re{con-3} into \re{corfun} and \re{s=-1}
one can define the basis of the wave functions $\varphi_{l_1,l_2}$,
in which the operator $q_3$ can be represented as a
finite--dimensional matrix $(q_3)_{l_1,l_2;l_1',l_2'}$. The eigenvalues
of this matrix give the quantized values of the charge $q_3$ and
the corresponding eigenstates define the decomposition of the
interpolating operator over the basis of conformal operators \re{con-3}.
It is clear that quantized values of $q_3$ can be parameterized in the same
form as the diagonal elements of the matrix
$$
q_3=q_3(l_1,l_2)\,,\quad\quad\quad l_1,\,l_2\ge 0\,\quad l_1+l_2=h-3\,.
$$
Comparing this relation with \re{para} we find that they are identical
provided that we identify integers as $N_1=l_1$ and $N_2=l_2+3$.
It can be easily seen that the same correspondence holds for the
higher Reggeon states, $N_1=l_1$, $...$, $N_{n-2}=l_{n-2}$ and
$N_{n-1}=l_{n-1}+n$.

\appendix{C}{``Practical'' asymptotical approximation}
 
\noindent
Suppose $f(z)$ is a function of $z$ having the asymptotic expansion
\be
f(z)=a_0 +\frac{a_1}{z}+\frac{a_2}{z^2}+ \cdots \,.
\lab{A1}
\ee
Let us assume that for a given $z$ the absolute values of the
first $n$ successive terms of the series decrease and then they increase
starting from $(n+1)-$th term. Then, the asymptotic expansion is expecting
to give the best approximation to $f(z)$ when it is truncated at about
$n=n(z)$ terms \ci{as1}. The corresponding remainder term
$$
R_n(z)\equiv f(z) - \sum_{k=0}^{n-1} \frac{a_k}{z^{k+1}}
$$
is called the optimum remainder term. The accuracy of the approximation is
controlled by the value of $R_n(z)$.
 
\bigskip
 
\noindent {\it Euler transformation\/}.
 
\bigskip
 
\noindent
There is a simple method, the so--called Euler's transformation, which allows
us to increase the accuracy of the asymptotic approximation \ci{as2,as1}.
Let us perform
the identical transformation of the series \re{A1} by expanding its each term
in a series of inverse powers of $z+x_0$ as
$$
f(z)=\sum_{k=0}^{\infty} \frac{a_k}{z^{k+1}}
    =\sum_{k=0}^{\infty} \frac{b_k}{(z+x_0)^{k+1}}\,,
\qquad
b_k=\sum_{j=0}^k \lr{{k\atop j}}\ x_0^j\ a_{k-j}
$$
with $x_0$ being a parameter of the transformation.
It turns out that under proper choice of the parameter $x_0$ the sum of the
first $n'=n'(z,x_0)$ terms of a new series converges more rapidly and the
corresponding optimum remainder term $R_{n'}$ becomes smaller then
$R_n$ for the original series.
 
\bigskip

\noindent {\it Estimation of the remainder terms\/}.
 
\bigskip
 
\noindent
Suppose that $S_n=\sum_{k=0}^n a_n$ is the partial sum of
sign--alternating series  $S_\infty=\sum_{k=0}^\infty a_n$
and $R_n$ is the corresponding remainder term. Then,
$S_n$ overestimate the $S_\infty$ when
the last term included is positive, $a_n>0$, and underestimate it when
$a_n<0$. This means that the remainder term $R_n$ has a sign opposite to $a_n$
and it follows from the relation $R_n-R_{n+1}=a_{n+1}$ that
$$
|R_n| < |a_n|\,, \qquad |R_n| < |a_{n+1}|\,.
$$
Thus, the absolute value of the optimum remainder term is smaller
than the last term included into the approximant and the first term
excluded from the approximant \ci{as3}.

\bb{99}
\bi{Col}  S.C. Frautschi, {\it Regge poles and S-matrix theory\/},
          New York, W.A. Benjamin, 1963;
\\        V. de Alfaro and T. Regge, {\it Potential scattering},
          Amsterdam, North-Holland, 1965;
\\        P.D.B. Collins, {\it An introduction to Regge theory
          and high energy physics\/}, Cambridge University Press, 1977.
\bi{softP}P.V. Landshoff, {\it The two Pomerons\/}, DAMTP preprint,
          hep-ph/9410250.
\bi{bfkl} E.A. Kuraev,  L.N. Lipatov and V.S. Fadin,
          Phys. Lett. B60 (1975) 50;
          Sov. Phys. JETP 44 (1976) 443; 45 (1977) 199;
\\        Ya.Ya. Balitsky and L.N. Lipatov, Sov. J. Nucl. Phys. 28 (1978) 822.
\bi{onium}A.H. Mueller and B. Patel, Nucl. Phys. B425 (1994) 471;
\\        A.H. Mueller, Nucl. Phys. B437 (1995) 107;
\\        Z. Chen and A.H. Mueller, Columbia U. preprint,
          CU--TP--691, Apr 1995.
\bi{Lip1} L.N. Lipatov, {\it Pomeron in quantum chromodynamics\/},
          in ``Perturbative QCD'', pp.411--489, ed. A.H. Mueller,
          World Scientific, Singapore, 1989.
\bi{Gr}   V.N. Gribov, Sov. Phys. JETP 26 (1968) 414.
\bi{CW}   H. Cheng, J. Dickinson, C.Y. Lo and K. Olaussen,
          Phys. Rev. D23 (1981) 534;
\\        H. Cheng and T.T. Wu, {\it Expanding Protons: Scattering at
          High Energies\/}, MIT Press, Cambridge, Massachusetts, 1987.
\bi{Bar}  J. Bartels, Nucl. Phys. B175 (1980) 365.
\bi{KP}   J. Kwiecinski and M. Praszalowicz, Phys. Lett. B94 (1980) 413.
\bi{odd2} B. Nicolescu, Nucl. Phys. (Proc.Suppl.) 25B (1992) 142.
\bi{odd}  P. Gauron, L.N. Lipatov and B. Nicolescu,
          Z. Phys. C63 (1994) 253; Phys. Lett. B304 (1993) 334;
\\        R. Janik, Jagellonian Univ. preprint TPJU--18/95, 
          Nov 1995 [hep-th/9511210].
\bi{odd1} P. Gauron and B. Nicolescu, Phys. Lett. B260 (1991) 407;
\\        P. Gauron, L. Lukaszuk and B. Nicolescu, Phys. Lett. B294 (1992) 298.
\bi{Ven}  G. Veneziano, Nucl. Phys. B74 (1974) 365;
          Phys. Lett. 52B (1974) 220;
\\        A. Schwimmer and G. Veneziano, Nucl. Phys. B81 (1974) 445;
\\        M. Ciafaloni, G. Marchesini and G. Veneziano, Nucl. Phys. B98 (1975)
          472; 493.
\bi{Lip2} L.N. Lipatov, Phys. Lett. B251 (1990) 284;  B309 (1993) 394.
\bi{FK}   L.D. Faddeev and G.P. Korchemsky,
          Stony Brook preprint ITP--SB--94--14, Apr 1994 [hep-th/9404173];
          Phys. Lett. B 342 (1995) 311.
\bi{Lip}  L.N. Lipatov, Padova preprint DFPD-93-TH-70, Oct 1993 
          [hep-ph/9311037]; JETP Lett. 59 (1994) 596.
\bi{Q}    R.J. Baxter, {\it Exactly Solved Models in Statistical
          Mechanics\/}, Academic Press, London, 1982.
\bi{BA1}  E.K. Sklyanin, L.A. Takhtajan and L.D.Faddeev,
          Theor. Math. Phys. 40 (1980) 688.
\bi{BA2}  L.A. Takhtajan and L.D. Faddeev, Russ. Math. Survey 34 (1979) 11.
\bi{BA3}  L.D. Faddeev,
          Stony Brook preprint, ITP-SB-94-11, Mar 1994 [hep-th/9404013];
          in Nankai Lectures on Mathematical Physics,
          Integrable Systems, ed. by X.-C.Song, pp.23-70,
          Singapore: World Scientific, 1990.
\bi{BA4}  V.E. Korepin, N.M.Bogoliubov and A.G. Izergin, {\it Quantum
          inverse scattering method and correlation functions\/},
          Cambridge Univ. Press, 1993.
\bi{K}    G.P. Korchemsky, Nucl. Phys. B443 (1995) 255.
\bi{MW}   Z. Maassarani and S. Wallon, Saclay preprint, Saclay-SPhT-95-081,
          Jun 1995 [hep-th/9507056].
\bi{BPZ}  A.A. Belavin, A.M. Polyakov and A.B. Zamolodchikov,
          Nucl. Phys. B241 (1984) 333.
\bi{Sl}   V.P. Spiridonov, private communication.
\bi{GP}   M. Gaudin and V. Pasquier, J. Phys. A25 (1992) 5243.
\bi{Toda} M. Kac and P. Van Moerbeke, Proc. Nat. Acad. Sci. USA 72 (1975) 1627,
          2879;
\\        H. Flaschka and D.W. McLaughlin, Prog. Theor. Phys. 55 (1976) 438.
\bi{Guz}  M.C. Gutzwiller, Ann. Phys. 124 (1980) 347; 133 (1981) 304.
\bi{Skl}  E.K. Sklyanin, {\it The quantum Toda chain\/},
          Lecture Notes in Physics (Springer) 226 (1985) 196;
          {\it Quantum Inverse Scattering Method. Selected Topics\/},
          in ``Quantum Group and Quantum Integrable Systems'' (Nankai
          Lectures in Mathematical Physics), ed. Mo-Lin Ge, Singapore:
          World Scientific, 1992, pp.63--97 [hep-th/9211111].
\bi{as2}  K. Knopp, {\it Infinite sequences and series\/}, Dover Publications,
          New York, 1956.
\bi{as3}  R.B. Dingle, {\it Asymptotic expansion: their derivation and
          interpretation\/}, Academic Press, New York and London, 1973.
\bi{as1}  F.W.J. Olver, {\it Asymptotics and special functions\/},
          Academic Press, New York and London, 1974.
\bi{ZJ}  J. Zinn-Justin, Phys. Rep. 70 (1981) 109.
\bi{ER}  A.V. Efremov and A.V. Radyushkin, Theor. Mat. Phys. 42 (1980) 147.
\bi{M}   Yu.M. Makeenko, Sov. J. Nucl. Phys. 33 (1981) 440.
\bi{Ohr} T. Ohrndorf, Nucl. Phys. B198 (1982) 26.
\bi{Lam} C.S. Lam and M.V. Tratnik, Can. J. Phys. 63 (1985) 1427.
\bi{BDFL}B.J. Brodsky, P. Damgaard, Y. Frishman and G.P. Lepage,
         Phys. Rev. D33 (1986) 1881.
\bi{BF}  V.M. Braun and I.B. Filyanov, Z. Phys. C48 (1990) 239.
\eb
\end{document}